\documentclass[12pt]{article}
\usepackage{pearl-sit}
\begin{document}
\title{An Information Theoretic Measure of
\\
Judea Pearl's Identifiability and Causal Influence
}

\author{Robert R. Tucci\\
        P.O. Box 226\\
        Bedford,  MA   01730\\
        tucci@ar-tiste.com}

\date{\today}
\maketitle
\vskip2cm
\section*{Abstract}
In this paper,
we define a
new
information theoretic
measure that we
call the ``uprooted information".
We show that a necessary
and sufficient condition for
a probability $P(s|do(t))$
to be ``identifiable" (in the sense of Pearl)
in a graph $G$ is that
its uprooted information
be non-negative for all
models of the graph $G$.
In this paper,
we also give a new algorithm
for deciding,
for a Bayesian net that is
semi-Markovian, whether
a probability $P(s|do(t))$
is identifiable,
and, if it is identifiable,
for expressing it without allusions to
confounding variables.
Our algorithm is
closely based on a previous algorithm by Tian
and Pearl, but
seems to
correct a small flaw in theirs.
In this paper,
we also find a {\it necessary and sufficient
graphical condition} for a probability
$P(s|do(t))$ to be identifiable
when $t$ is a singleton set.
So far, in the prior literature,
it appears that
only a {\it sufficient graphical
condition} has been given for this. By ``graphical" we mean
that it is directly
based on Judea Pearl's 3 rules of
do-calculus.

\newpage
\section{Introduction}
For a good textbook on Bayesian networks,
see, for example, the one by Koller and
Friedman, Ref.\cite{KF}.
We will henceforth abbreviate
``Bayesian networks" by ``B-nets".

In a seminal 1995 paper (Ref.\cite{P95}),
Judea Pearl defined
his $do()$ operator. Then he
stated and proved his 3 Rules
of do-calculus. In that
paper, he also
defined for the first time
those probabilities $P(s|do(t))$
that are identifiable (where $s$ and $t$
denote disjoint sets of visible nodes,
for a given
B-net whose
nodes are of two kinds, either visible or
unobserved.)
Pearl also gave
various
examples of identifiable
and non-identifiable probabilities
$P(s|do(t))$.

Identifiable probabilities $P(s|do(t))$
can be expressed as a function
of the probability distribution
$P(v)$ of visible nodes. Call the act
of doing this $P(v)$ expressing
$P(s|do(t))$.

Later on, in Refs.\cite{R290L}
and \cite{R290A},
Tian and Pearl gave
an algorithm for $P(v)$ expressing
any identifiable
$P(s|do(t))$, for a
special type of
B-net called
a semi-Markovian net.
They also consider B-nets that are not semi-Markovian,
but that won't concern us here as
this paper will only deal with
semi-Markovian nets.

Ref.\cite{HV}
by Huang and Valtorta
and Ref.\cite{SP}
by Shpitser and Pearl
have further validated
the algorithm of
Tian and Pearl
by proving that
the 3 rules of do-calculus
are enough to prove
the algorithm.

In this paper,
we define a
new, as far as we know (but
read the comments about
Ref.\cite{Rag} below)
information theoretic
measure that we
call the ``uprooted information".
We show that a necessary
and sufficient condition for
a probability $P(s|do(t))$
to be identifiable
in a graph $G$ is that
its uprooted information
be non-negative for all
models of the graph $G$.

In Ref.\cite{Rag},
Raginsky introduced
an
 information
 theoretic measure
 that he called ``directed
 information" and he
 related it, in a loose way, to
 Pearl's do-calculus.
 In this paper,
 besides the uprooted information,
 we also define a different quantity
 which we call
 the ``information
 loss". Our ``information loss"
is exactly
 equal to Raginsky's  directed-information.
 Thus, the uprooted information and
 Raginsky's  directed-information
 are different quantities,
 although they are related.

 This paper connects
 the fields of information theory and
 Pearl's identifiability
 in a strong way, by means of an if-and-only-if
 theorem,
 whereas Ref.\cite{Rag} by Raginsky
 has very little to say about identifiability.
Ref.\cite{Rag} only mentions
 identifiability in its 5th and last
 section, and there only
 to connect information theory
  with one of the simplest possible examples
 of identifiability,
 what Pearl
 calls the back-door formula.

 Note that Pearl's do-calculus rules
 are a {\it direct
 offshoot} of d-separation.
 The Raginsky paper
 spends most of its time deriving
 some rules that are {\it less} general
 than Pearl's do-calculus rules and
 are not stated in terms of d-separation.
 In fact, the Raginsky paper
 mentions the word ``d-separation"
  for the first time, in italics, in the last
  paragraph of the paper.
Contrary to the Raginsky paper,
 our paper will put Pearl's do-calculus rules
 and d-separation front and center,
 ad-nauseam. In fact,
 this paper contains
more than a dozen d-separation arguments
 with accompanying figures.

In this paper,
we also give a new algorithm
that does the same thing
as the algorithm by Tian and Pearl
that was mentioned above.
Our algorithm is closely
based on the one by Tian
and Pearl, but
seems to correct a small flaw in theirs.
This paper includes 9
examples of B-nets
to which we apply our algorithm. All examples
are placed at the end
of the paper, as appendices.

We also prove (in Section
\ref{sec-tp-fig9} of this paper)
that
an example given in Ref.\cite{R290L}
by Tian and Pearl (viz., the example
illustrated by Fig.9 of Ref.\cite{R290L})
is actually NOT identifiable,
contrary to what Ref.\cite{R290L}
claims! Our algorithm
doesn't get stumped by
this example but
the Tian and Pearl algorithm
apparently does.

We also find a {\it necessary and sufficient
graphical condition} for a probability
$P(s|do(t))$ to be identifiable
when $t$ is a singleton set.
So far, in the prior literature,
it appears that
only a {\it sufficient graphical
condition} has been given
for this. By ``graphical" we mean
that it is directly
based on Judea Pearl's 3 rules of
do-calculus.

In a future paper,
we hope to generalize
the measure of
uprooted information
to quantum mechanics
by using the nowadays
standard prescription
of replacing
probability
distributions by
density matrices.

\section{Some Basic Notation}
\label{sec-basic}

In this section,
we will define some notation that is used throughout the paper.

Ref.\cite{Tuc-intro}
is a short, pedagogical
introduction to Judea Pearl's do-calculus
written by Tucci, the same author as the
present paper.
The reader of the present
paper is expected to
have read Ref.\cite{Tuc-intro} first,
and to be thoroughly familiar
with the notation of that previous paper.

As usual, $\ZZ,\RR, \CC$
will denote the integers, real numbers,
and complex numbers, respectively.
We will sometimes
add superscripts to these
symbols to indicate subsets of
these sets. For instance,
we'll use $\RR^{\geq 0}$
to denote the set of non-negative reals.
For $a,b\in\ZZ$ such that $a\leq b$,
let
 $Z_{a,b}=\{a,a+1,a+2,\ldots, b\}$.

Let $Bool=\{0,1\}$.
Suppose $x,y\in Bool$.
Let $\mynot{x} = 1-x$.
Let $\wedge$ denote AND,
$\vee$ denote OR, and $\oplus$
denote mod 2 addition (a.k.a.
XOR). Hence
\beq
\begin{array}{c|c||c|c|c|c|}
x&y&x+y&x\wedge y&x\vee y&x\oplus y
\\\hline\hline
0&0&0&0&0&0
\\\hline
0&1&1&0&1&1
\\\hline
1&0&1&0&1&1
\\\hline
1&1&2&1&1&0\\
\hline
\end{array}
\;
\eeq
Note that one can express some of these operations
in terms of others. For example, $x\wedge y =xy$,
$x\vee y = x+y-xy=x\oplus y \oplus xy$,
$x\oplus y = x+y-2xy$,
etc.

Suppose we are
given a set $(a_j)_{\forall j\in S}$.
If $T\subset S$, we will
sometimes use $a_{T}$
to
denote the
set
$(a_j)_{\forall j\in T}$.
For example,
$a_{1,2,3}=(a_1,a_2,a_3)$.
If $a.=(a_1,a_2,a_3,\ldots a_N)$,
and $j\in Z_{1,N}$,
let
$a_{<j}= (a_1,a_2,\ldots,a_{j-1})$,
$a_{\leq j}= (a_1,a_2,\ldots,a_{j})$.
$a_{>j}$ and $a_{\geq j}$
are defined
in the obvious way.

 Let
$\delta^x_{y}=\delta(x,y)$
denote the Kronecker delta function:
it equals 1 if $x=y$ and 0 if $x\neq y$.

In cases where $f(x)$
is a complicated expression of $x$,
we will often use the abbreviation

\beq
\frac{f(x)}{\sum_x num} = \frac{f(x)}{\sum_x f(x)}
\;.
\eeq

Random variables will be denoted
by underlined letters; e.g.,
$\rva$.
The (finite) set of values (a.k.a. states) that
$\rva$ can assume will be denoted
by $S_\rva$. Let $N_\rva=|S_\rva|$.
The probability that
$\rva=a$ will be denoted by $P(\rva=a)$
or $P_\rva(a)$, or simply by $P(a)$
if the latter will not lead to confusion
in the context it is being used.

Given a known probability
distribution $\{P(x)\}_{\forall x\in S_\rvx}$, we will
use the following shorthand
to denote the $P(x)$-weighted average of a
function $f(x)$:

\beq
\av{f(x)}_x = \sum_{x \in S_\rvx} P(x) f(x)
\;.
\eeq
In cases where
we are dealing with
several probability
distributions
$\{P(x)\}_{\forall x\in S_\rvx}$ and $\{Q(x)\}_{\forall x\in S_\rvx}$, and we want to make
clear which one
 of them we are averaging over,
we might replace $\av{f(x)}_x$
by the more explicit notations
$\av{f(x)}_{P(x)}$ or $\av{f(x)}_{P_\rvx}$.

Given two probability distributions
$\{P(x)\}_{\forall x\in S_\rvx}$ and $\{Q(x)\}_{\forall x\in S_\rvx}$,
the relative entropy of $P(x)$
over $Q(x)$ is defined
as

\beq
D(P(x)//Q(x))_{\forall x\in S_\rvx}=
\sum_{x\in S_\rvx} P(x)\ln \frac{P(x)}{Q(x)}
\;.
\eeq

Consider a graph $G$ with nodes $\rvx.$.
Suppose $\rvb.\subset\rv{B}.
\subset \rvc.\subset \rvx.$.
Using notation
 which we used previously
 in Ref.\cite{Tuc-intro},
when $\rv{B}.$
 contains $\rvb.$
 and all the ancestors of
 $\rvb.$
 in the graph $G_{\rvc.}$,
 we write\footnote{The line over ``an"
in Eq.(\ref{eq-over-under-line}) means that
the set $\rv{B}.$
includes $\rvb.$ and the line under ``an" means that
the set $\rv{B}.$ is a
random variable.}

\beq
\rv{B}. = \ol{\rv{an}}(\rvb., G_{\rvc.})
\;.
\label{eq-over-under-line}
\eeq
In this paper,
we will say that $\rv{B}.$ is an
{\bf ancestral set} in $G_{\rvc.}$ if

\beq
\rv{B}. = \ol{\rv{an}}(\rv{B}., G_{\rvc.})
\;.
\eeq

Given a B-net with
nodes $\rvx.=(\rvx_1,\rvx_2,\ldots,\rvx_N)$,
suppose

\beq
\rvx_{j(N)}\larrow\ldots\rvx_{j(2)}\larrow\rvx_{j(1)}
\;
\label{eq-top-order-jfun}
\eeq
is a {\bf topological ordering} (top-ord) of $\rvx.$. Therefore,
$j(\cdot):Z_{1,N}\rarrow Z_{1,N}$ is a permutation map. The
argument of $j(\cdot)$ labels time.
Hence,
$j(2)$ occurs after or concurrently with $j(1)$,
$j(3)$ occurs after or concurrently with $j(2)$,
and so on.
We will set $j(t)=\av{t}$
and represent Eq.(\ref{eq-top-order-jfun})
by

\beq
\rvx\av{N}\larrow\ldots\rvx\av{2}\larrow\rvx\av{1}
\;
\eeq
or just by $\{\rvx{\av{t}}\}_{\forall t}$.
Likewise, if $\rva.\subset \rvx.$,
we will
represent a top-ord of $\rva.$ by
$\{\rva{\av{t}}\}_{\forall t}$.

The Pauli matrices will be denoted by

\beq
\sigma_X=
\left[
\begin{array}{cc}
0 & 1\\
1 & 0
\end{array}
\right]
\;
\;,\;\;
\sigma_Y=
\left[
\begin{array}{cc}
0 & -i\\
i & 0
\end{array}
\right]
\;
\;,\;\;
\sigma_Z=
\left[
\begin{array}{cc}
1 & 0\\
0 & -1
\end{array}
\right]
\;.
\eeq
We will also have occasion to use the following 2X2
matrix, which we call the
averaging matrix:

\beq
\cala = \frac{1}{2}
\left[
\begin{array}{cc}
1 & 1\\
1 & 1
\end{array}
\right]
\;.
\eeq

If we define $\Omega$ to be the following
orthogonal matrix (real space rotation)

\beq
\Omega= \frac{1}{\sqrt{2}}
\left[
\begin{array}{cc}
1 & 1\\
-1 & 1
\end{array}
\right]
=
e^{i\frac{\pi}{4}\sigma_Y}
\;,
\eeq
then $\cala$ can be
diagonalized as follows:

\beq
\cala = \Omega
\left[
\begin{array}{cc}
0 & 0\\
0 & 1
\end{array}
\right]
\Omega^T
\;.
\eeq
More generally,
if we consider the
effect of
$\Omega(\cdot)\Omega^T$ on
$
\left[
\begin{array}{cc}
c&f\\
g&(1+d)
\end{array}
\right]
$ where $c,d,f,g\neq 0$,
we get

\beq
\Omega
\left[
\begin{array}{cc}
c&f\\
g&(1+d)
\end{array}
\right]
\Omega^T
=
(1+d)\cala
+
\frac{1}{2}
\left[
\begin{array}{cc}
f+g&f-g\\
g-f&-(f+g)
\end{array}
\right]
+
\frac{c}{2}
\left[
\begin{array}{cc}
1&-1\\
-1&1
\end{array}
\right]
\;.
\eeq

$\cala$ is obviously
real, Hermitian and a projector
($\cala^2 = \cala$).
It projects $\sigma_X$ to
itself and the other two Pauli matrices to zero:

\beq
\cala
\left\{
\begin{array}{c}
\sigma_X\\
\sigma_Y\\
\sigma_Z
\end{array}
\right\}
\cala
=
\left\{
\begin{array}{c}
\sigma_X\\
0\\
0
\end{array}
\right\}
\;.
\eeq

\section{Visible and Unobserved Variables,
\\ Identifiability}

In this section, we will define
what Judea
Pearl calls ``identifiability"
of a quantity associated
with a B-net.
To define identifiability,
we first have to partition
the nodes of a B-net
into visible and unobserved
ones.

Recall our notation from Ref.\cite{Tuc-intro}.
A B-net with graph $G$ and nodes
$\rvx.$ has a full probability distribution
\beq
P(x.)=
\prod_j P(x_j|pa(\rvx_j))
\;.
\label{eq-def-bnet}
\eeq

Henceforth, we will refer to
all B-nets with the same graph
$G$  but different
 probability distributions $P_{\rvx.}$,
as different {\bf models of $G$}.
Let $\calp(G)$ be
the set of all $P_{\rvx.}$
that can be assigned to a graph $G$.
$\calp(G)$ will be called the
set of possible models for $G$.

Assume that $\rvx.$ equals the union of two disjoint
sets $\rvu.$ and $\rvv.$. We will call the $\rvu.$
the unobserved or hidden or confounding
variables.
We will call the $\rvv.$ the visible or observed variables.

A function $F(x.)$ (for instance,
$F(x.)=P(s.|\myhat{t}.)$)
is said to be {\bf identifiable
or $P_{\rvv.}$ expressible} if
it can be expressed as a function of
$P_{\rvv.}=\{P(v.)\}_{\forall v.}$. Equivalently,
$F(x.)$ is identifiable if
for any two
probability distributions $P^{(1)}(x.)$ and
$P^{(2)}(x.)$ for the same graph $G$,

\beq
(\forall v.)(P^{(1)}(v.)=P^{(2)}(v.))
\implies
(\forall x.)(F^{(1)}(x.)=F^{(2)}(x.))
\;.
\label{eq-def-ident12}
\eeq
If we define
$\delta P(v.)= P^{(1)}(v.)-P^{(2)}(v.)$
and
$\delta F(x.)=
F^{(1)}(x.)-F^{(2)}(x.)$,
then Eq.(\ref{eq-def-ident12}) can be written as

\beq
(\forall v.)(\delta P(v.)=0)
\implies
(\forall x.)(\delta F(x.)=0)
\;.
\eeq
Henceforth, if a quantity $F(x.)$
is identifiable in $G$,
we will refer to the act of
calculating an expression for
it as a function of $P_{\rvv.}$
as {\bf $P_{\rvv.}$ expressing $F(x.)$}.

\begin{claim}\label{cl-inherit-id}
(Lemma 13 in Ref.\cite{R290L})
Suppose $G$ is a subgraph of graph $G^+$.
Let graph $G$ (resp., $G^+$)
have nodes $\rvx.=(\rvv.,\rvu.)$
(resp., $\rvx.^+=(\rvv.^+,\rvu.^+)$).
Suppose $\rvs.$ and $\rvt.$
are disjoint subsets of $\rvv.$.
Then
\beq
P(s.|\myhat{t}.)\mbox{ is identifiable in }G^+
\implies
P(s.|\myhat{t}.)\mbox{ is identifiable in }G
\;
\eeq
or, equivalently,
\beq
P(s.|\myhat{t}.)\mbox{ is not identifiable in }G
\implies
P(s.|\myhat{t}.)\mbox{ is not identifiable in }G^+
\;.
\eeq
In other words,
the identifiability
of $P(s.|\myhat{t}.)$ in a graph $G$ is
inherited by the
sub-graphs of $G$
(whereas un-identifiability
is inherited by super-graphs).
\end{claim}
\proof

Suppose we are given
models $P^{(1)}_G$, $P^{(2)}_G$ for graph $G$
such that
\beq
(\forall v.)(\delta P_G(v.)=0)\mbox{ and }
(\exists (s.,t.))(\delta P_G(s.|\myhat{t}.)\neq 0)
\;.
\label{eq-delta-probs-pre}
\eeq
For each $\lambda\in\{1,2\}$,
define model $P^{(\lambda)}_{G^+}$
by setting

\beq
P^{(\lambda)}_{G^+}(x_j|pa(\rvx_j, G^+))=
\left\{
\begin{array}{l}
P^{(\lambda)}_{G}(x_j|pa(\rvx_j, G))
\mbox{ if $\rvx_j$ is old node; i.e., if}
\rvx_j\in G
\\
\delta_{x_j}^0
\mbox{ if  $\rvx_j$ is new node; i.e., if }\rvx_j\in G^+-G
\end{array}
\right.
\;.
\eeq
Since the new nodes are always constant, frozen
at the same state,
and there are no arrows between the new
and old nodes,
we can conclude from
Eq.(\ref{eq-delta-probs-pre}) that

\beq
(\forall v.^+)(\delta P_{G^+}(v.^+)=0)\mbox{ and }
(\exists (s., t.))(\delta P_{G^+}(s.|\myhat{t}.)\neq 0)
\;.
\eeq
\qed

\begin{claim}\label{cl-pruning-g}
$P(s.|\myhat{t}.)$ is identifiable
in $G_{\rvv.}$ if and only if
$P(s.|\myhat{t}.)$ is identifiable
in $G_{\rvv.^-}$ where
$\rvv.^-= \ol{\rv{an}}(\rvs.\cup\rvt.,G_{\rvv.})$
\end{claim}
\proof

\beqa
P(s.|\myhat{t}.)&=&
\underbrace{\sum_{v.-s.}}_{
\sum_{v.^- -s.}\sum_{v.-v.^-}}
\av{\prod_{j:\rvv_j\in \rvv.-\rvt.}
P(v_j|pa(\rvv_j,G_{\rvv.}),u.)
}_{u.}
\label{eq-gv}
\\
&=&
\sum_{v.^- -s.}
\av{\prod_{j:\rvv_j\in \rvv.^- -\rvt.}
P(v_j|pa(\rvv_j,G_{\rvv.^-}),u.)
}_{u.}
\label{eq-gv-minus}
\;.
\eeqa
Note that
Eq.(\ref{eq-gv-minus})
is identical to
Eq.(\ref{eq-gv})
except that $\rvv.$ is replaced
by $\rvv.^-$. Going from Eq.(\ref{eq-gv}) to
Eq.(\ref{eq-gv-minus}) is possible
because none of the $P(v_j|.)$
factors
make any allusion to $v.-v.^-$
in their ``second compartment",
the one for parents.
\qed

\begin{claim}(Lemma 2 in Ref.\cite{R290L})
When $t.=t$ is a
singleton, the previous claim is true
with
$\rvv.^-$ replaced by
$\ol{\rv{an}}(\rvs.,G_{\rvv.})$.
\end{claim}
\proof

Either $\rvt\in \ol{\rv{an}}(\rvs.,G_{\rvv.})$,
in which case
$\ol{\rv{an}}(\rvs.\cup \rvt,G_{\rvv.})=
\ol{\rv{an}}(\rvs.,G_{\rvv.})$,
or
$\rvt\notin \ol{\rv{an}}(\rvs.,G_{\rvv.})$,
in which case
$P(s.|\myhat{t})=P(s.)$
and is thus identifiable
in both
$G_{\ol{\rv{an}}(\rvs.\cup \rvt,G_{\rvv.})}$
and
$G_{
\ol{\rv{an}}(\rvs.,G_{\rvv.})}$.
\qed

\begin{claim}
\label{cl-cond-vminus-c-hat}
$P(s.|\myhat{t}.)=P(s.|\myhat{t}., (v.^-)^{c\wedge})$
, where
$\rvv.^- = \ol{\rv{an}}(\rvs.\cup\rvt.,G_{\rvv.})$
and $(\rvv.^-)^{c\wedge}=
(\rvv.-\rvv.^-)^\wedge$.
\end{claim}
\proof

\begin{figure}[h]
    \begin{center}
    \epsfig{file=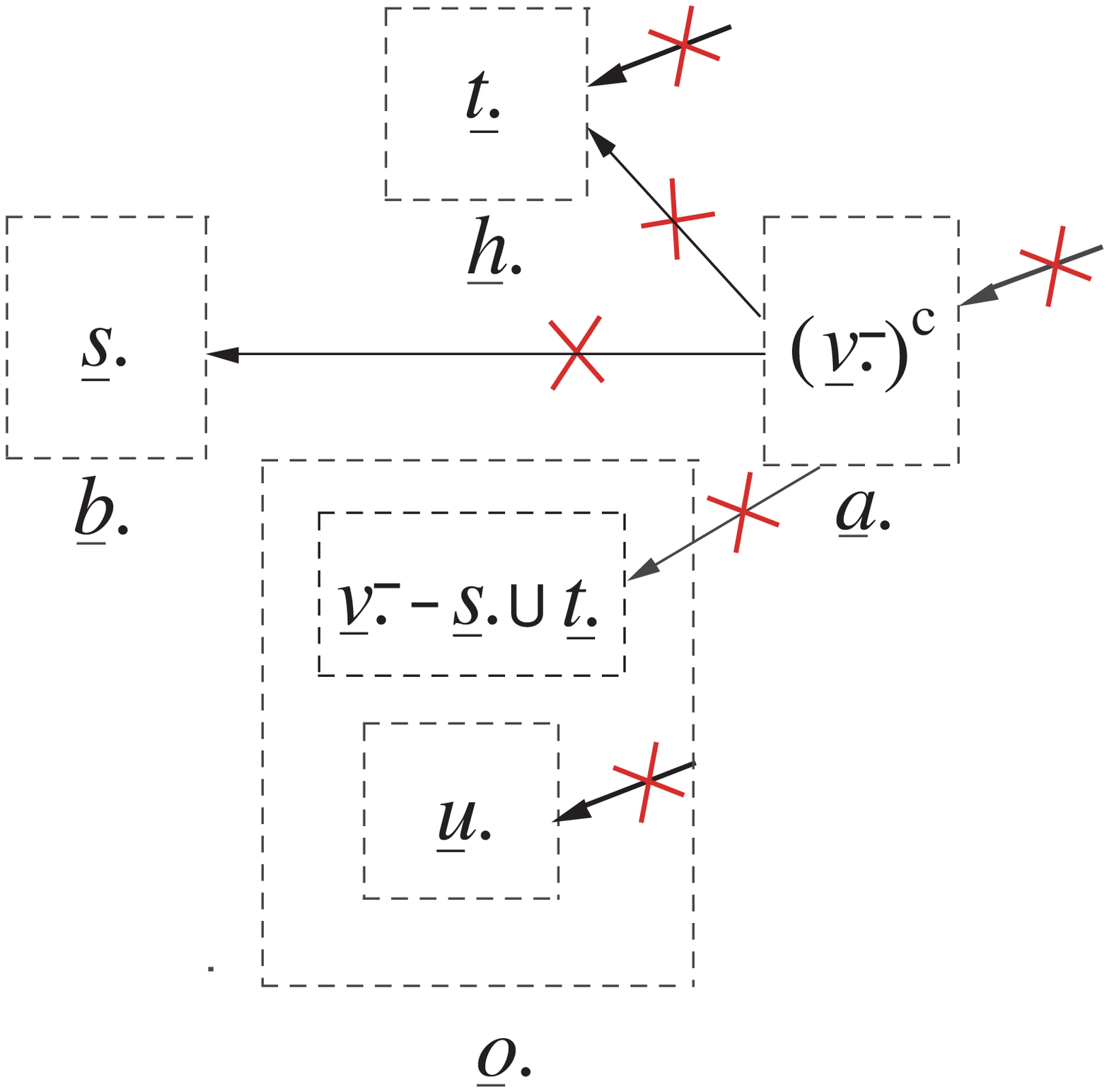, height=2.3in}
    \caption{
    A portrait of $G_{\myhat{\rvh}.,\myhat{\rva}.}$,
    alluded to in Claim \ref{cl-cond-vminus-c-hat}
    }
    \label{fig-vminus}
    \end{center}
\end{figure}

See Ref.\cite{Tuc-intro}
where the 3 Rules
of Judea Pearl's do-calculus
are stated. Using the notation there,
let
$
\rvb.=\rvs,\rva.=(\rvv.^-)^c,\rvh.=\rvt.,
\rvi.=\emptyset,\rvo.=(\rvu.,\rvv.^--\rvs.\cup\rvt.)$.
Note that
$\rva.^-=\rva.-
\rv{an}(\rvi.,G_{\myhat{\rvh}.})=\rva.$ so
$G_{\myhat{\rvh}.,(\rva.^-)^\wedge}=
G_{\myhat{\rvh}.,\myhat{\rva}.}$.
Fig.\ref{fig-vminus} portrays
$G_{\myhat{\rvh}.,\myhat{\rva}.}$.
Apply Rule 3 to that figure.
\qed

\section{Uprooted Information}

In this section,
we will define what we call
an uprooted information, and various
associated quantities.
In later sections, we will
show that there is an intimate
connection between uprooted
information and identifiability.

Throughout this section,
let $\rvb.,\rva.$ and $\rve.$
be disjoint subsets of the set
$\rvx.$ of nodes of a graph $G$.
We will use
the following abbreviations:
$I=$ information,
$M=$ mutual,
$C=$ conditional
and $\wedge=$ uprooted.
Thus, for instance, ``$\wedge CMI$"
will stand for
``uprooted conditional mutual information".

For all $a.,b., e.$, we define
\beq
\begin{array}{|c|c|}
\hline
\begin{array}{c}
(C\mbox{ Probability })\\
P(b.|a.)
\end{array}
&
\begin{array}{c}
(\wedge\mbox{ Probability })\\
P(b.|\myhat{a}.)
\end{array}
\\
\hline
\begin{array}{c}
(MI)\\
P(b.:a.)
=\frac{P(b.|a.)}{P(b.)}
\end{array}
&
\begin{array}{c}
(\wedge MI)\\
P(b.:\myhat{a}.)
=\frac{P(b.|\myhat{a}.)}{P(b.)}
\end{array}
\\
\hline
\begin{array}{c}
(CMI)\\
P(b.:a.|e.)
=\frac{P(b.|a.,e.)}{P(b.|e.)}
\end{array}
&
\begin{array}{c}
(\wedge CMI)\\
P(b.:\myhat{a}.|e.)
=\frac{P(b.|\myhat{a}.,e.)}{P(b.|e.)}
\end{array}
\\
\hline
\end{array}
\;.
\label{eq-defs-p-infos}
\eeq
For the case of
$\wedge$ CMI, recall from Ref.\cite{Tuc-intro} that
$P(b.|\myhat{a}.,e.)=\frac{P(b.,e.|\myhat{a}.)}
{P(e.|\myhat{a}.)}$. We also define what
we call ``losses" as follows:

\beq
\begin{array}{|c|}
\hline
\begin{array}{c}
(\wedge MI\mbox{ loss })\\
\frac{P(b.:a.)}{P(b.:\myhat{a}.)}
\end{array}
\\
\hline
\begin{array}{c}
(\wedge CMI\mbox{ loss })\\
\frac{P(b.:a.|e.)}{P(b.:\myhat{a}.|e.)}
\end{array}
\\
\hline
\end{array}
\;.
\label{eq-defs-h-infos}
\eeq

We will
also refer by the {\it same}
name to the weighted
averages (over $P(x.)$)
of the quantities defined
in Eqs.(\ref{eq-defs-p-infos})
and (\ref{eq-defs-h-infos}),
as long as it is clear
from context which of
the two we are referring to.
So define

\beq
\begin{array}{|c|c|}
\hline
\begin{array}{c}
(C \mbox{ Entropy })\\
H(\rvb.|\rva.)=\av{1/P(b.|a.)}_{a.,b.}
\end{array}
&
\begin{array}{c}
(\wedge \mbox{ Entropy })\\
H(\rvb.|\myhat{\rva}.)=
\av{1/P(b.|\myhat{a}.)}_{a.,b.}
\end{array}
\\
\hline
\begin{array}{c}
(MI)\\
H(\rvb.:\rva.)=\av{P(b.:a.)}_{a.,b.}
\end{array}
&
\begin{array}{c}
(\wedge MI)\\
H(\rvb.:\myhat{\rva}.)=
\av{P(b.:\myhat{a}.)}_{a.,b.}
\end{array}
\\
\hline
\begin{array}{c}
(CMI)\\
H(\rvb.:\rva.|\rve.)=
\av{P(b.:a.|e.)}_{a.,b.,e.}
\end{array}
&
\begin{array}{c}
(\wedge CMI)\\
H(\rvb.:\myhat{\rva}.|\rve.)=
\av{P(b.:\myhat{a}.|e.)}_{a.,b.,e.}
\end{array}
\\
\hline
\end{array}
\;,
\eeq
and

\beq
\begin{array}{|c|}
\hline
\begin{array}{c}
(\wedge MI\mbox{ loss })\\
H_{loss}(\rvb.:\myhat{\rva}.)=
\av{\frac{P(b.:a.)}{P(b.:\myhat{a}.)}}_{a.,b.}
\end{array}
\\
\hline
\begin{array}{c}
(\wedge CMI\mbox{ loss })\\
H_{loss}(\rvb.:\myhat{\rva}.|\rve.)=
\av{\frac{P(b.:a.|e.)}{P(b.:\myhat{a}.|e.)}}_{a.,b.,e.}
\end{array}
\\
\hline
\end{array}
\;.
\eeq

As is well known,
$H(\rva.:\rvb.)$
an
$H(\rva.:\rvb.|\rve.)$
must be non-negative. However,
$H(\rva.:\myhat{\rvb.})$
(and thus
$H(\rva.:\myhat{\rvb}.|\rve.$ too)
can be negative. For
example,
in Section \ref{sec-2-3-nodes},
we give a graph that we call
INDEF and a model for
that graph such that
$H(\rvy:\myhat{\rvx})=-\ln(2)$.
On the other hand, what we call losses
are always non-negative
because they can be expressed
as weighted averages of
relative entropies. Indeed,
\beq
H_{loss}(\rvb.:\myhat{\rva}.|\rve.)=
\sum_{a.,e.}P(a.,e.)
D[P(b.|a.,e.)//P(b.|\myhat{a}.,e.)]_{\forall b.}\geq 0
\;.
\eeq
Note also that the $\wedge$CMI,
, $\wedge$CMI loss and
CMI
are related by

\beq
H(\rvb.:\myhat{\rva}.|\rve.)
+
\underbrace{H_{loss}(\rvb.:\myhat{\rva}.|\rve.)}_{\geq 0}
=
\underbrace{H(\rvb.:\rva.|\rve.)}_{\geq 0}
\;.
\label{eq-hhat-hloss-h}
\eeq

\begin{claim}\label{cl-zero-h}
For any graph $G$, there exists
a model of $G$ such that $H(\rvs.:\myhat{\rvt}.)=H(\rvs.:\rvt.)=0$.
\end{claim}
\proof

Consider any model of $G$
that satisfies:
for all $j$ such that
$\rvx_j\in \rvx.$, $P(x_j|pa(\rvx_j))=P(x_j|pa(\rvx_j)-\rvt.)$.
For such a model,
all arrows exiting all nodes
in node set $\rvt.$
can be erased.
 Hence,
$H(\rvs.:\myhat{\rvt}.)=H(\rvs.:\rvt.)=0$.
\qed

The sign of the uprooted information
$H(\rvs.|\myhat{\rvt}.)$
obeys the following
simple inheritance
property
analogous to
the inheritance property
(Claim \ref{cl-inherit-id})
for identifiability.

\begin{claim}\label{cl-inherit-hminus}
Suppose $G$ is a subgraph of graph $G^+$.
Let graph $G$ (resp., $G^+$)
have nodes $\rvx.=(\rvv.,\rvu.)$
(resp., $\rvx.^+=(\rvv.^+,\rvu.^+)$).
Suppose $\rvs.$ and $\rvt.$
are disjoint subsets of $\rvv.$.
Let $P_G$ (resp., $P_{G^+}$)
denote a model for $G$ (resp., $G^+$).
Then
\beq
(\forall P_{G^+})
( H_{P_{G^+}}(\rvs.|\myhat{\rvt}.)\geq 0)
\implies
(\forall P_{G})
( H_{P_{G}}(\rvs.|\myhat{\rvt}.)\geq 0)
\;
\eeq
or, equivalently,
\beq
(\exists P_{G})
( H_{P_G}(\rvs.|\myhat{\rvt}.)< 0)
\implies
(\exists P_{G^+})
( H_{P_{G^+}}(\rvs.|\myhat{\rvt}.)< 0)
\;
\eeq
\end{claim}
\proof

Suppose we are given a
$G$ model $P_{G}(x.)$
such that
$H_{P_G}(\rvs.|\myhat{\rvt}.)< 0$.
Define a $G^+$ model $P_{G^+}(x.)$
by setting

\beq
P_{G^+}(x_j|pa(\rvx_j, G^+))=
\left\{
\begin{array}{l}
P_{G}(x_j|pa(\rvx_j, G))
\mbox{ if  $\rvx_j$ is old node; i.e., if}
\rvx_j\in G
\\
\delta_{x_j}^0
\mbox{ if  $\rvx_j$ is new node; i.e., if }\rvx_j\in G^+-G
\end{array}
\right.
\;.
\eeq
Since the new nodes are always frozen
at the same state,
and
all the arrows between the old
and new nodes can be erased,
we can conclude that
$H_{P_{G^+}}(\rvs.|\myhat{\rvt}.)< 0$.
\qed

Even though $H(\rvb_2,\rvb_1:\rva.)\geq
H(\rvb_1:\rva.)$, note that

\beq
\mbox{ Not true: } H(\rvb_2,\rvb_1:\myhat{\rva}.)\geq H(\rvb_1:\myhat{\rva}.)
\;.
\eeq
For example, for
the graph of Fig.\ref{fig-frontdoor},
$H(\rvy,\rvz:\myhat{\rvx})$
is not identifiable so
it is negative for some models
of the graph. However, for the same graph, the frontdoor formula proven in
Section \ref{sec-frontdoor}
implies that $H(\rvy:\myhat{\rvx})\geq 0$
for all models of the graph.

\section{Uprooted Information of
2 and 3 Node Graphs}\label{sec-2-3-nodes}
In this section, we will consider
the uprooted information
$H(\rvy:\myhat{\rvx})$
where $\rvy$ and $\rvx$
are two of the nodes of
a graph that has a
total number of either 2 or 3 nodes.
These are trivial examples, but
I find them instructive.
For one thing, they illustrate
the connection
between the identifiability
of $P(y|\myhat{x})$ and the sign
of $H(\rvy:\myhat{\rvx})$.

\begin{figure}[h]
    \begin{center}
    \epsfig{file=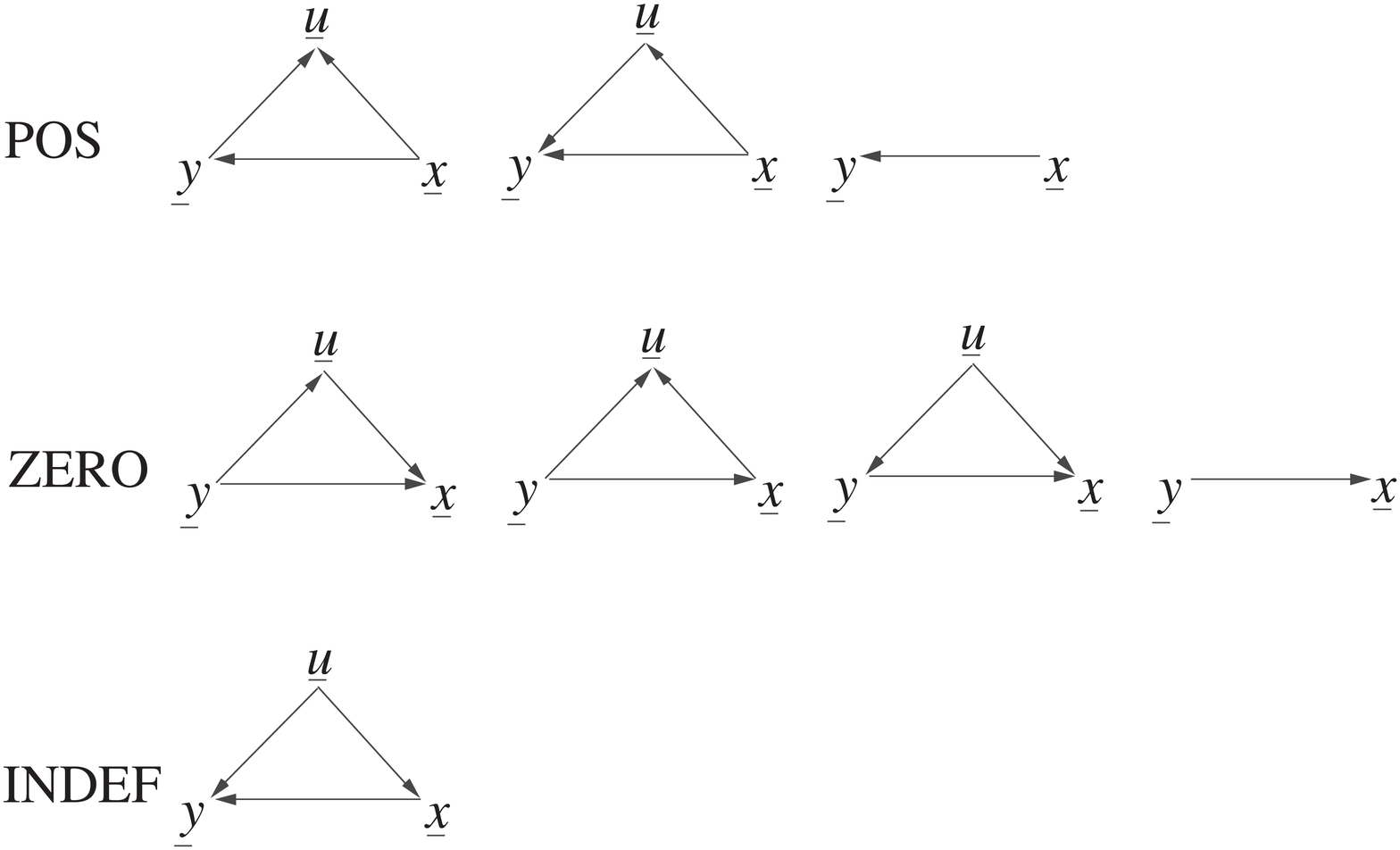, height=2.75in}
    \caption{3 sets of graphs with 2 or 3 nodes
    that are considered in
    Section \ref{sec-2-3-nodes}.
    }
    \label{fig-2-3-nodes}
    \end{center}
\end{figure}

Fig.\ref{fig-2-3-nodes}
 defines 3 graph sets that I call
POS, ZERO and INDEF.

\begin{claim}\label{eq-pos-type-h}
For graphs of type
POS defined in Fig.\ref{fig-2-3-nodes},
$H(\rvy:\myhat{\rvx})=H(\rvy:\rvx)$.
\end{claim}
\proof

\begin{figure}[h]
    \begin{center}
    \epsfig{file=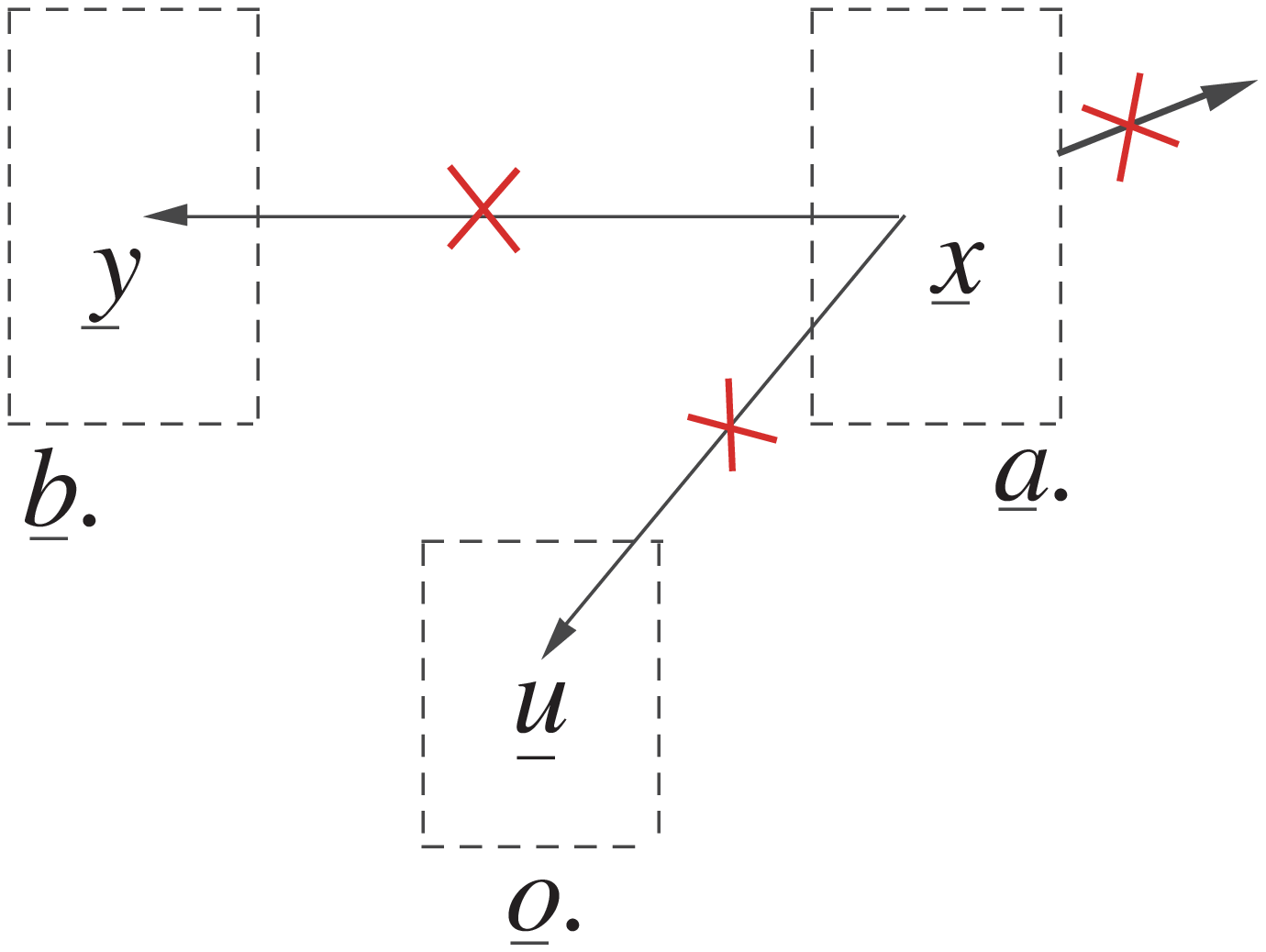, height=1.5in}
    \caption{
    A portrait of $G_{\myhat{\rvh}.,\myvee{\rva}.}$,
    alluded to in Claim \ref{eq-pos-type-h}
    }
    \label{fig-pos}
    \end{center}
\end{figure}

We want to prove that $P(y|\myhat{x})=P(y|x)$.
See Ref.\cite{Tuc-intro}
where the 3 Rules
of Judea Pearl's do-calculus
are stated. Using the notation there,
let
$\rvb.=\rvy,\rva.=\rvx,\rvh.=\emptyset,
\rvi.=\emptyset,\rvo.=\rvu$.
Fig.\ref{fig-pos} portrays
$G_{\myhat{\rvh}.,\myvee{\rva}.}$.
Apply Rule 2 to that figure.
\qed

\begin{claim}\label{eq-zero-type-h}
For graphs of type
ZERO defined in Fig.\ref{fig-2-3-nodes},
$H(\rvy:\myhat{\rvx})=0$.
\end{claim}
\proof

\begin{figure}[h]
    \begin{center}
    \epsfig{file=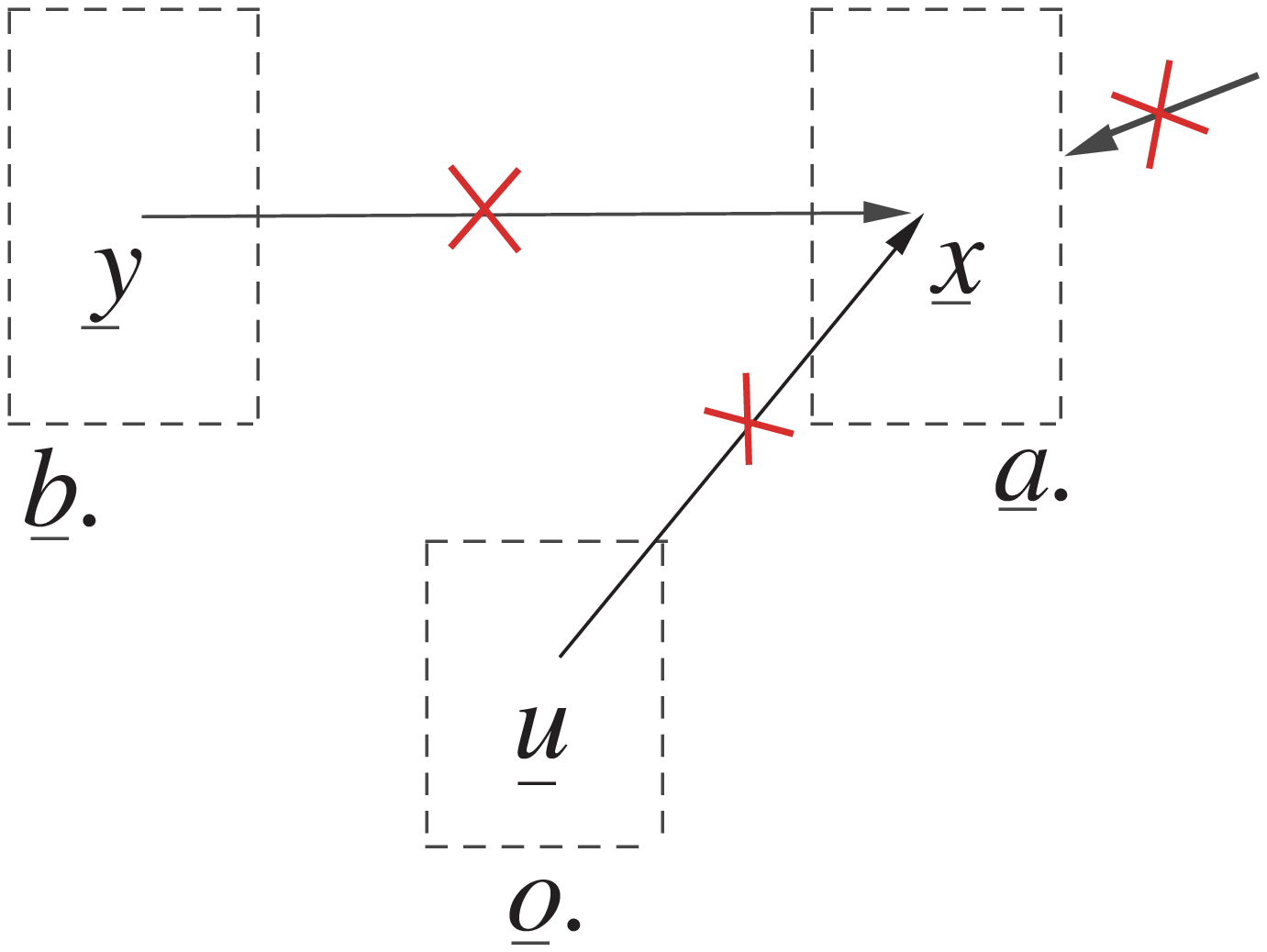, height=1.5in}
    \caption{
    A portrait of $G_{\myhat{\rvh}.,\myhat{\rva}.}$,
    alluded to in Claim \ref{eq-zero-type-h}
    }
    \label{fig-zero}
    \end{center}
\end{figure}

We want to prove that $P(y|\myhat{x})=P(y)$.
See Ref.\cite{Tuc-intro}
where the 3 Rules
of Judea Pearl's do-calculus
are stated. Using the notation there,
let
$
\rvb.=\rvy,\rva.=\rvx,\rvh.=\emptyset,
\rvi.=\emptyset,\rvo.=\rvu$.
Note that
$\rva.^-=\rva.-
\rv{an}(\rvi.,G_{\myhat{\rvh}.})=\rva.$ so
$G_{\myhat{\rvh}.,(\rva.^-)^\wedge}=
G_{\myhat{\rvh}.,\myhat{\rva}.}$.
Fig.\ref{fig-zero} portrays
$G_{\myhat{\rvh}.,\myhat{\rva}.}$.
Apply Rule 3 to that figure.
\qed

\begin{claim}\label{eq-indef-type-h}
Let $I=\left[-H(\rvx|\rvy),H(\rvy:\rvx)\right]$.
For graphs of type
INDEF defined in Fig.\ref{fig-2-3-nodes},
and for every $a\in I$,
there exists a model
with $H(\rvy:\myhat{\rvx}.)=a$.
Note that
the lower endpoint of $I$ is
$ -H(\rvx|\rvy)=H(\rvy:\rvx)-H(\rvx)$.
\end{claim}
\proof

$H(\rvy:\myhat{\rvx})\leq H(\rvy:\rvx)$ follows
immediately from Eq.(\ref{eq-hhat-hloss-h}).
To prove the lower bound
on $H(\rvy:\myhat{\rvx})$, note that

\beq
P(y|\myhat{x})=
\sum_{u}P(y|x,u)P(u)
\geq
\sum_{u}P(y|x,u)P(x|u)P(u)
=P(x,y)
\;.
\eeq
Hence

\beq
H(\rvy:\myhat{\rvx})=
\av{\ln(
\frac{P(y|\myhat{x})}{P(y)})
}_{x,y}
\geq
\av{\ln P(x|y)}_{x,y}=-H(\rvx|\rvy)
\;
\eeq

Next we give a model that achieves the
left endpoint of the interval $I$, and another that
achieves the right one.

If for all $x,y,u$, one has
$P(x|u)=P(x)$ and
$P(y|x,u)=P(y|x)$, then
the arrows
between $\rvu$ and $(\rvx,\rvy)$
can be erased, so
the graph INDEF behaves just like the
graph $\rvy\larrow \rvx$, for which
$H(\rvy:\myhat{\rvx})=H(\rvy:\rvx)$.

To get a
model for which $H(\rvy:\myhat{\rvx})=-H(\rvx|\rvy)$
let's
assume $S_\rvu= S_\rvx=S_\rvy= Z_{0,N-1}$.
Let $\oplus$ denote addition mod $N$.
For all $u,x,y\in Z_{0,N-1}$, let

\beq
\left\{
\begin{array}{l}
P(u)=\frac{1}{N}
\\
P(x|u)=\delta_x^u
\\
P(y|x,u)=\delta_y^{x\oplus u}
\end{array}
\right.
\;.
\eeq
Then

\beq
P(x,y)=\frac{1}{N}\sum_u \delta_y^{x\oplus u} \delta_x^u=
\frac{\delta_y^0}{N}
\;
\eeq
and

\beq
P(y|\myhat{x})=\frac{1}{N}\sum_u \delta_y^{x\oplus u}=
\frac{1}{N}
\;.
\eeq
Hence,

\beqa
H(\rvy:\myhat{\rvx})
&=&\sum_{x,y}
P(x,y)\ln \frac{P(y|\myhat{x})}{P(y)}
\\
&=&
\sum_{x,y}\frac{\delta_y^0}{N}
\ln \frac{\frac{1}{N}}{\delta_y^0}
\\
&=&
-\ln N
\;
\eeqa
and

\beqa
-H(\rvx|\rvy)
&=&\sum_{x,y}
P(x,y)\ln P(x|y)
\\
&=&
\sum_{x,y}\frac{\delta_y^0}{N}
\ln \frac{\frac{1}{N}\delta_y^0}{\delta_y^0}
\\
&=&
-\ln N
\;.
\eeqa
\qed

Eq.(\ref{eq-3graph-types-table})
summarizes in tabular form the results of
the last 3 claims.

\beq
\begin{array}{c||c|c|c|c}
\hline
\begin{array}{c}
\mbox{graph}\\
\mbox{set}
\end{array}\downarrow\setminus H(\rvy:\myhat{\rvx})\rightarrow
&
[-H(\rvx|\rvy),0)
&
0
&
(0,H(\rvy:\rvx))
&
H(\rvy:\rvx)
\\
\hline\hline
POS&&&&\checkmark
\\
\hline
ZERO&&\checkmark&&
\\
\hline
INDEF&\checkmark&\checkmark&\checkmark&\checkmark
\\
\hline
\end{array}
\;\label{eq-3graph-types-table}
\eeq

\begin{claim}
If $\rvv.=(\rvy,\rvx)$ and $\rvu.=\rvu$, then
$P(y|\myhat{x})$ is identifiable
(resp., not identifiable)
for the graphs POS and ZERO (resp., INDEF)
\end{claim}
\proof

In the proof of
Claim \ref{eq-pos-type-h} (resp.,
Claim \ref{eq-zero-type-h}),
we showed that $P(y|\myhat{x})=
P(y|x)$ (resp.,
$P(y|\myhat{x})=P(y)$) so $P(y|\myhat{x})$
is identifiable for the POS (resp., ZERO) graphs.
Claim \ref{cl-id-1tooth}
shows that
$P(y|\myhat{x})$ is not identifiable
for INDEF graphs.
\qed

\section{Semi-Markovian Net, C-components}

In this section,
we define
what Pearl and co-workers
call
a semi-Markovian net
and its associated c-components.
Semi-Markovian nets
are a special type of B-net
for which the theory of identifiability is simpler
than for general B-nets.

A semi-Markovian net is a B-net for
which
the unobserved nodes $\rvu.$
are all root nodes (i.e., have no parents).
Furthermore, for each $j$, $\rvu_j$
has exactly two elements of the set $\rvv.$ as
children. The node $\rvu_j$
and its two outgoing arrows
will be called, as in Ref.\cite{R290L},
a ``bi-directed arc".

For a semi-Markovian net,
Eq.(\ref{eq-def-bnet}) for the $P(x.)$
of a general B-net  reduces to

\beq
P(x.)=
\prod_j\left\{
P(v_j|pa(\rvv_j))\right\}
\prod_k\left\{
P(u_k)\right\}
\;.
\eeq
Therefore, for a semi-Markovian net,

\beq
P(v.)=
\av{
\prod_j
P(v_j|v.\cap pa(\rvv_j),u.\cap pa(\rvv_j))
}_{u.}
\;.
\eeq
Note that $v.\cap pa(\rvv_j)=
pa(\rvv_j, G_{\rvv.})$.

Henceforth, given a set $\rva.\subset \rvv.$
where $\rvv.$ are the visible nodes of graph $G$,
we will use the notations

\beq
[\rva.]^c= \rvv.-\rva.
\;,
\eeq
for the {\bf complement} (in $\rvv.$)
of the set $\rva.$,
and

\beq
\qq{a.}=P(a.|[a.]^{c\wedge})
=P(a.|[v.-a.]^{\wedge})
\;
\eeq
for the {\bf probability of $\rva.$
with uprooted complement}.
This notation is idiosyncratic
to this paper. In Ref.\cite{R290L}, Tian and Pearl
denote $\qq{a.}$ by $Q[\rva.]$.

By the definition of the uprooting operator,

\beq
\qq{a.}
=
\av{\prod_{j:\rvv_j\in\rva.}
P(v_j|pa(\rvv_j, G_{\rvv.}), u.)}_{\rvu.}
\;.
\eeq

Given any two
elements $\rvv_{j_1}$ and
$\rvv_{j_2}$ of $\rvv.$,
we will write $\rvv_{j_1}\sim \rvv_{j_1}$
and say
$\rvv_{j_1}$ and
$\rvv_{j_2}$ are equivalent if
there is an undirected path from
$\rvv_{j_1}$ to
$\rvv_{j_2}$ along arrows
all of which emanate from $\rvu.$
nodes. This is an equivalence relation,
and it partitions $\rvv.$ into
equivalence classes. We will
call
such classes the {\bf c-components} (connected or
confounding components)
of $\rvv.$ and we will denote them by
$\rvvcc{\gamma}$ for
$\gamma=0,1,2,\ldots,N_\rv{\gamma}-1$.
For each $\gamma$,
we can also find a set $(\rvu.)_\gamma\subset \rvu.$
such that $(\rvu.)_\gamma=\rvu.\cap \rv{pa}(\rvvcc{\gamma})$.
Just like the sets
$\{\rvvcc{\gamma}\}_{\forall \gamma}$
give a disjoint partition of $\rvv.$,
the sets
$\{(\rvu.)_\gamma\}_{\forall \gamma}$
give a disjoint partition of $\rvu.$.
Thus, we can write

\begin{subequations}
\label{eq-semi-Markovian}
\beq
\rvv.=\bigcup_{\gamma}\rvvcc{\gamma}
\;,\;\;
\rvu.=\bigcup_{\gamma}(\rvu.)_{\gamma}
\;
\eeq

It is easy to see
that $P(v.)$ can be
expressed as follows, as a product of
factors labeled by
the c-component label $\gamma$:

\beq
P(v.)=
\prod_\gamma
\qq{\vcc{\gamma}}
\;,
\label{eq-pv-is-prod-pvcc}
\eeq
where

\beq
\qq{\vcc{\gamma}}=
\av{
\prod_{j:\rvv_j\in\rvvcc{\gamma}}
P(v_j|pa(\rvv_j, G_{\rvv.}), (u.)_{\gamma})
}_{(u.)_\gamma}
\;.
\eeq
\end{subequations}

We end this section
by proving various properties
of semi-Markovian
nets that are useful in
the theory of identifiability.

\begin{claim}\label{cl-pv-cc}
(Lemma 1 in Ref.\cite{R290L})
Consider a semi-Markovian net so that
Eqs.(\ref{eq-semi-Markovian}) apply.
Suppose $\{\rvv\av{j}\}_{\forall j}$
is a topological ordering of the set $\rvv.$
in the graph $G$.
Let $\rvv.$ have the c-component
decomposition

\beq
\rvv.=\bigcup_{\gamma}\rvvcc{\gamma}
\;.
\eeq
Then

\beq
P(v.)=
\prod_\gamma
\qq{\vcc{\gamma}}
\;
\eeq
where

\beq
\qq{\vcc{\gamma}}=
\prod_{j:\rvv\av{j}\in \rvvcc{\gamma}}
P(v\av{j}|v\av{< j})
\;.
\label{eq-pvcc-top-ord}
\eeq
\end{claim}
\proof

Since the $\rvu.$ are
all root nodes, a top-ord of $G$
is given by

\beq
\rvv\av{|\rvv.|}\larrow
\ldots\larrow
\rvv\av{2}\larrow
\rvv\av{1}\larrow
\rvu\av{|\rvu.|}\larrow
\ldots\larrow
\rvu\av{2}\larrow
\rvu\av{1}
\;.
\eeq
Now remember that
if $\rvx.=(\rvx_1,\rvx_2,\ldots\rvx_N)$
are the nodes of the graph,
and $\{\rvx\av{j}\}_{\forall j}$
is a top-ord of them,
then one can use the chain rule with
conditioning on
past nodes or
one can use it with conditioning on future nodes:

\beqa
P(x.)&=&\prod_{j=1}^N P(\rvx_j|\rvx\av{<j})
\\
&=&\prod_{j=1}^N P(\rvx_j|\rvx\av{>j})
\;
\eeqa
where $\rvx\av{<1}=\rvx\av{>N}=1$.
If we use the chain rule which
conditions on the future nodes,
then we
get

\beq
P(x.)=P(u.|v.)\prod_{j=1}^{|\rvv.|}
P(v\av{j}|v\av{>j})
\;.
\eeq
Summing over $u.$ then gives
\beqa
P(v.)&=&
\prod_{j=1}^{|\rvv.|}
P(v\av{j}|v\av{>j})
\\
&=&
\prod_{j=1}^{|\rvv.|}
P(v\av{j}|v\av{<j})
\;.
\label{eq-pv-top-ord}
\eeqa
Eq.(\ref{eq-pvcc-top-ord}) follows
by applying $\delta_{[\rvv.-\rvvcc{\gamma}]^\wedge}$
to both sides of
Eq.(\ref{eq-pv-top-ord}).
\qed

\begin{claim}\label{cl-ph-cc}
(Lemma 4 in Ref.\cite{R290L})
Consider a semi-Markovian net so that
Eqs.(\ref{eq-semi-Markovian}) apply.
Suppose $\rvh.\subset \rvv.$ and
$\{\rvh\av{j}\}_{\forall j}$
is a topological ordering of the set $\rvh.$
in the graph $G$.
Let $\rvh.$ have the c-component
decomposition

\beq
\rvh.=\bigcup_{\gamma}\rvhcc{\gamma}
\;.
\eeq
Then

\beq
\qq{h.}=
\prod_\gamma
\qq{\hcc{\gamma}}
\;
\eeq
where

\beq
\qq{\hcc{\gamma}}=
\prod_{j:\rvh\av{j}\in \rvhcc{\gamma}}
P(h\av{j}|h\av{< j},h.^{c\wedge})
\;
\eeq
\end{claim}
\proof

Note that
this claim reduces to
Claim \ref{cl-pv-cc}.
when $\rvh.=\rvv.$ because
$\rvv.^c=\emptyset$. The proof of this claim
is very similar to the proof of
Claim \ref{cl-pv-cc}.
\qed

\begin{claim}\label{cl-rule3-ancestral}
(Lemma 3 in Ref.\cite{R290L})
Consider a semi-Markovian net so that
Eqs.(\ref{eq-semi-Markovian}) apply.
Suppose $\rva.\subset \rvc.\subset \rvv.$
and $\rva.$ is ancestral in $G_{\rvc.}$.
Then

\beq
\underbrace{
\sum_{c.-a.}
\qq{c.}
}_{=P(a.|c.^{c\wedge})}
=
\qq{a.}
\;.
\eeq
In particular, if $\rvc.=\rvv.$, then

\beq
\underbrace{\sum_{v.-a.}
P(v.)}_{=P(a.)}
=
\qq{a.}
\;.
\eeq
\end{claim}
\proof

Just note that
\beqa
P(a.|c.^{c\wedge})
&=&\sum_{c.-a.}
\av{
\prod_{j:\rvv_j\in \rvc.-\rva.}
\left\{
P(v_j|pa(\rvv_j))
\right\}
\prod_{j:\rvv_j\in \rva.}
\{
P(v_j|\underbrace{pa(\rvv_j)}_{\subset \rva.\cup \rvu.})
\}
}_{u.}
\\
&=&
\av{
\prod_{j:\rvv_j\in \rva.}
P(v_j|pa(\rvv_j))
}_{u.}
\\
&=&
\qq{a.}
\;.
\eeqa

\begin{figure}[h]
    \begin{center}
    \epsfig{file=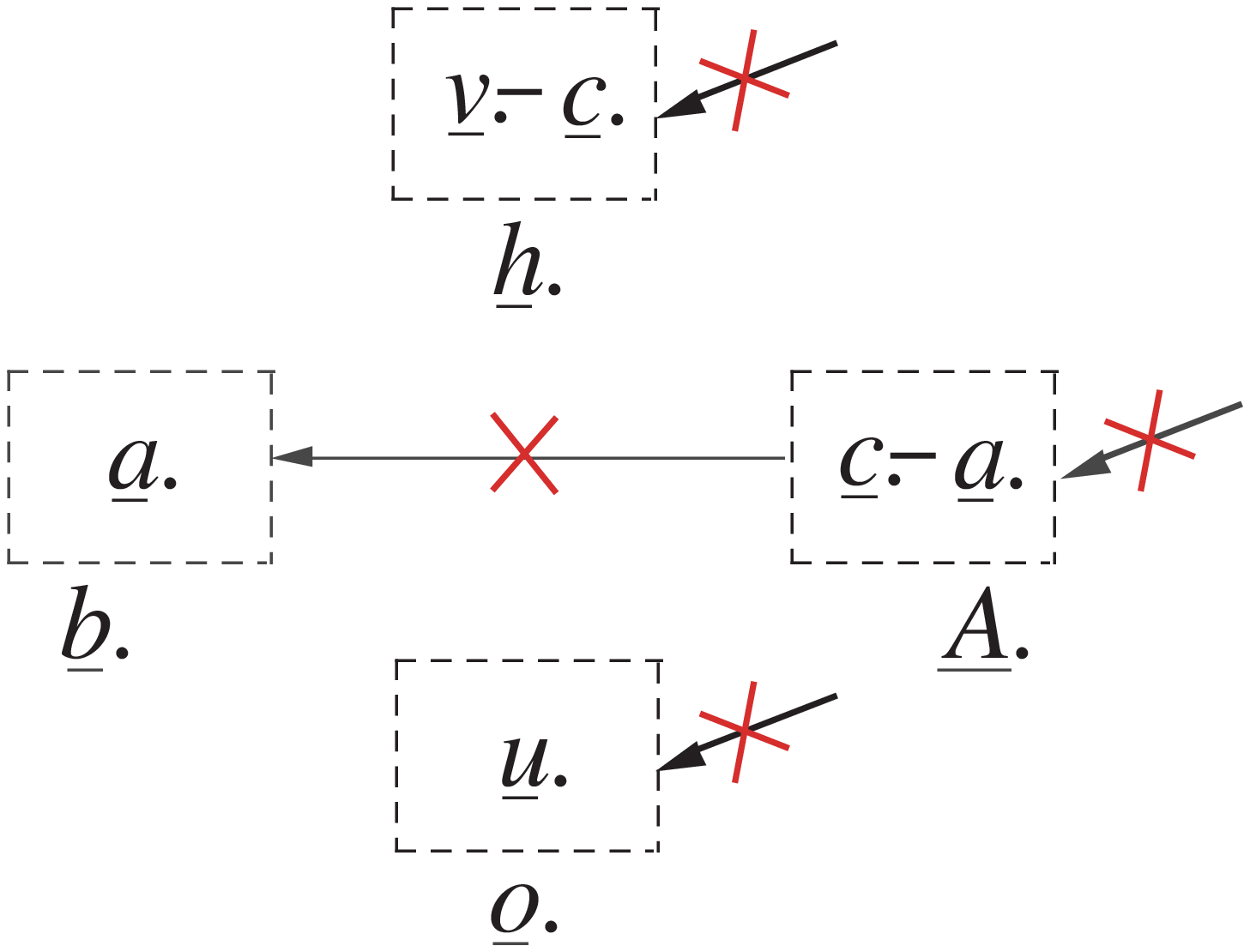, height=1.6in}
    \caption{
    A portrait of
    $G_{\myhat{\rvh}.,\myhat{\rv{A}}.}$
    alluded to in Claim \ref{cl-rule3-ancestral}.
    }
    \label{fig-ancestral}
    \end{center}
\end{figure}

An alternative proof, based on the do-calculus
rules, is as follows. We want to show that

\beq
P(a.|c.^{c\wedge})
=\underbrace{
\qq{a.}
}_{P(a.|(c.-a.)^{\wedge},c.^{c\wedge})}
\;.
\eeq
See Ref.\cite{Tuc-intro}
where the 3 Rules
of Judea Pearl's do-calculus
are stated. Using the notation there,
let
$\rvb.=\rva.$,
$\rv{A}.= \rvc.-\rva.$,
$\rvh.=\rvv.-\rvc.$,
$\rvi.=\emptyset$,
$\rvo.=\rvu.$.
Note that
$\rv{A}.^-=\rv{A}.-
\rv{an}(\rvi.,G_{\myhat{\rvh}.})=\rv{A}.$
so $G_{\myhat{\rvh}.,(\rv{A}.^-)^\wedge}=
G_{\myhat{\rvh}.,\myhat{\rv{A}}.}$.
Fig.\ref{fig-ancestral} portrays
$G_{\myhat{\rvh}.,\myhat{\rv{A}}.}$.
Apply Rule 3 to that figure.
\qed

\section{
$P(s.|\myhat{t}.)$
when $\rvt.=\rvt$ is a singleton}

In Section \ref{sec-algo-singleton},
we will give an algorithm
for $P_{\rvv.}$ expressing
$P(s.|\myhat{t}.)$
where $\rvt.=\rvt$ is a singleton.
In section \ref{sec-iff-singleton},
we will prove that the algorithm
fails iff
$P(s.|\myhat{t})$
is not identifiable in $G$.

Appendices
\ref{app-id} and \ref{app-not-id}
contain
several examples of
graphs and of quantities $P(s.|\myhat{t}.)$
in those graphs,
with $\rvt.=\rvt$ singleton.
In the examples of
Appendix \ref{app-id},
we show that
$P(s.|\myhat{t})$
is identifiable and we proceed to
$P_{\rvv.}$ express it,
using the algorithm given below.
In the examples of
Appendix \ref{app-not-id},
we show that
$P(s.|\myhat{t})$
is not identifiable
by giving two different
models of the graph $G$ that have the
same $P_{\rvv.}$
but different
$P(s.|\myhat{t})$.

\subsection{Algorithm for $P_{\rvv.}$ expressing $P(s.|\myhat{t})$}
\label{sec-algo-singleton}

In this section,
we will give an algorithm
for $P_{\rvv.}$ expressing
$P(s.|\myhat{t}.)$
where $\rvt.=\rvt$ is a singleton.

Suppose $\rvd.$ is the ancestral set
of $\rvs.$ in $G_{\rvv.-\rvt}$ so

\beq
\rvd.= \rv{\ol{an}}(\rvs.,G_{\rvv.-\rvt})
\;,
\eeq
and let

\beq
\rvr.=\rvv.-\rvd.
\;.
\eeq

Note that

\beqa
P(s.|\myhat{t})&=&P(s.|[(v.-t)-d.]^\wedge,\myhat{t})
\label{eq-use-ancestral}
\\
&=&
P(s.|[v.-d.]^{\wedge})
\\
&=& P(s.|d.^{c\wedge})
\label{eq-pst-pd}
\;,
\eeqa
where Eq.(\ref{eq-use-ancestral}) follows from
Claim \ref{cl-rule3-ancestral}.

Let $\rvv.=\bigcup_\gamma \rvvcc{\gamma}$
be the c-component decomposition of $\rvv.$
in $G_{\rvv.}$. For each $\gamma$,
let

\beq
(\rvd.)_{\gamma}=\rvd.\cap\rvvcc{\gamma}
\;
\eeq
and

\beq
(\rvr.)_\gamma =
\rvr.\cap\rvvcc{\gamma}=
\rvvcc{\gamma} - (\rvd.)_\gamma
\;.
\eeq
Note that
$\rvd.=\bigcup_\gamma (\rvd.)_\gamma$
and the
$(\rvd.)_\gamma$ are mutually disjoint
but they are not c-components. That's
why we denote them as $(\rvd.)_\gamma$
instead of $(\rvd.)_{cc\;\gamma}$.

\begin{claim}
\beq
\qq{d.}=
\prod_\gamma
\qq{(d.)_\gamma}
\;.
\label{eq-pd-pdgamma}
\eeq
\end{claim}
\proof

Let LHS and RHS denote the left and right hand
sides of Eq.(\ref{eq-pd-pdgamma}). Then

\beqa
LHS
&=&
\delta_{[\rvv.-\rvd.]^\wedge}
P(v.)
\\
&=&
\delta_{[\rvv.-\rvd.]^\wedge}
\prod_\gamma
\qq{\vcc{\gamma}}
\\
&=&
\prod_\gamma
\left\{\delta_{[\rvvcc{\gamma}-(\rvd.)_\gamma]^\wedge}
\right\}
\prod_\gamma
\left\{\qq{\vcc{\gamma}}\right\}
\\
&=&
\prod_\gamma
\left\{\delta_{[\rvvcc{\gamma}-(\rvd.)_\gamma]^\wedge}
\qq{\vcc{\gamma}}\right\}
\\
&=&
RHS
\;
\eeqa
\qed

Define
$\gamma_t$ to be the $\gamma$ such that
$\rvt\in \rvvcc{\gamma}$. We will
also use the following
shorthand notations

\beq
\rv{\calv}.=\rvvcc{\gamma_t}
\;,\;\;
\rv{\cald}.=(\rvd.)_{\gamma_t}
\;,\;\;
\rv{\calr}.=(\rvr.)_{\gamma_t}=\rv{\calv}.-\rv{\cald}.
\;
\eeq

\begin{claim}\label{cl-ancestral-cc}
For all $\gamma\neq \gamma_t$,

\beq
\qq{(d.)_\gamma}=
P((d.)_\gamma|\vcc{\gamma}^{c\wedge})
\;.
\label{eq-pdgamma-not-t-simp}
\eeq
\end{claim}
\proof

\begin{figure}[h]
    \begin{center}
    \epsfig{file=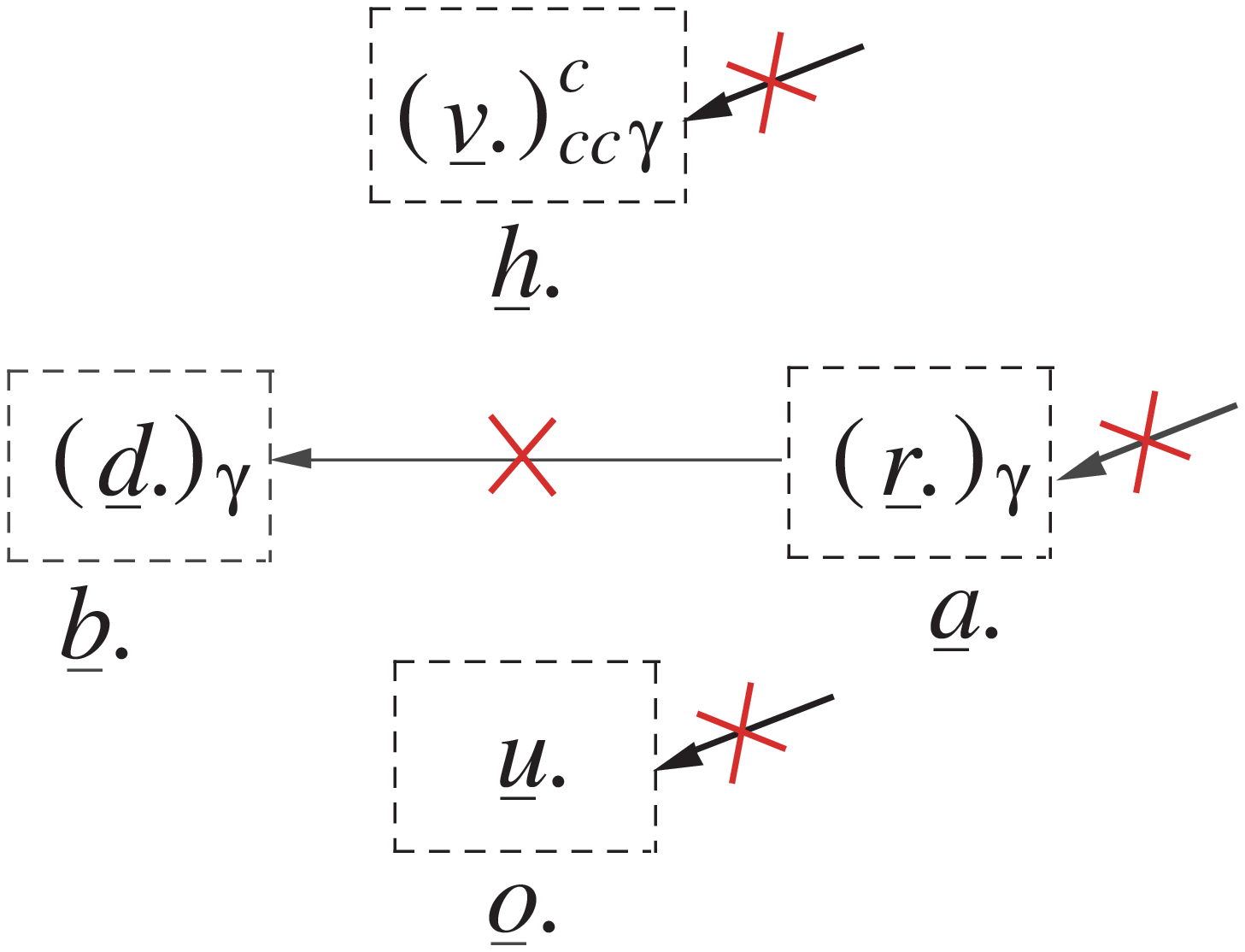, height=1.6in}
    \caption{
    A portrait of
    $G_{\myhat{\rvh}.,\myhat{\rva}.}$
    alluded to in Claim \ref{cl-ancestral-cc}.    }
    \label{fig-ancestral-cc}
    \end{center}
\end{figure}

We want to show that
\beq
\underbrace{P((d.)_\gamma|(r.)_\gamma^\wedge,
\vcc{\gamma}^{c\wedge})}_
{=\qq{(d.)_\gamma}}
=
P((d.)_\gamma|\vcc{\gamma}^{c\wedge})
\;.
\eeq
See Ref.\cite{Tuc-intro}
where the 3 Rules
of Judea Pearl's do-calculus
are stated. Using the notation there,
let
$\rvb.=(\rvd.)_\gamma$,
$\rva.=(\rvr.)_\gamma$,
$\rvh.=\rvvcc{\gamma}^{c}$,
$\rvi.=\emptyset$,,
$\rvo.=\rvu.$.
Note that
$\rva.^-=\rva.-
\rv{an}(\rvi.,G_{\myhat{\rvh}.})=\rva.$ so
$G_{\myhat{\rvh}.,(\rva.^-)^\wedge}=
G_{\myhat{\rvh}.,\myhat{\rva}.}$.
Fig.\ref{fig-ancestral-cc} portrays
$G_{\myhat{\rvh}.,\myhat{\rva}.}$.
Apply Rule 3 to that figure.
\qed

\begin{claim}\label{cl-ancestral-cc-t}
\beq
\qq{\cald.}=
P(\cald.|\calv.^{c\wedge},\myhat{t})
\;.
\label{eq-pdgamma-simp}
\eeq
\end{claim}
\proof

\begin{figure}[h]
    \begin{center}
    \epsfig{file=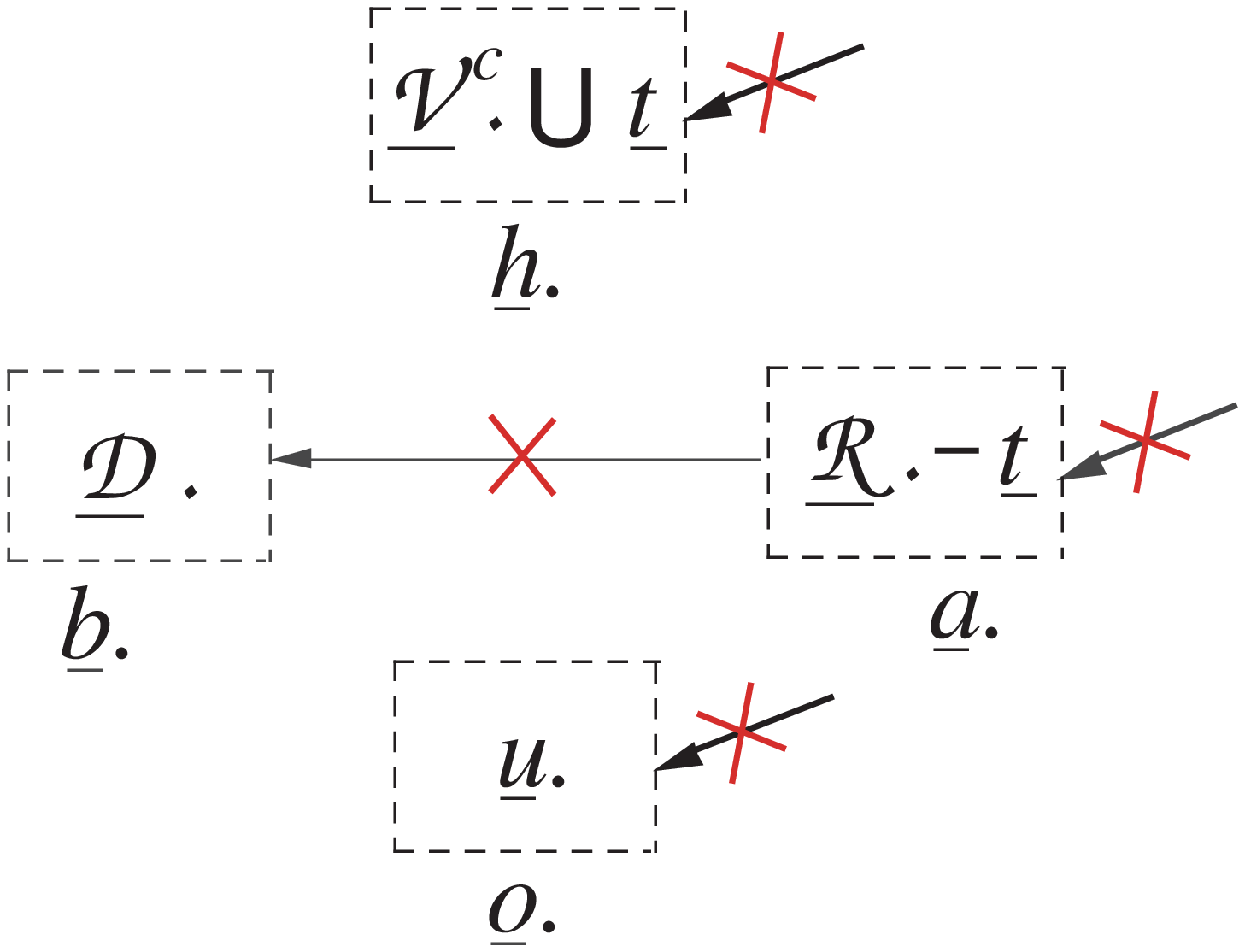, height=1.6in}
    \caption{
     A portrait of
    $G_{\myhat{\rvh}.,\myhat{\rva}.}$
    alluded to in Claim \ref{cl-ancestral-cc-t}.
    }
    \label{fig-ancestral-cc-t}
    \end{center}
\end{figure}

We want to show that
\beq
\underbrace{P(\cald.|[\calr.-t]^\wedge,
\calv.^{c\wedge},\myhat{t})}_
{=\qq{\cald.}}
=
P(\cald.|\calv.^{c\wedge},\myhat{t})
\;.
\eeq
See Ref.\cite{Tuc-intro}
where the 3 Rules
of Judea Pearl's do-calculus
are stated. Using the notation there,
let
$\rvb.=\rv{\cald}.$,
$\rva.=\rv{\calr}.-\rvt$,
$\rvh.=(\rv{\calv}.^{c},\rvt)$,
$\rvi.=\emptyset$,,
$\rvo.=\rvu.$.
Note that
$\rva.^-=\rva.-
\rv{an}(\rvi.,G_{\myhat{\rvh}.})=\rva.$ so
$G_{\myhat{\rvh}.,(\rva.^-)^\wedge}=
G_{\myhat{\rvh}.,\myhat{\rva}.}$.
Fig.\ref{fig-ancestral-cc-t} portrays
$G_{\myhat{\rvh}.,\myhat{\rva}.}$.
Apply Rule 3 to that figure.
\qed

Now we
can combine Eqs.
(\ref{eq-pst-pd}),
(\ref{eq-pd-pdgamma}),
(\ref{eq-pdgamma-not-t-simp}),
(\ref{eq-pdgamma-simp})
to get
\beq
P(s.|\myhat{t})
=
\sum_{d.-s.}
P(\cald.|\calv.^{c\wedge},\myhat{t})
\prod_{\gamma\neq \gamma_t}
P((d.)_\gamma |\;\vcc{\gamma}^{c\wedge})
\;.
\label{eq-pre-upsilons}
\eeq

\begin{figure}[h]
    \begin{center}
    \epsfig{file=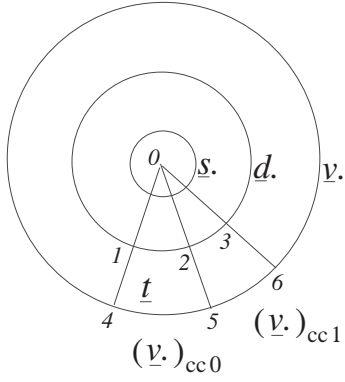, height=2in}
    \caption{
     Pie-cutting analogy
     for Eq.(\ref{eq-pre-upsilons}).
    }
    \label{fig-pie-cutting}
    \end{center}
\end{figure}

Eq.(\ref{eq-pre-upsilons})
is reminiscent of cutting a
pie. Fig.\ref{fig-pie-cutting}
explains this analogy further.
In this figure,
$\rvs.$, $\rvd.$ and $\rvv.$
are circular regions
nested this way:
$\rvs.\subset\rvd.\subset\rvv.$.
Let $0,1,2,\ldots,6$ denote
points on the pie, and
let $(0,1,2)$ be the
pie slice with corners $0,1,2$.
Then $\cald.=(0,1,2)$,
$\calv.=(0,4,5)$.
Note that $\rvt \in \calv.$.
For some $\gamma$ different from $\gamma_t$,
$(\rvd.)_\gamma =(0,2,3)$
and $\rvvcc{\gamma}=(0,5,6)$.

Eq.(\ref{eq-pre-upsilons})
suggests the following
iterative algorithm.
To $P_{\rvv.}$ express
$P(s.|\myhat{t})$, call
${\tt PV\_EXPRESS\_ONE}(\rvs.,\rvt,\rvv.)$,
where
\begin{tabbing}
\qquad\=
\qquad\=
\qquad\=
\qquad\=
\\
{\bf Subroutine} ${\tt PV\_EXPRESS\_ONE}(\rv{\sigma}.,\rvt,\rv{\beta}.)$ \{
\\
\>{\bf inputs}:$(\rv{\sigma}.,\rvt,\rv{\beta}.)$,
where must have $\rv{\sigma}.\cap\rvt=\emptyset$ and
$\rv{\sigma}.\cup\rvt\subset \rv{\beta}.\subset \rvv.$
\\
\>Set continue\_flag = true.
\\
\>Do while (continue\_flag == true) \{
\\
\>\>Find c-components $\{\rvbcc{\gamma}\}_{\forall \gamma}$ of $\rv{\beta}.$ in $G_{\rv{\beta}.}$,
$\rv{\beta}. =\bigcup_\gamma
\rvbcc{\gamma}$
\\
\>\>Find $\rvd.=\ol{\ul{an}}(\rv{\sigma}.,G_{\rv{\beta}.-\rvt})$
\\
\>\>For all $\gamma$ \{ Set
$(\rvd.)_\gamma=\rvvcc{\gamma}\cap\rvd.\}$
\\
\>\>Let $\gamma_t$ be $\gamma$ such that
$\rvt\in (\rvd.)_\gamma$.
\\
\>\>Set $\rv{\cald}.=(\rvd.)_{\gamma_t}$ and
$\rv{\calv}.=\rvvcc{\gamma_t}$.
\\
\>\>Store expression
$P(\sigma.|\myhat{t})
=
\sum_{d.-\sigma.}
P(\cald.|\calv.^{c\wedge},\myhat{t})
\prod_{\gamma\neq \gamma_t}
P((d.)_\gamma |\;\vcc{\gamma}^{c\wedge})$
\\
\>\>For all $\gamma \neq \gamma_t$ \{
Express
$P((d.)_\gamma |\;\bcc{\gamma}^{c\wedge})$ without hats
via Claim \ref{cl-ph-cc} \}
\\
\>\>Apply do-calculus Rules 2 and 3 to $P(\cald.|
(\calv_.)^{c\wedge},\myhat{t})$
to see if $\myhat{t}=\tau\in\{1,t\}$
\\
\>\>If Rule 2 or 3 succeeds \{
\\
\>\>\>Express $P(\cald.|
(\calv.)^{c\wedge},\tau)$
without hats via Claim \ref{cl-ph-cc}.
\\
\>\>\>Set continue\_flag = false
\\
\>\>\} else \{
\\
\>\>\>
Prune graph: Replace graph $G_{\rv{\beta}.}$
by
$G_{\rv{\beta}.^-}$,
where $\rv{\beta}.^-=
\ol{\rv{an}}(\rv{\sigma}.\cup\rvt,G_{\rv{\beta}.})$. \\
\>\>\>Set $\rv{\beta}.\larrow\rv{\beta}.^-$
\\
\>\>\>If $\rv{\cald.}$
is a c-component of $G_{\rv{\beta}.}$
\{
\\
\>\>\>\>Return FAIL message
\\
\>\>\>\>Exit program
\\
\>\>\>\}
\\
\>\>\}
\\
\>\}
\\
\>Do loop must store
information with each step.
\\
\>Collect information from each step
of the sequence\\
\>to assemble expression without hats\\
\>for the $P(s.|\myhat{t})$ considered at the beginning of the sequence.
\\
\}
\end{tabbing}

Note that this algorithm ``prunes" the graph before looping back again.
Pruning the graph is justified by virtue
of Claim \ref{cl-pruning-g}.
It is
a convenient step that
gets rid of superfluous nodes.
It also turns out to be
a necessary step.
As illustrated by the example
of Section \ref{sec-miss-teeth},
the algorithm
${\tt PV\_EXPRESS\_ONE}()$
may fail if this step is not performed.

The algorithm
${\tt PV\_EXPRESS\_ONE()}$
applies Eq.\ref{eq-pre-upsilons}
once in each loop
step.
The first application uses
$(\rvs.^{(1)},\rvv.^{(1)})=(\rvs.,\rvv.)$
and generates $(\rv{\cald}.,\rv{\calv}.)$
which becomes
$(\rvs.^{(2)},\rvv.^{(2)})$
for the next step.
The algorithm
thus generates a
sequence
$\{(\rvs.^{(j)},\rvv.^{(j)})\}_{j=1}^N$.
The sequence terminates when
$\rvs.^{(N)}$ is a c-component
of the current graph.

\subsection{Necessary and
Sufficient Conditions for Identifiability of $P(s.|\myhat{t})$}
\label{sec-iff-singleton}
In this section,
we will prove that the algorithm
given in Section \ref{sec-iff-singleton}
fails iff
$P(s.|\myhat{t})$
is not identifiable in $G$.

\begin{claim}\label{cl-hpos-s-tsingle}
If $P(\cald.|\calv.^{c\wedge},\myhat{t})=
P(\cald.|\calv.^{c\wedge},\tau)$
where $\tau\in\{1,t\}$, then
$H(\rvs.:\myhat{\rvt})\geq 0$.
\end{claim}
\proof

Assume the premise of the claim.
Combine that with Eqs.
(\ref{eq-pv-is-prod-pvcc})
and (\ref{eq-pre-upsilons}) to get

\beqa
P(s.:\myhat{t})
&=&
\frac{1}{P(s.)}
\sum_{d.-s.}
P(\cald.|\calv.^{c\wedge},\tau)
\prod_{\gamma\neq \gamma_t}
P((d.)_\gamma|\vcc{\gamma}^{c\wedge})
\\
&=&
\frac{1}{P(s.)}
\sum_{d.-s.}P(v.)
\frac{P(\cald.|\calv.^{c\wedge},\tau)}
{P(\calv.|\calv.^{c\wedge})}
\prod_{\gamma\neq \gamma_t}
\left\{
\frac{P((d.)_\gamma|\vcc{\gamma}^{c\wedge})}
{P((v.)_\gamma|\vcc{\gamma}^{c\wedge})}
\right\}
\;.
\label{eq-pst-long-form}
\eeqa
Next note that

\beq
\frac{P(\cald.|\calv.^{c\wedge},\tau)}
{P(\calv.|\calv.^{c\wedge})}
=
\frac{1}{
P(\tau|\calv.^{c\wedge})
P(\calr.-\tau|\calv.^{c\wedge},\cald.,\tau)
}
\;,
\eeq
and

\beq
\prod_{\gamma\neq \gamma_t}
\left\{
\frac{P((d.)_\gamma|\vcc{\gamma}^{c\wedge})}
{P((v.)_\gamma|\vcc{\gamma}^{c\wedge})}
\right\}
=
\frac{1}
{\prod_{\gamma\neq \gamma_t}
P((r.)_\gamma|\vcc{\gamma}^{c\wedge},(d.)_\gamma)}
\;.
\eeq
Defining a conditional probability distribution
$Q(r.|v.-r.)$ by

\beq
Q(r.|v.-r.)=
P(\tau|\calv.^{c\wedge})
P(\calr.-\tau|\calv.^{c\wedge},\cald.,\tau)
\prod_{\gamma\neq \gamma_t}
P((r.)_\gamma|\vcc{\gamma}^{c\wedge},(d.)_\gamma)
\;
\eeq
allows us to write Eq.(\ref{eq-pst-long-form})
more succinctly as

\beq
P(s.:\myhat{t})
=
\sum_{d.-s.}
\frac{P(v.)}
{P(s.)Q(r.|v.-r.)
}
\;.
\label{eq-sum-pv-over-q}
\eeq
Define

\beq
Q(v.)=
P(d.-s.|r.\cup s.)P(s.)Q(r.|v.-r.)
\;.
\eeq
Now note that

\beqa
H(\rvs.:\myhat{\rvt})
&=&
\av{\ln P(s.:\myhat{t})}_{s.,t}
\\
&=&
\av{\ln P(s.:\myhat{t})}_{s.,t,r.-t}
\label{eq-add-r-minus-t-dep}
\\
&=&
\av{\ln(
\sum_{d.-s.}
\frac{P(d.-s.|r.\cup s.)}{P(d.-s.|r.\cup s.)}
\frac{P(v.)}
{P(s.)Q(r.|v.-r.)}
)
}_{s.\cup r.}
\\
&=&
\av{\ln(
\sum_{d.-s.}
P(d.-s.|r.\cup s.)
\frac{P(v.)}
{Q(v.)}
)
}_{s.\cup r.}
\\
&\geq&
\av{\ln(\frac{P(v.)}{Q(v.)})}_{P(v.)}
\label{eq-apply-ln-concavity}
\\
&=&
D(P(v.)//Q(v.))_{\forall v.}
\geq 0
\;.
\eeqa
Eq.(\ref{eq-add-r-minus-t-dep})
follows because the quantity being averaged,
$P(s.|\myhat{t})$, depends only on $s.$ and $t$.
Since it is
independent of
$r.-t$, we may do a weighted average
over $r.-t$ also without changing the
final average.
Inequality Eq.(\ref{eq-apply-ln-concavity})
follows from the concavity of the $\ln(\cdot)$
function. Indeed, $\ln(x)$ is a concave function
over $x\in \RR^{\geq 0}$
so if $P(a)$ is a probability distribution
over $S_\rva$ and $f(a)\geq 0$ for all $a\in S_\rva$, then
\beq
\ln\left(\sum_a P(a) f(a)\right)
\geq
\sum_a P(a) \ln\left(f(a)\right)
\;.
\eeq
\qed

Next we give one of the most important
claims of this paper. The claim might
even come close to the exalted level
of being called a theorem.
It gives
two
separate conditions,
one ``graphical", and
one ``informational",
such that either of them
alone is necessary and sufficient
for $P(s.|\myhat{t})$
to be identifiable in $G$.

\begin{claim}\label{cl-main-theo}
For any graph $G$, the following
are equivalent:\footnote{The labels $ID,GR,H+$
stand for ``identifiability",
``graphical"
and ``$H$ positive", respectively.}
\begin{description}
\item[(ID)] $P(s.|\myhat{t})$ is identifiable in $G$.
\item[(GR)] $P(s.^{(N)}|(v.^{(N)})^{c\wedge},\myhat{t})=
P(s.^{(N)}|(v.^{(N)})^{c\wedge},\tau)$ where $\tau\in\{1,t\}$.
$(\rvs.^{(N)},\rvv.^{(N)})$
is the last term
in the sequence
$\{(\rvs.^{(j)},\rvv.^{(j)})\}_{j=1}^N$
generated by the algorithm {\tt PV\_EXPRESS\_ONE()}.
\item[(H+)] $H(\rvs.:\myhat{\rvt})\geq 0$ for all models of $G$.
\end{description}
\end{claim}
\proof

\begin{description}
\item[(GR$\implies$ ID)]
Assume GR. Then after applying
Eq.(\ref{eq-pre-upsilons})
multiple times,
we get a product of $P_{\rvv.}$ expressible probabilities
times
$P(s.^{(N)}|(v.^{(N)})^{c\wedge},\myhat{t})$. The latter is itself equal to
$P(s.^{(N)}|(v.^{(N)})^{c\wedge},\tau)$
by GR.
Thus ID is true.
\item[(GR$\implies$ H+)] This follows from
Claim \ref{cl-hpos-s-tsingle}.
\item[(not(GR)$\implies$ not(ID) and not(H+))]
Using the notation
of Ref.\cite{Tuc-intro}, for $j\in\{1,2,3\}$,
call $(\rvb.\perp \rva.|\myhat{\rvh}.,\rvi.)_{G_j}$
the ``premise" of Rule $j$ .
Assume not(GR). Then the Rule 2 premise
and the Rule 3
premise are both false,
where $\rvb.=\rvs.^{(N)}$,$\rva.=\rvt$,
$\rvh.=(\rvv.^{(N)})^c$, $\rvi.=\emptyset$,
$\rvo.=(\rvo.^{(a)},\rvo.^{(b)})$,
$\rvo.^{(a)}=\rvu.$,
$\rvo.^{(b)}=(\rvv.^{(N)}-\rvs.^{(N)}\cup\rvt)$.
Note that
$\rva.^-=\rva.-
\rv{an}(\rvi.,G_{\myhat{\rvh}.})=\rva.$ so
$G_{\myhat{\rvh}.,(\rva.^-)^\wedge}=
G_{\myhat{\rvh}.,\myhat{\rva}.}$.
Fig.\ref{fig-id-iff-bans} portrays
$G_{\myhat{\rvh}.,\myvee{\rva}.}$
for Rule 2 and
$G_{\myhat{\rvh}.,\myhat{\rva}.}$
for Rule 3. Arrows
with an ``X R2" (resp., ``X R3")
on them
are banned by Rule 2 (resp., Rule 3).

Note that Fig.\ref{fig-id-iff-bans}
places a ban on arrows
pointing from $\rvo.^{(b)}$
to $\rvb.$. This is justified
because $\rvs.^{(N)}$ equals
the last $\cald.$
and $\rvv.^{(N)}$
equals the last $\calv.$.
With $\rvd.=\ol{\rv{an}}
(\sigma.,G_{\rv{\beta}.-\rvt})$,
we have
(1) $\rvb.=\cald.$ is inside $\rvd.$,
(2)
$\rvo.^{(b)}=\calv.-\cald.\cup\rvt$
is disjoint from $\rvd.$,
and (3) $\rvd.$ is ancestral in
$G_{\rv{\beta}.-\rvt}$.
\begin{figure}[h]
    \begin{center}
    \epsfig{file=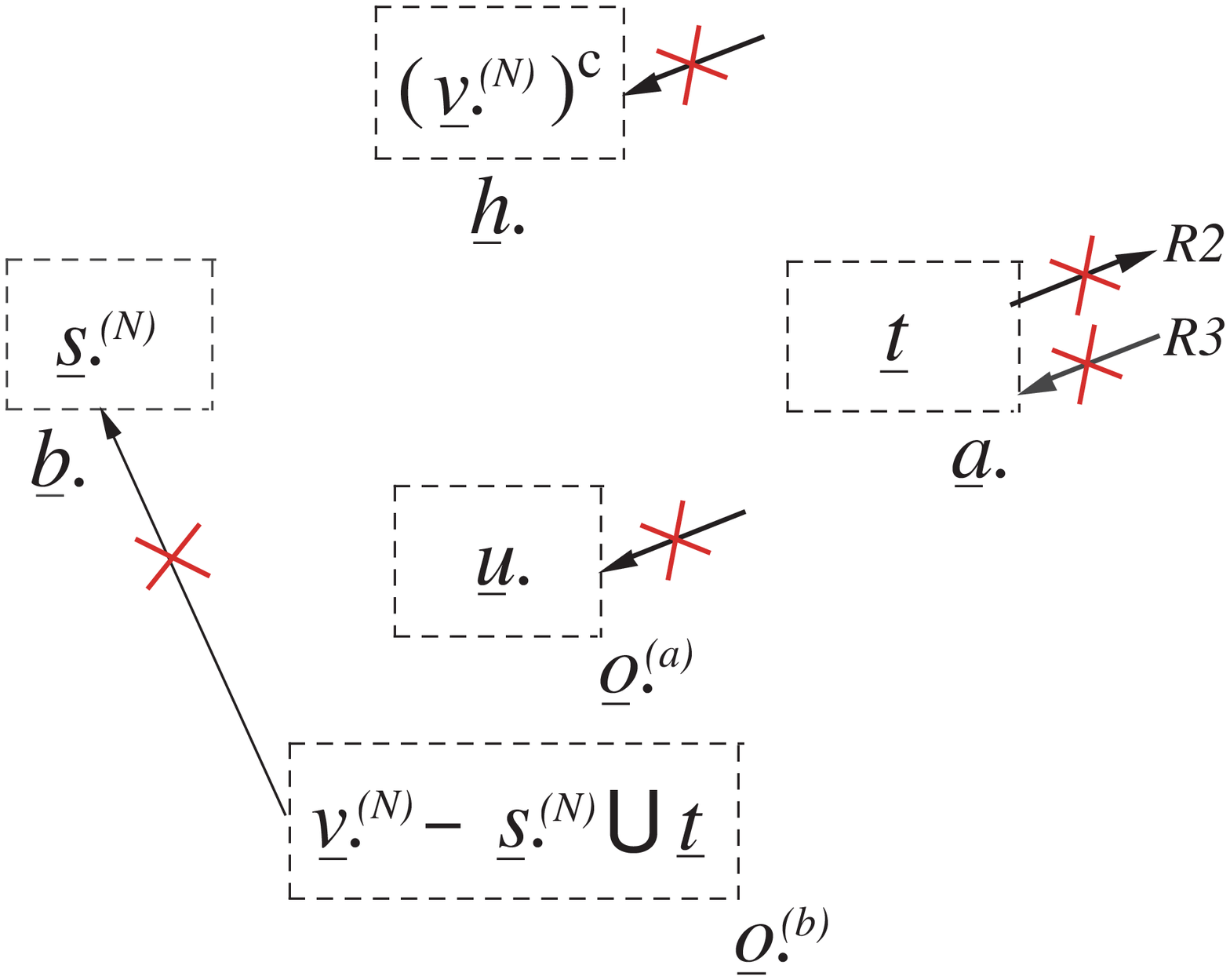, height=2.3in}
    \caption{
    A portrait of
    $G_{\myhat{\rvh}.,\myvee{\rva}.}$
    for Rule 2
    and
    $G_{\myhat{\rvh}.,\myhat{\rva}.}$
    for Rule 3, alluded to in
    Claim \ref{cl-main-theo}.
    There is also a ban
    on arrows from $\rvo.^{(b)}$
    to $\rvb.$.
    }
    \label{fig-id-iff-bans}
    \end{center}
\end{figure}

\begin{figure}[h]
    \begin{center}
    \epsfig{file=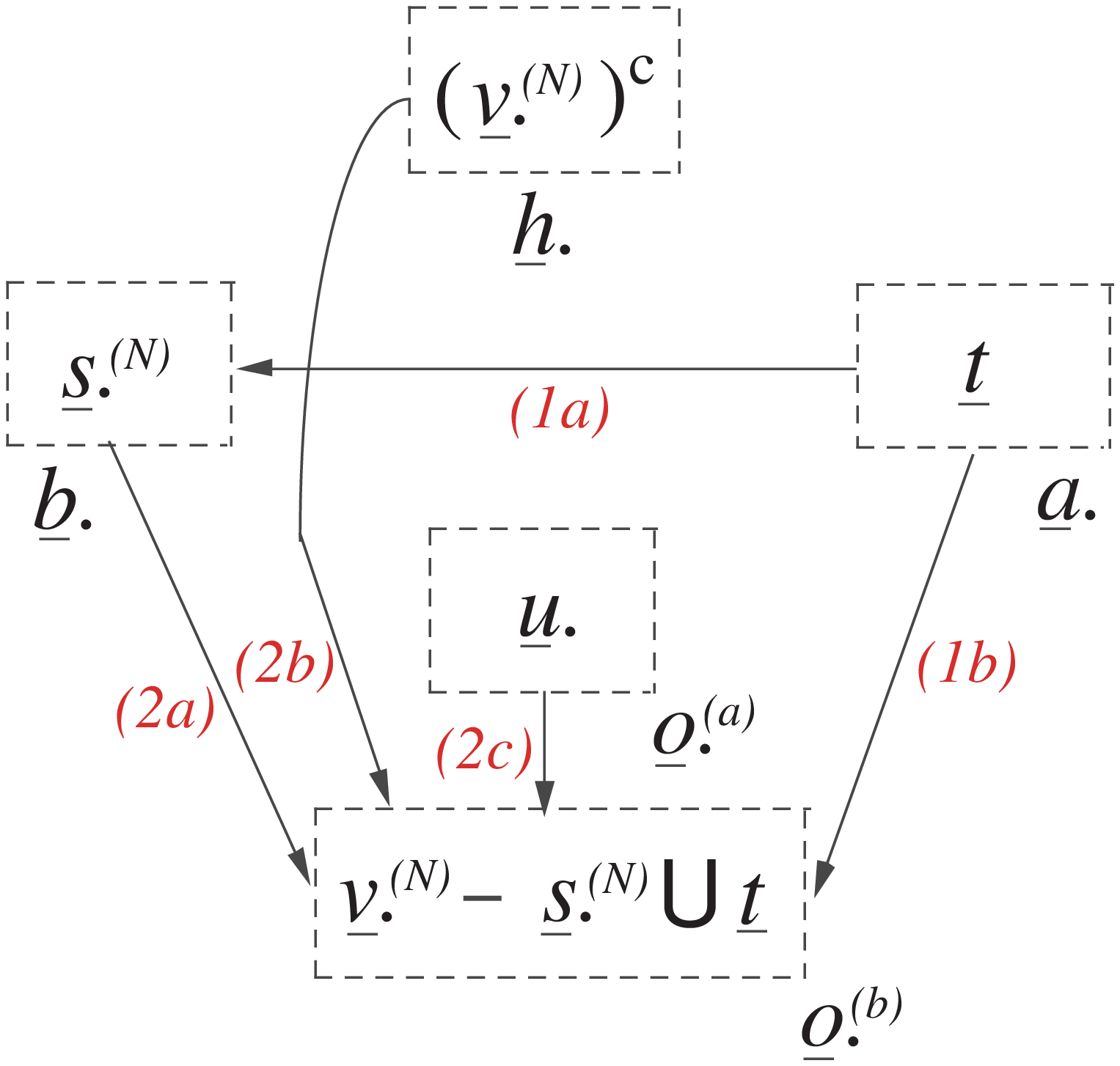, height=2.3in}
    \caption{
    Possible behaviors of path $\gamma_3$
    alluded to in
    Claim \ref{cl-main-theo}.
    }
    \label{fig-id-iff-path3}
    \end{center}
\end{figure}
\begin{figure}[h]
    \begin{center}
    \epsfig{file=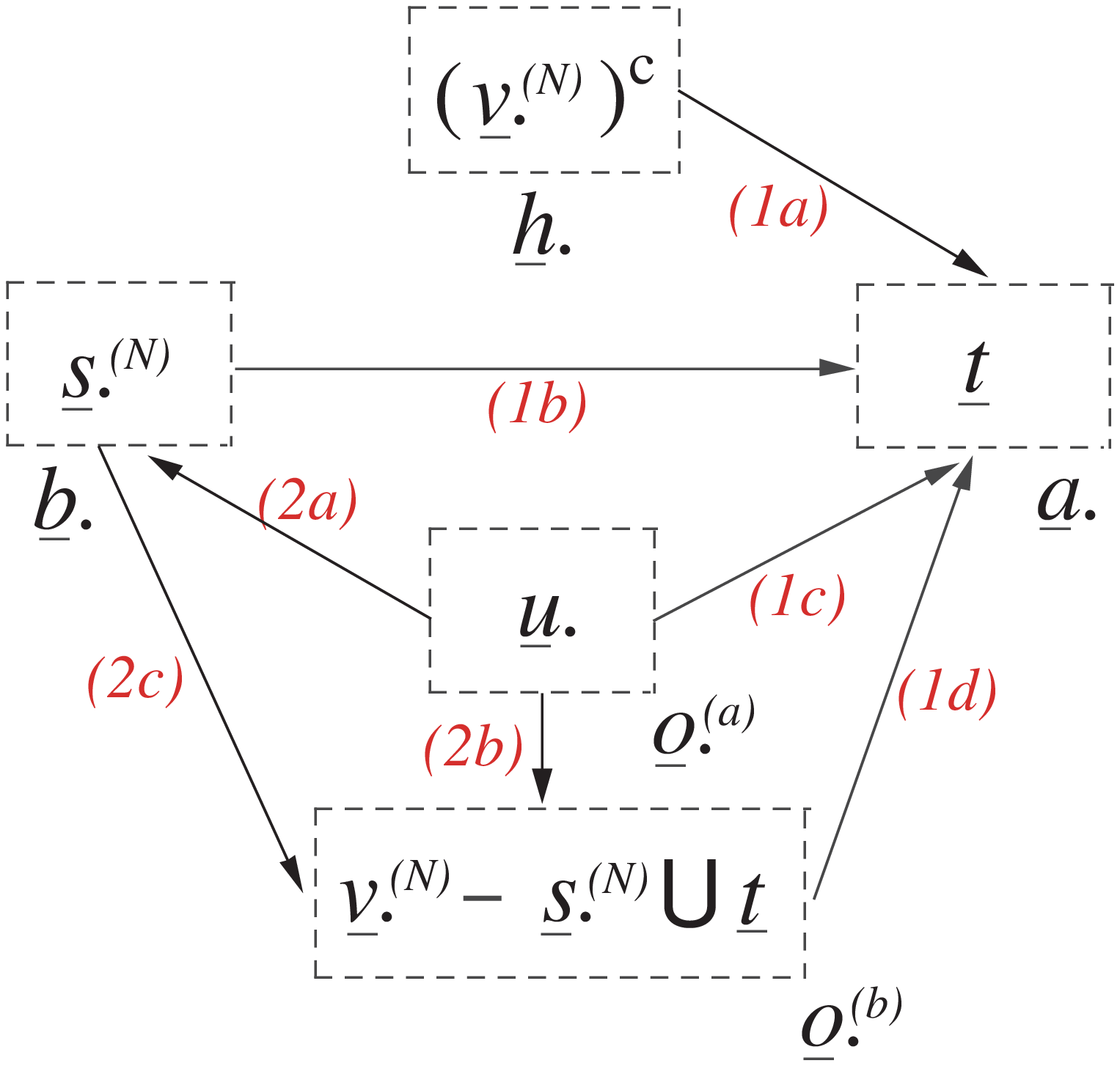, height=2.3in}
    \caption{
    Possible behaviors of path $\gamma_2$
    alluded to in
    Claim \ref{cl-main-theo}.
    }
    \label{fig-id-iff-path2}
    \end{center}
\end{figure}
\begin{figure}[h]
    \begin{center}
    \epsfig{file=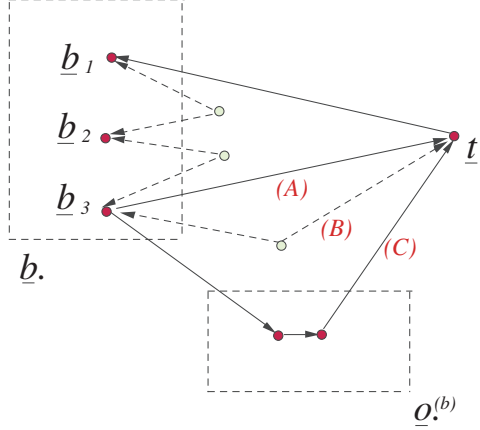, height=2.3in}
    \caption{
    In Claim \ref{cl-main-theo},
    when not(GR) is assumed,
    there must be a closed path
    of either type (A), (B) or (C).
    All 3 types are
    either standard or modified
    shark teeth graphs
    of the kind discussed in Section
    \ref{sec-3teeth}.
    }
    \label{fig-id-iff-fin}
    \end{center}
\end{figure}

Since the premise of Rule 3
is false,
there must exist an undirected path
$\gamma_3$
from $\rva.$ to $\rvb.$
that is unblocked at fixed $(\rvh.,\rvi.)$.
Figure \ref{fig-id-iff-path3}
illustrates possible behaviors
of path $\gamma_3$.
$\gamma_3$ must contain an arrow exiting node $\rva.=\rvt$. This
means $\gamma_3$ must contain
either an arrow
$(1a)$ or an arrow $(1b)$.
If path $\gamma_3$
contains arrow $(1b)$,
then it must also contain
at least one of the following
 arrows: $(2a)$
or $(2b)$ or $(2c)$.
Let $\gamma_3\supset (1b,2a)$
mean that path $\gamma_3$
contains arrows $(1b)$ and $(2a)$.
Thus, $\gamma_3$ must satisfy
one of the following 4 cases.
\beq
\gamma_3 \supset
\left\{
\begin{array}{ll}
1a & \mbox{OK}
\\
(1b,2a) & \mbox{blocked}
\\
(1b,2b) & \mbox{blocked}
\\
(1b,2c) & \mbox{blocked}
\end{array}
\right.
\;.
\eeq
As indicated, the last 3 cases
are not really possible
because in all 3 cases $\gamma_3$
would have to have a collider
outside $\rvh.$,
so in order for $\gamma_3$ to remain unblocked,
that collider would have to have a descendant in
$\rvh.$.
But that can't happen since there is a
ban on arrows entering $\rvh.$.

Since the premise of Rule 2
is false,
there must exist an undirected path
$\gamma_2$
from $\rva.$ to $\rvb.$
that is unblocked at fixed $(\rvh.,\rvi.)$.
Figure \ref{fig-id-iff-path2}
illustrates possible behaviors
of path $\gamma_2$.
$\gamma_2$ must contain an arrow entering node $\rva.=\rvt$. This
means $\gamma_2$ must contain
one of the following arrows:
$(1a),(1b),(1c)$ or $(1d)$.
If path $\gamma_2$
contains arrow $(1c)$,
then it must also contain
at least one of the following
 arrows: $(2a)$
or $(2b)$.
If path $\gamma_2$
contains arrow $(1d)$,
then it must also contain
 arrow $(2c)$.
Thus, $\gamma_2$ must satisfy
one of the following 4 cases.
\beq
\gamma_2 \supset
\left\{
\begin{array}{lll}
1a& \mbox{blocked} &
\\
1b & \mbox{OK} & (A)
\\
(1c,2a) & \mbox{OK} & (B)
\\
(1c,2b) & \mbox{blocked} &
\\
(1d,2c) & \mbox{OK} & (C)
\end{array}
\right.
\;
\eeq
As indicated, the first and
fourth cases
are not really possible.
For the first case,
$\gamma_2$
would have to have a non-collider
inside $\rvh.$
and that would block the path.
For the fourth case, $\gamma_2$
would have to have a collider
outside $\rvh.$,
so in order for $\gamma_2$ to remain unblocked,
that collider would have to have a descendant in
$\rvh.$.
But that can't happen since there is a
ban on arrows entering $\rvh.$.

Fig.\ref{fig-id-iff-fin}
combines the OK cases
for path $\gamma_3$
with the OK cases for
path $\gamma_2$.
Let $\rvb_1$
be the node where
$\gamma_3$
first makes contact
with $\rvb.$.
Let $\rvb_3$
be the node where
$\gamma_2$
first makes contact
with $\rvb.$.
There must
be path between $\rvb_1$
and $\rvb_3$
that is composed
of a sequence of visible nodes
(for example,
the visible nodes $\rvb_1,\rvb_2,\rvb_3$
in Fig.\ref{fig-id-iff-fin})
connected
pairwise
by bidirected arcs,
and those
visible nodes
must
all lie inside $\rvs.^{(N)}$.
This follows because,
by construction,
$\rvs.^{(N)}$
is a c-component
of the current graph.

Thus, the full graph $G$ must
contain a subgraph, call it $G^-$,
isomorphic to
the shark teeth
graph discussed in
Section \ref{sec-3teeth}.
$G^-$ is not identifiable
and there exists a model
for it
with $H(\rvs.:\myhat{\rvt})<0$.
Therefore, by virtue of Claims
\ref{cl-inherit-id} and \ref{cl-inherit-hminus},
$G$ is not identifiable and
there exists a model for it
with
$H(\rvs.:\myhat{\rvt})<0$.

\end{description}\mbox{}
\qed

\section{
$P(s.|\myhat{t}.)$
when $\rvt.$ is NOT a singleton}

\subsection{Algorithm for $P_{\rvv.}$ expressing $P(s.|\myhat{t}.)$}
\label{sec-algo-not-singleton}

In Section \ref{sec-algo-not-singleton},
we gave an algorithm
called {\tt PV\_EXPRESS\_ONE()} for
$P_{\rvv.}$ expressing $P(s.|\myhat{t}.)$
when $\rvt.$ is a singleton.
Below we give
a recursive algorithm called {\tt PV\_EXPRESS()}
that
handles the $\rvt.$ non-singleton
case by
calling {\tt PV\_EXPRESS\_ONE()} repeatedly.

To $P_{\rvv.}$ express
$P(s.|\myhat{t}.)$, call
${\tt PV\_EXPRESS}(\rvs.,\rvt.,\rvv.)$,
where
\begin{tabbing}
\qquad\=
\qquad\=
\qquad\=
\qquad\=
\\
{\bf Subroutine} ${\tt PV\_EXPRESS}(\rv{\sigma}.,\rvt.,\rv{\beta}.)$ \{
\\
\>{\bf inputs}:$(\rv{\sigma}.,\rvt.,\rv{\beta}.)$,
where must have $\rv{\sigma}.\cap\rvt.=\emptyset$ and
$\rv{\sigma}.\cup\rvt.\subset \rv{\beta}.\subset \rvv.$
\\
\>
Prune graph: Replace graph $G_{\rv{\beta}.}$
by
$G_{\rv{\beta}.^-}$,
where $\rv{\beta}.^-=
\ol{\rv{an}}(\rv{\sigma}.\cup\rvt.,G_{\rv{\beta}.})$. \\
\>Set $\rv{\beta}.\larrow\rv{\beta}.^-$
\\
\>Find c-components $\{\rvbcc{\gamma}\}_{\forall \gamma}$ of $\rv{\beta}.$ in $G_{\rv{\beta}.}$,
$\rv{\beta}. =\bigcup_\gamma
\rvbcc{\gamma}$
\\
\>Find $\rvd.=\ol{\ul{an}}(\rv{\sigma}.,G_{\rv{\beta}.-\rvt.})$
\\
\>For all $\gamma$ \{ Set
$(\rvd.)_\gamma=\rvbcc{\gamma}\cap\rvd.$
and
$(\rvt.)_\gamma=\rvbcc{\gamma}\cap\rvt.$ \}
\\
\>Store expression
$P(\sigma.|\myhat{t}.)
=
\sum_{d.-\sigma.}
\prod_{\gamma}
P((d.)_\gamma |\;\bcc{\gamma}^{c\wedge},
(t.)_\gamma^\wedge)$
\\
\>For all $\gamma $ \{
\\
\>\>If $|(\rvt.)_\gamma|=0$  \{
\\
\>\>\>Express
$P((d.)_\gamma |\;\bcc{\gamma}^{c\wedge})$ without hats
via Claim \ref{cl-ph-cc}
\\
\>\> \} else if $|(\rvt.)_\gamma|=1$  \{
\\
\>\>\>Call ${\tt PV\_EXPRESS\_ONE}((\rvd.)_\gamma,\rvt,\rvbcc{\gamma})$
\\
\>\>\} else if $|(\rvt.)_\gamma|>1$  \{
\\
\>\>\>Call ${\tt PV\_EXPRESS}((\rvd.)_\gamma,(\rvt.)_\gamma,\rvbcc{\gamma})$
\\
\>\>\}\\
\>\}
\\
\>Revisit all nodes of the recursion tree, and
\\
\>collect information from each node of tree\\
\>to assemble expression without hats\\
\>for the $P(s.|\myhat{t}.)$ at the root node of tree.
\\
\}
\end{tabbing}
The above subroutine appears to
be consistent. It appears to
fail if and only if $P(s.|\myhat{\rvt.})$
is identifiable. Furthermore, it is
explicitly
based entirely on the do-calculus rules
(and standard identities from probability
theory
such as conditioning and chain rules.)
\subsection{Necessary and
Sufficient Conditions for Identifiability of $P(s.|\myhat{t}.)$}
\label{sec-iff-not-singleton}

One suspects that
Claim \ref{cl-main-theo}
can be generalized to also
encompass cases where $\rvt.$
is not a singleton. Here is
one partial generalization:

\begin{claim}
Consider a graph $G$
with nodes $\rvx.=(\rvv.,\rvu.)$.
Suppose $\rvs.$ and
$\rvt.$ are disjoint subsets of $\rvv.$.
Then the following are equivalent
\begin{description}
\item[(ID)] $P(\rvs.|\myhat{\rvt}.)$
is identifiable in $G$.
\item[(H+)] $H(\rvs.:\myhat{\rvt}.)\geq 0$
for all models of $G$
\end{description}
\end{claim}
\proof We won't give a rigorous proof of this,
just a plausibility argument.
\begin{description}
\item[(ID $\implies$ H+)]
From how $P(s.|\myhat{t}.)$
is defined
and the fact that
$P(s.|\myhat{t}.)$
is $P_{\rvv.}$ expressible, it should be
possible to prove that

\beq
P(s.|\myhat{t}.)=
\sum_{d.-s.}
\frac{P(v.)}{Q(v.-d.|d.)}
\;,
\eeq
for some set $\rvd.$ such that
$\rvs.\subset \rvd. \subset \rvv.-\rvt.$
and some conditional probability
distribution $Q(v.-d.|d.)$.
But this implies Eq.(\ref{eq-sum-pv-over-q})
so the proof following
Eq.(\ref{eq-sum-pv-over-q})
showing that $H(\rvs.:\myhat{\rvt})\geq 0$
applies here too with the small modification
that all $\rvt$ are replaced by $\rvt.$.
\item[(not(ID) $\implies$ not(H+))]
Assume that initially, our model of $G$
satisfies $H(\rvs.|\myhat{\rvt}.)=0$
(According to Claim \ref{cl-zero-h}
such a model exists).
Consider an infinitesimal
displacement of the probability
distribution $P(x.)$ of this model.
Let the displacement satisfy $\delta P(v.)=0$
for all $v.$. Then
$\delta H(\rvs.:\myhat{\rvt}.)=
\av{
\delta \ln P(s.|\myhat{t}.)}_{s.,t.}$.
Since not(ID), $P(s.|\myhat{t}.)$
is not $P_{\rvv.}$ expressible.
Hence, even with $\delta P(v.)=0$,
we can find
a $\delta \ln P(s.|\myhat{t}.)<0$
which makes
$H(\rvs.:\myhat{\rvt}.)$ infinitesimally negative.
\end{description}\mbox{}
\qed

\appendix

\section{Appendix- Examples of \\
identifiable probabilities}\label{app-id}
In this appendix, we present
several examples of identifiable
uprooted probabilities $P(s.|\myhat{t})$.
Almost all of the examples
that we will give have been considered before
by Pearl and Tian in
Refs.\cite{P95} and \cite{R290L}.
However, we analyze these examples using
our own algorithm,
the one proposed in Section
\ref{sec-algo-not-singleton},
 instead of
 the algorithm proposed by
Pearl and Tian in
Refs.\cite{P95} and \cite{R290L}.

For each example, we will
give a graph,
specify the value of $P(s.|\myhat{t})$
that we seek for that graph,
and $P_{\rvv.}$ express
$P(s.|\myhat{t})$.
This calculation will
rely on the following formula,
which comes from Eq.(\ref{eq-pre-upsilons}).

\beq
\underbrace{
P(s.|\myhat{t})}_{\Upsilon_1}
=
\underbrace{\sum_{d.-s.}
}_{\Upsilon_2}
\underbrace{
P(\cald.|\calv.^{c\wedge},\myhat{t})
}_{\Upsilon_3}
\underbrace{
\prod_{\gamma\neq \gamma_t}
P((d.)_\gamma |\;\vcc{\gamma}^{c\wedge})
}_{\Upsilon_4}
\;.
\label{eq-under-upsilons}
\eeq
When using Eq.(\ref{eq-under-upsilons}), we will
give the special values of the upsilon
terms $\Upsilon_j$ defined above.

Below, we will often use
tables of the form:

\beq
\begin{array}{c||c|c|c||c||c||}
&\multicolumn{3}{c||}{\ul{\calv}.=\rvvcc{0}}
&\rvvcc{1}&\rvvcc{2}
\\ \cline{2-6}
 &\rvv_1&\rvv_2&\rvv_3&\rvv_4&\rvv_5
\\ \hline
\rva.&\checkmark & & &\checkmark &\checkmark
\\ \hline
\rvb.&&&&&\checkmark
\\\hline
\end{array}
\;.
\eeq
In such tables, we will
label the rows  by various node sets (in this case $\rva.$
 and $\rvb.$), and
the columns by all the
$\rvv.$ nodes of the graph. A
check mark is put at the intersection
of a row $R$ and column $C$ if node set $R$ contains
node $C$. Such tables also indicate
for each element of $\rvv.$, what
c-component $\rvvcc{\gamma}$
it belongs to.

\subsection{Example of backdoor formula
(see Ref.\cite{P95})}\label{sec-backdoor}

\begin{figure}[h]
    \begin{center}
    \epsfig{file=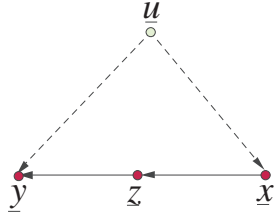, height=1.25in}
    \caption{Graph $G$ for Sections \ref{sec-backdoor}
    and \ref{sec-frontdoor}.
    For this graph,
    $P(y|\myhat{z})$ (resp.,
    $P(y|\myhat{x})$) is expressible
    in terms of $P(v.)$ using
    what Pearl calls the backdoor
    (resp., frontdoor) formula.
    }
    \label{fig-frontdoor}
    \end{center}
\end{figure}

In this example,
we want to
$P_{\rvv.}$ express
$P(y|\myhat{z})$
for the graph of Fig.\ref{fig-frontdoor}.

For this example,
the following table applies.

\beq
\begin{array}{c||c||c|c||}
&\ul{\calv}.=\rvvcc{0}
&\multicolumn{2}{c||}{\rvvcc{1}}
\\ \cline{2-4}
 &\rvz&\rvy&\rvx
\\ \hline
\rvt&\checkmark & &
\\ \hline
\rvs.& &\checkmark &
\\ \hline
\rvd.& &\checkmark &
\\\hline
\end{array}
\;
\label{table-bd}
\eeq

One possible topological ordering for the
visible nodes $\rvv.$
of this graph is

\beq
\rvy\larrow\rvz\larrow\rvx
\;
\label{eq-bd-topo}
\eeq

According to Claim \ref{cl-pv-cc},
\beq
P(v.)= \underbrace{\qq{z}}_{P(z|x)}
\qq{y,x}
\;,
\label{eq-bd-pv}
\eeq
where

\beqa
\qq{y,x} &=&\av{P(y|z,u)P(x|u)}_u
\\
&=&
P(y|z,x)P(x)
\;.
\label{eq-bd-qo}
\eeqa

Eq.(\ref{eq-under-upsilons})
can be specialized using
the data from table Eq.(\ref{table-bd})
to get the following values for the upsilon terms:

\beq
\Upsilon_1 = \Upsilon_4= P(y|\myhat{z})
\;,
\eeq

\beq
\Upsilon_2=\Upsilon_3=1
\;.
\eeq
Note that
\beqa
P(y|\myhat{z})&=&
\sum_x \qq{y,x}
\\
&=&
\sum_x P(y|z,x)P(x)
\;.
\eeqa

\subsection{Example of frontdoor formula
(see Ref.\cite{P95})}\label{sec-frontdoor}

In this example,
we want to
$P_{\rvv.}$ express $P(y|\myhat{x})$
for
the same graph (Fig.\ref{fig-frontdoor})
used in the previous example.

For this example,
the following table applies.

\beq
\begin{array}{c||c|c||c||}
&\multicolumn{2}{c||}{\ul{\calv}.=\rvvcc{0}}&\rvvcc{1}
\\ \cline{2-4}
 &\rvx&\rvy&\rvz
\\ \hline
\rvt&\checkmark & &
\\ \hline
\rvs.& &\checkmark &
\\ \hline
\rvd.& &\checkmark & \checkmark
\\\hline
\end{array}
\;
\label{table-fd}
\eeq

Eqs.(\ref{eq-bd-topo}),
(\ref{eq-bd-pv}),
(\ref{eq-bd-qo}) from the
previous example
apply for this example also.

Eq.(\ref{eq-under-upsilons})
can be specialized using
the data from table Eq.(\ref{table-fd})
to get the following values for the upsilon terms:

\beq
\Upsilon_1 = P(y|\myhat{x})
\;,
\eeq

\beq
\Upsilon_2 = \sum_z
\;,
\eeq

\beq
\Upsilon_3= P(y|\myhat{z},\myhat{x})
\;,
\eeq

\beq
\Upsilon_4= P(z|x)
\;.
\eeq

\begin{claim}\label{cl-fd}
\beq
P(y|\myhat{z},\myhat{x})
=P(y|\myhat{z})
\;.
\label{eq-fd}
\eeq
\end{claim}
\proof

\begin{figure}[h]
    \begin{center}
    \epsfig{file=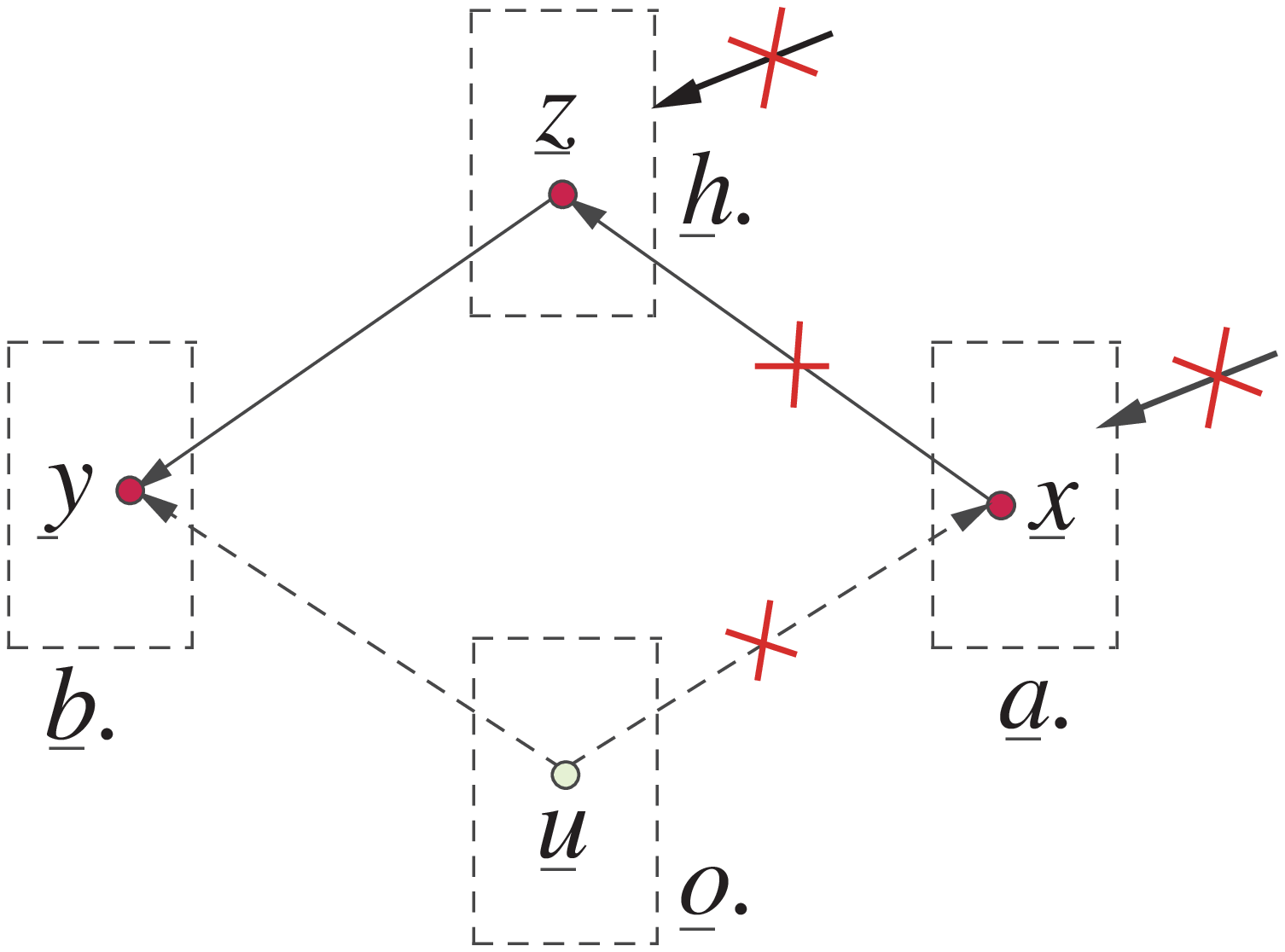, height=1.75in}
    \caption{A portrait of
    $G_{\myhat{\rvh}.,\myhat{\rva}.}$,
alluded to in Claim \ref{cl-fd}.
    }
    \label{fig-d-sep-frontdoor}
    \end{center}
\end{figure}

See Ref.\cite{Tuc-intro}
where the 3 Rules
of Judea Pearl's do-calculus
are stated. Using the notation there,
let
$\rvb.=\rvy,
\rva.=\rvx,
\rvh.=\rvz,
\rvi.=\emptyset,
\rvo.=\rvu $.
Note that
$\rva.^-=\rva.-
\rv{an}(\rvi.,G_{\myhat{\rvh}.})=\rva.$ so
$G_{\myhat{\rvh}.,(\rva.^-)^\wedge}=
G_{\myhat{\rvh}.,\myhat{\rva}.}$.
Fig.\ref{fig-d-sep-frontdoor} portrays
$G_{\myhat{\rvh}.,\myhat{\rva}.}$.
Apply Rule 3
to that figure.
\qed

Note that
\beqa
P(y|\myhat{z})&=&
\sum_x \qq{y,x}
\\
&=&
\sum_x P(y|z,x)P(x)
\;.
\label{eq-fin-q0-fd}
\eeqa

Combining the
upsilon values given, Eq.(\ref{eq-fd}) and
Eq.(\ref{eq-fin-q0-fd}),
 we conclude that
Eq.(\ref{eq-under-upsilons}), when
fully specialized to this example,
becomes

\beqa
P(y|\myhat{x})
&=&
\sum_z
\left[\sum_{x'} \qq{y,x'}\right]
 P(z|x)
 \\
 &=&
 \sum_z
 \left[
 \sum_{x'}
 P(y|z,x')P(x')
 \right]
 P(z|x)
\;.
\eeqa

\subsection{Example from Ref.\cite{R290L}-Fig.2}
\label{sec-tp-fig2}

\begin{figure}[h]
    \begin{center}
    \epsfig{file=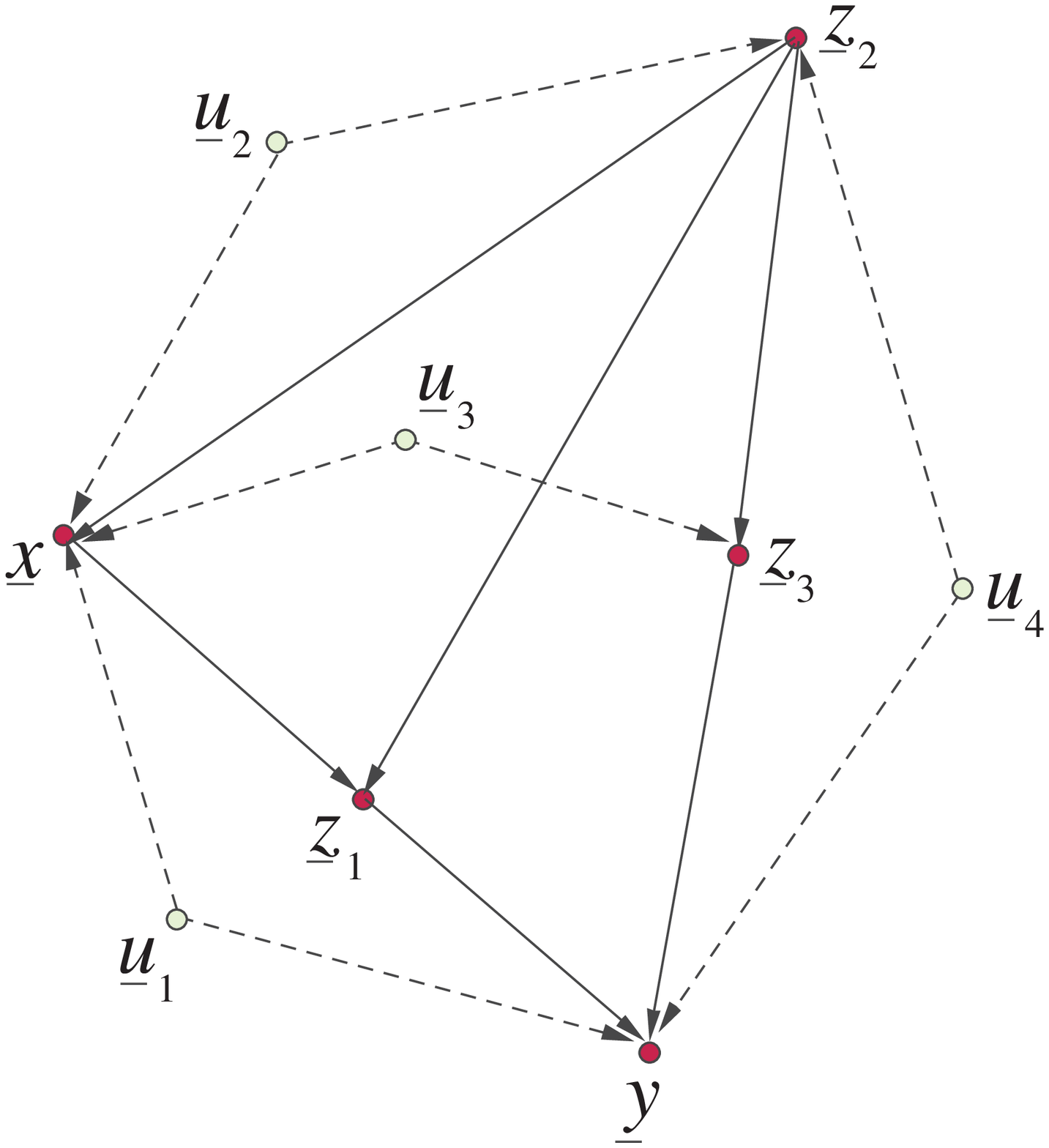, height=2.75in}
    \caption{
    Graph $G$ for Section \ref{sec-tp-fig2}.
    }
    \label{fig-tp-fig2}
    \end{center}
\end{figure}

In this example,
we want to
$P_{\rvv.}$ express
$P(y|\myhat{x})$
for the graph of Fig.\ref{fig-tp-fig2}.

For this example,
the following table applies.

\beq
\begin{array}{c||c|c|c|c||c||}
&\multicolumn{4}{c||}{\ul{\calv}.=\rvvcc{0}}&\rvvcc{1}
\\ \cline{2-6}
 &\rvx&\rvy&\rvz_3&\rvz_2&\rvz_1
\\ \hline
\rvt&\checkmark & & & &
\\ \hline
\rvs.&&\checkmark  &&&
\\ \hline
\rvd.& &\checkmark&\checkmark&\checkmark&\checkmark
\\\hline
\end{array}
\;
\label{table-tp-fig2}
\eeq

One possible topological ordering for the
visible nodes $\rvv.$
of this graph is

\beq
\rvy\larrow\rvz_3\larrow\rvz_1\larrow\rvx\larrow\rvz_2
\;
\eeq

According to Claim \ref{cl-pv-cc},
\beq
P(v.)=\qq{y,x,z_3,z_2}  \underbrace{\qq{z_1}}_{P(z_1|x,z_2)}
\;,
\eeq
where

\beqa
\qq{y,x,z_3,z_2} &=&\av{
{ P(y|z_{1,3},u_{1,4})P(z_3|z_2,u_3)
P(x|z_2,u_{1,2,3})P(z_2|u_{2,4})}}_{u.}
\\
&=&
P(y|z_3,z_1,x,z_2)P(z_3|z_1,x,z_2)P(x|z_2)P(z_2)
\;.
\eeqa

Eq.(\ref{eq-under-upsilons})
can be specialized using
the data from table Eq.(\ref{table-tp-fig2})
to get the following values for the upsilon terms:

\beq
\Upsilon_1 = P(y|\myhat{x})
\;,
\eeq

\beq
\Upsilon_2 = \sum_{z.}
\;,
\eeq

\beq
\Upsilon_3= P(y,z_3,z_2|\myhat{z}_1,\myhat{x})
\;,
\eeq

\beq
\Upsilon_4= P(z_1|x,z_2)
\;.
\eeq

\begin{claim}\label{cl-tp-fig2}
\beq
P(y,z_3,z_2|\myhat{z}_1,\myhat{x})
=P(y,z_3,z_2|\myhat{z}_1)
\;.
\label{eq-tp-fig2}
\eeq
\end{claim}
\proof

\begin{figure}[h]
    \begin{center}
    \epsfig{file=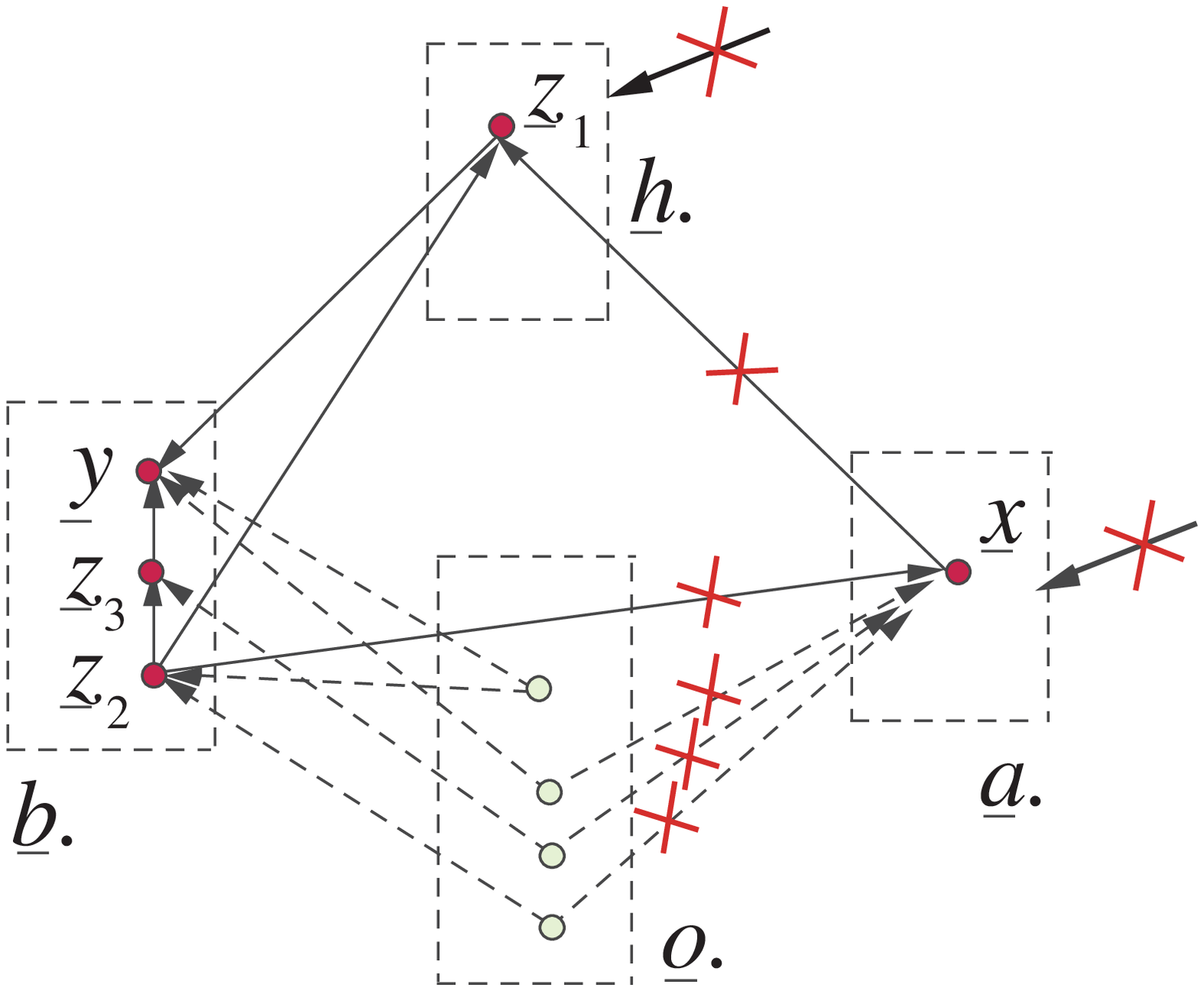, height=2.0in}
    \caption{A portrait of
    $G_{\myhat{\rvh}.,\myhat{\rva}.}$,
alluded to in Claim \ref{cl-tp-fig2}.
    }
    \label{fig-d-sep-tp-fig2}
    \end{center}
\end{figure}

See Ref.\cite{Tuc-intro}
where the 3 Rules
of Judea Pearl's do-calculus
are stated. Using the notation there,
let
$\rvb.=(\rvy,\rvz_3,\rvz_2),
\rva.=\rvx,
\rvh.=\rvz_1,
\rvi.=\emptyset,
\rvo.=\rvu. $.
Note that
$\rva.^-=\rva.-
\rv{an}(\rvi.,G_{\myhat{\rvh}.})=\rva.$
so
$G_{\myhat{\rvh}.,(\rva.^-)^\wedge}=
G_{\myhat{\rvh}.,\myhat{\rva}.}$.
Fig.\ref{fig-d-sep-tp-fig2} portrays
$G_{\myhat{\rvh}.,\myhat{\rva}.}$.
Apply Rule 3
to that figure.
\qed

Note that
\beq
P(y,z_3,z_2|\myhat{z}_1)=
\sum_x \qq{y,x,z_3,z_2}
\;.
\label{eq-fin-q0-tp-fig2}
\eeq

Combining the
upsilon values given, Eq.(\ref{eq-tp-fig2})
and Eq.(\ref{eq-fin-q0-tp-fig2}), we conclude that
Eq.(\ref{eq-under-upsilons}), when
fully specialized to this example,
becomes

\beq
P(y|\myhat{x})=\sum_{z.}
 \left[\sum_{x'}\qq{y,x',z_3,z_2}\right]
 P(z_1|x,z_2)
\;.
\eeq

\subsection{Example from Ref.\cite{R290L}-Fig.3}
\label{sec-tp-fig3}

\begin{figure}[h]
    \begin{center}
    \epsfig{file=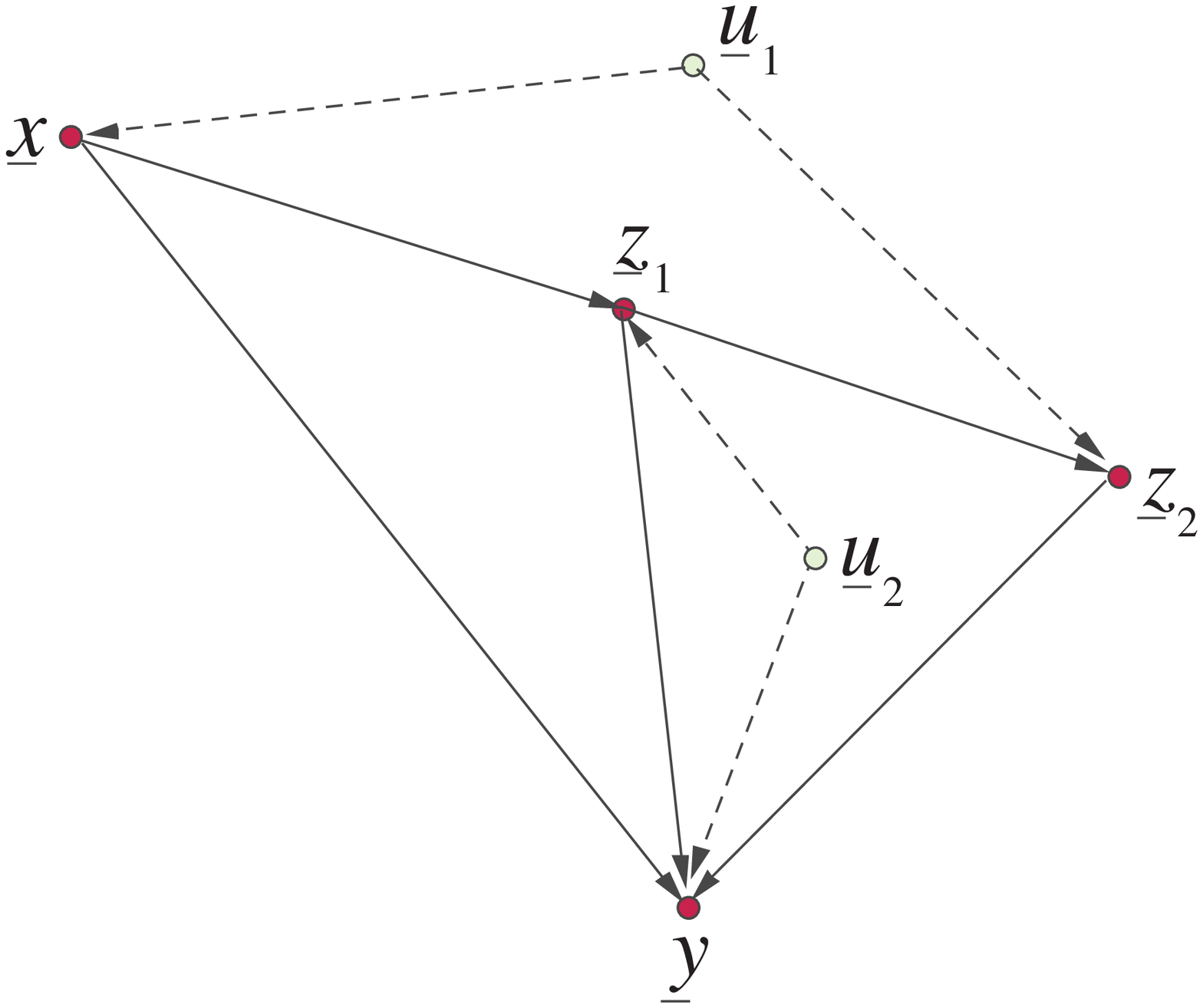, height=2.25in}
    \caption{
    Graph $G$ for Section \ref{sec-tp-fig3}.}
    \label{fig-tp-fig3}
    \end{center}
\end{figure}

In this example,
we want to
$P_{\rvv.}$ express
$P(y|\myhat{x})$
for the graph of Fig.\ref{fig-tp-fig3}.

For this example,
the following table applies.

\beq
\begin{array}{c||c|c||c|c||}
&\multicolumn{2}{c||}{\ul{\calv}.=\rvvcc{0}}
&\multicolumn{2}{c||}{\rvvcc{1}}
\\ \cline{2-5}
 &\rvx&\rvz_2&\rvz_1&\rvy
\\ \hline
\rvt&\checkmark & & &
\\ \hline
\rvs.&&&&\checkmark
\\ \hline
\rvd.& &\checkmark&\checkmark&\checkmark
\\\hline
\end{array}
\;
\label{table-tp-fig3}
\eeq

One possible topological ordering for the
visible nodes $\rvv.$
of this graph is

\beq
\rvy\larrow\rvz_2\larrow\rvz_1\larrow\rvx
\;
\eeq

According to Claim \ref{cl-pv-cc},
\beq
P(v.)= \qq{z_2,x}\qq{y,z_1}
\;,
\eeq
where

\beqa
\qq{z_2,x} &=&\av{
P(z_2|z_1,u_1)P(x|u_1)
}_{u_1}
\\
&=&
P(z_2|z_1,x)P(x)
\;,
\eeqa
and

\beqa
\qq{y,z_1} &=&\av{
P(y|x,z_2,z_1,u_2)P(z_1|x,u_2)
}_{u_2}
\\
&=&
P(y|z_2,z_1,x)P(z_1|x)
\;.
\eeqa

Eq.(\ref{eq-under-upsilons})
can be specialized using
the data from table Eq.(\ref{table-tp-fig3})
to get the following values for the upsilon terms:

\beq
\Upsilon_1 = P(y|\myhat{x})
\;,
\eeq

\beq
\Upsilon_2 = \sum_{z_1,z_2}
\;,
\eeq

\beq
\Upsilon_3= P(z_2|(z_1, y)^\wedge,\myhat{x})
\;,
\eeq

\beq
\Upsilon_4=\qq{y,z_1}
\;.
\eeq

\begin{claim}\label{cl-tp-fig3}
\beq
P(z_2|(z_1, y)^\wedge,\myhat{x})
=P(z_2|(z_1, y)^\wedge)
\;.
\label{eq-tp-fig3}
\eeq
\end{claim}
\proof

\begin{figure}[h]
    \begin{center}
    \epsfig{file=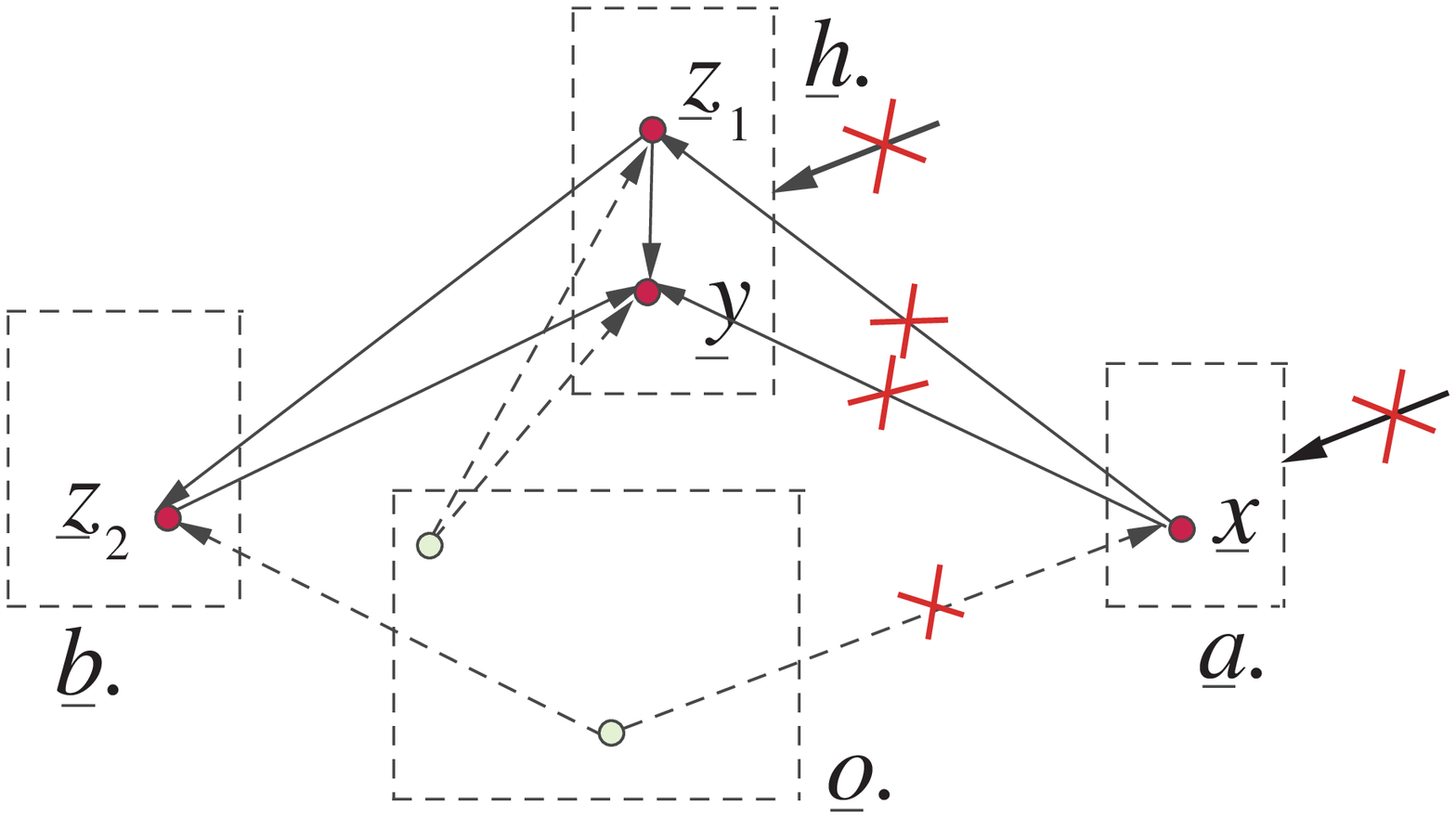, height=1.75in}
    \caption{A portrait of
    $G_{\myhat{\rvh}.,\myhat{\rva}.}$,
alluded to in Claim \ref{cl-tp-fig3}.
    }
    \label{fig-d-sep-tp-fig3}
    \end{center}
\end{figure}

See Ref.\cite{Tuc-intro}
where the 3 Rules
of Judea Pearl's do-calculus
are stated. Using the notation there,
let
$\rvb.=\rvz_2,
\rva.=\rvx,
\rvh.=(\rvz_1,\rvy),
\rvi.=\emptyset,
\rvo.=\rvu. $.
Note that
$\rva.^-=\rva.-
\rv{an}(\rvi.,G_{\myhat{\rvh}.})=\rva.$
so
$G_{\myhat{\rvh}.,(\rva.^-)^\wedge}=
G_{\myhat{\rvh}.,\myhat{\rva}.}$.
Fig.\ref{fig-d-sep-tp-fig3} portrays
$G_{\myhat{\rvh}.,\myhat{\rva}.}$.
Apply Rule 3
to that figure.
\qed

Note that
\beq
P(z_2|(z_1, y)^\wedge)=
\sum_x \qq{z_2,x}
\;.
\label{eq-fin-q0-tp-fig3}
\eeq

Combining the
upsilon values given, Eq.(\ref{eq-tp-fig3})
and Eq.(\ref{eq-fin-q0-tp-fig3}), we conclude that
Eq.(\ref{eq-under-upsilons}), when
fully specialized to this example,
becomes

\beq
P(y|\myhat{x})=\sum_{z_1,z_2}
\left[\sum_{x'}\qq{z_2,x'}\right] \qq{y,z_1}
\;.
\eeq

\subsection{Example from Ref.\cite{R290L}-Fig.6}
\label{sec-tp-fig6}

\begin{figure}[h]
    \begin{center}
    \epsfig{file=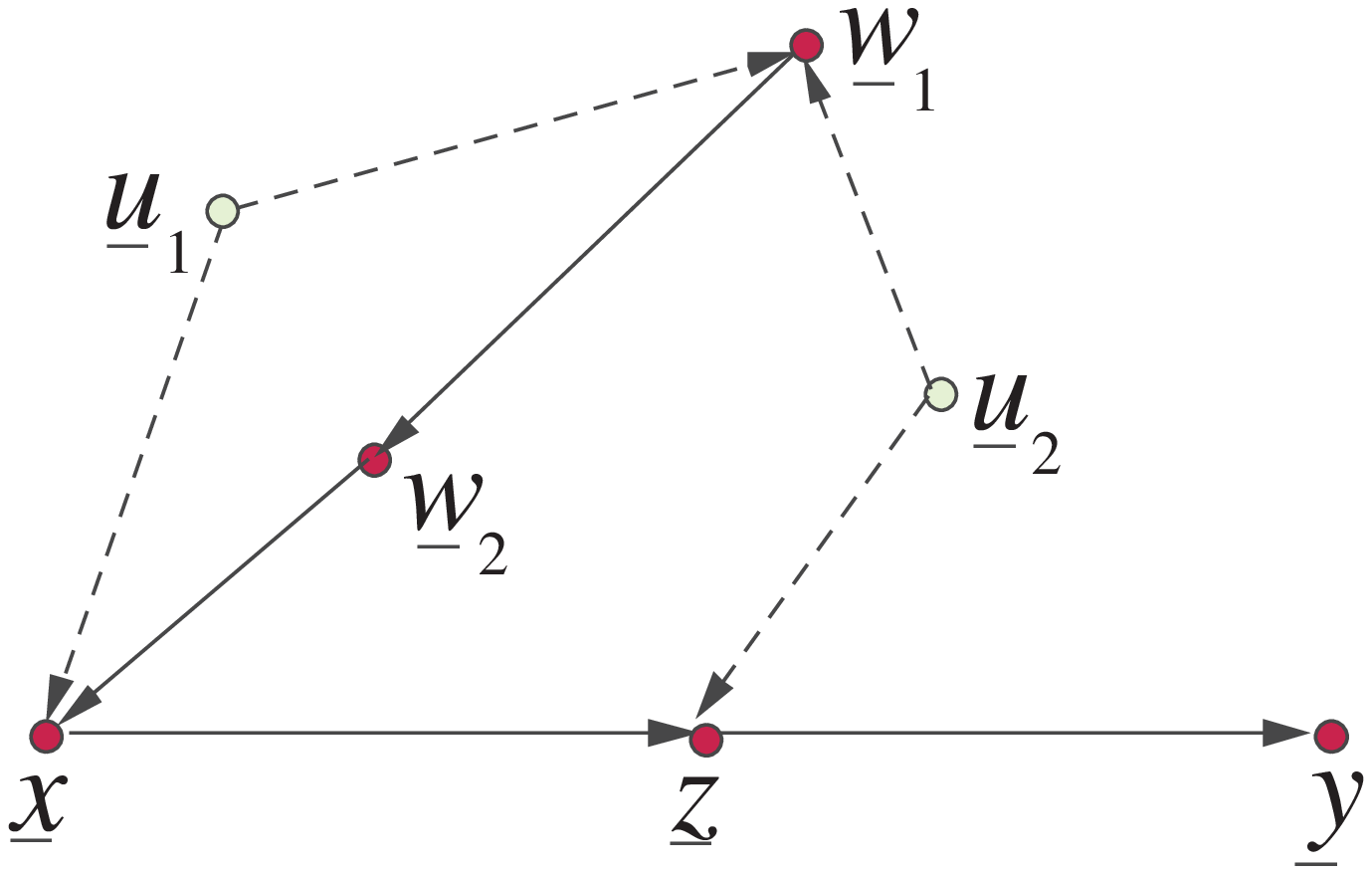, height=1.5in}
    \caption{
    Graph $G$ for Section \ref{sec-tp-fig6}.
    }
    \label{fig-tp-fig6}
    \end{center}
\end{figure}

In this example,
we want to
$P_{\rvv.}$ express
$P(y|\myhat{x})$
for the graph of Fig.\ref{fig-tp-fig6}.

For this example,
the following table applies.

\beq
\begin{array}{c||c|c|c||c||c||}
&\multicolumn{3}{c||}{\ul{\calv}.=\rvvcc{0}}
&\rvvcc{1}&\rvvcc{2}
\\ \cline{2-6}
 &\rvx&\rvz&\rvw_1&\rvw_2&\rvy
\\ \hline
\rvt&\checkmark & & & &
\\ \hline
\rvs.&&&&&\checkmark
\\ \hline
\rvd.& &\checkmark&&&\checkmark
\\\hline
\end{array}
\;
\label{table-tp-fig6}
\eeq

One possible topological ordering for the
visible nodes $\rvv.$
of this graph is

\beq
\rvy\larrow\rvz\larrow\rvx\larrow\rvw_2\larrow\rvw_1
\;
\eeq

According to Claim \ref{cl-pv-cc},
\beq
P(v.)= \qq{z,x,w_1}
\underbrace{\qq{w_2}}_{=P(w_2|w_1)}
\underbrace{\qq{y}}_{=P(y|z,x,w_{1,2})=P(y|z)}
\;,
\eeq
where

\beqa
\qq{z,x,w_1} &=&\av{
P(z|x,u_2)P(x|w_2,u_1)P(w_1|u_1,u_2)
}_{u_1,u_2}
\\
&=&
P(z|x, w_2, w_1)P(x|w_2,w_1)P(w_1)
\;.
\eeqa

Eq.(\ref{eq-under-upsilons})
can be specialized using
the data from table Eq.(\ref{table-tp-fig6})
to get the following values for the upsilon terms:

\beq
\Upsilon_1 = P(y|\myhat{x})
\;,
\eeq

\beq
\Upsilon_2 = \sum_{z}
\;,
\eeq

\beq
\Upsilon_3= P(z|(w_2, y)^\wedge,\myhat{x})
\;,
\eeq

\beq
\Upsilon_4=P(y|z)
\;.
\eeq

\begin{claim}\label{cl-tp-fig6}
\beq
P(z|(w_2, y)^\wedge,\myhat{x})
=P(z|(w_2, y)^\wedge,x)
\;.
\label{eq-tp-fig6}
\eeq
\end{claim}
\proof

\begin{figure}[h]
    \begin{center}
    \epsfig{file=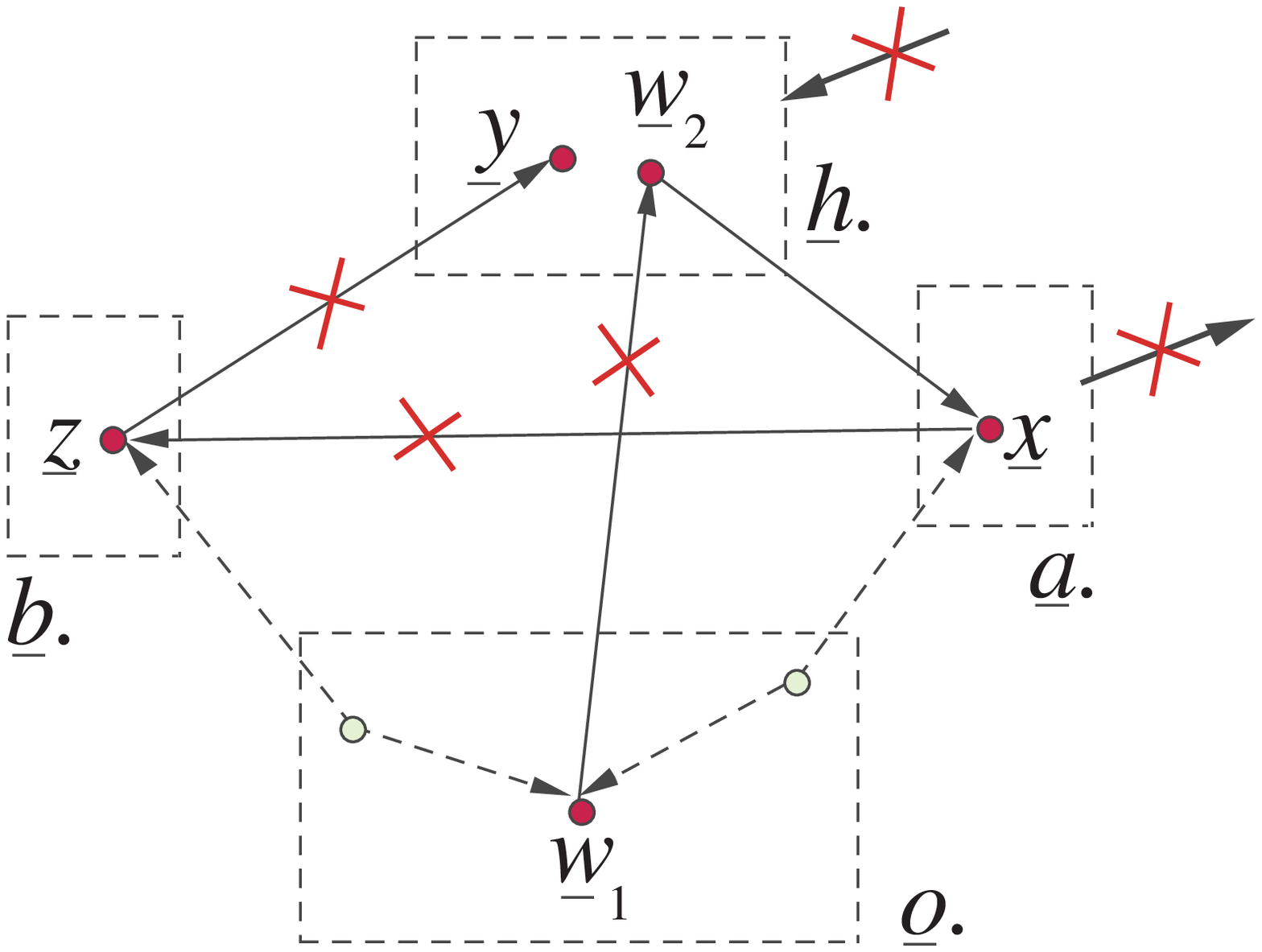, height=2.0in}
    \caption{A portrait of
    $G_{\myhat{\rvh}.,\myvee{\rva}.}$,
alluded to in Claim \ref{cl-tp-fig6}.
    }
    \label{fig-d-sep-tp-fig6}
    \end{center}
\end{figure}

See Ref.\cite{Tuc-intro}
where the 3 Rules
of Judea Pearl's do-calculus
are stated. Using the notation there,
let
$\rvb.=\rvz,
\rva.=\rvx,
\rvh.=(\rvw_2,\rvy),
\rvi.=\emptyset,
\rvo.=(\rvw_1,\rvu_1,\rvu_2) $.
Fig.\ref{fig-d-sep-tp-fig6} portrays
$G_{\myhat{\rvh}.,\myvee{\rva}.}$.
Apply Rule 2
to that figure.
\qed

Note that
\beq
P(z|(w_2, y)^\wedge,x)=
\frac{\sum_{w_1}\qq{z,x,w_1}}{\sum_{z}num}
\;.
\label{eq-fin-q0-tp-fig6}
\eeq

Combining the
upsilon values given, Eq.(\ref{eq-tp-fig6})
and Eq.(\ref{eq-fin-q0-tp-fig6}), we conclude that
Eq.(\ref{eq-under-upsilons}), when
fully specialized to this example,
becomes

\beq
P(y|\myhat{x})=\sum_{z}
\left[
\frac{\sum_{w_1}\qq{z,x,w_1}}{\sum_{z}num}
\right]
 P(y|z)
\;.
\eeq
Note
that the right hand
side of the last
equation
appears to depend on $w_2$ but doesn't.

\subsection{3 shark teeth graph
with middle tooth missing}
\label{sec-miss-teeth}

\begin{figure}[h]
    \begin{center}
    \epsfig{file=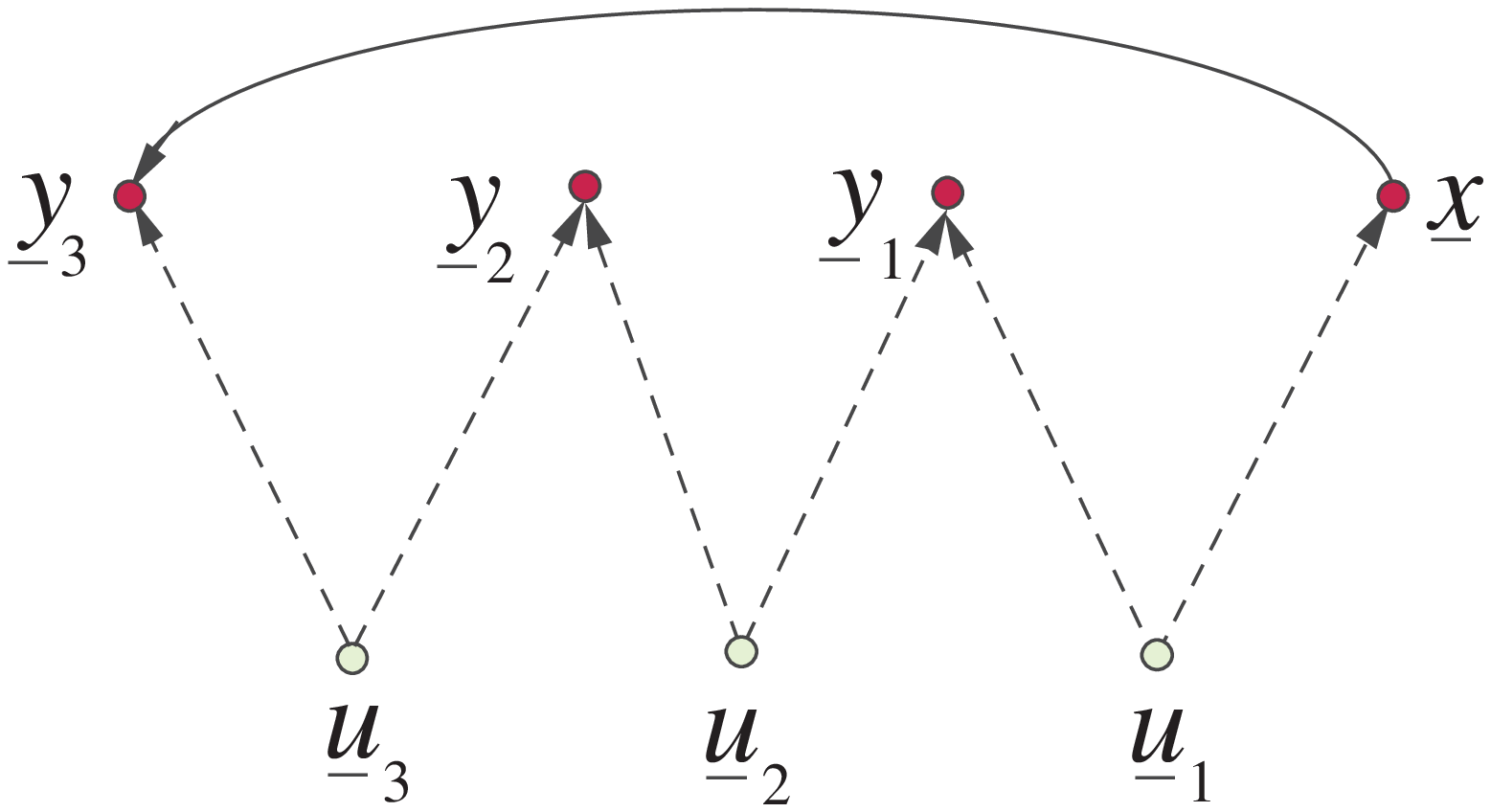, height=1.25in}
    \caption{
    Graph $G$ for Section \ref{sec-miss-teeth}.
    }
    \label{fig-miss-teeth}
    \end{center}
\end{figure}

In this example,
we want to
$P_{\rvv.}$ express
$P(y_{3,1}|\myhat{x})$
for the graph of Fig.\ref{fig-miss-teeth}\footnote{
Fig.\ref{fig-miss-teeth} is identical to Fig.\ref{fig-3teeth},
but we repeat it here for convenience.}. We refer to the set $\rvy.$
as teeth and to $\rvy_2$
as a missing tooth
in this example.

For this example,
the following table applies.

\beq
\begin{array}{c||c|c|c|c||}
&\multicolumn{4}{c||}{\ul{\calv}.=\rvvcc{0}}
\\ \cline{2-5}
 &\rvx&\rvy_1&\rvy_2&\rvy_3
\\ \hline
\rvt&\checkmark & & &
\\ \hline
\rvs.& &\checkmark&&\checkmark
\\ \hline
\rvd.& &\checkmark&&\checkmark
\\\hline
\end{array}
\;
\label{table-miss-teeth}
\eeq

One possible topological ordering for the
visible nodes $\rvv.$
of this graph is

\beq
\rvy_3\larrow\rvy_2\larrow\rvy_1\larrow\rvx
\;
\eeq

According to Claim \ref{cl-pv-cc},
\beq
P(v.)= \qq{y.,x}
\;,
\eeq
where

\beqa
\qq{y.,x} &=&\av{
P(y_3|x,u_3)P(y_2|u_3,u_2)P(y_1|u_2,u_1)P(x|u_1)
}_{u.}
\\
&=&
P(y_3|y_2,y_1)P(y_2|y_1,x)P(y_1|x)P(x)
\;.
\eeqa

Eq.(\ref{eq-under-upsilons})
can be specialized using
the data from table Eq.(\ref{table-miss-teeth})
to get the following values for the upsilon terms:

\beq
\Upsilon_1 = P(y_{3,1}|\myhat{x})
\;,
\eeq

\beq
\Upsilon_2 = 1
\;,
\eeq

\beq
\Upsilon_3= P(y_{3,1}|\myhat{x})
\;,
\eeq

\beq
\Upsilon_4=1
\;.
\eeq

\begin{claim}\label{cl-miss-teeth-r23-fail}
Rule 2 (resp., Rule 3) fails to prove that
$P(y_{3,1}|\myhat{x})$ equals
$P(y_{3,1}|x)$ (resp., $P(y_{3,1})$).
\end{claim}
\proof

\begin{figure}[h]
    \begin{center}
    \epsfig{file=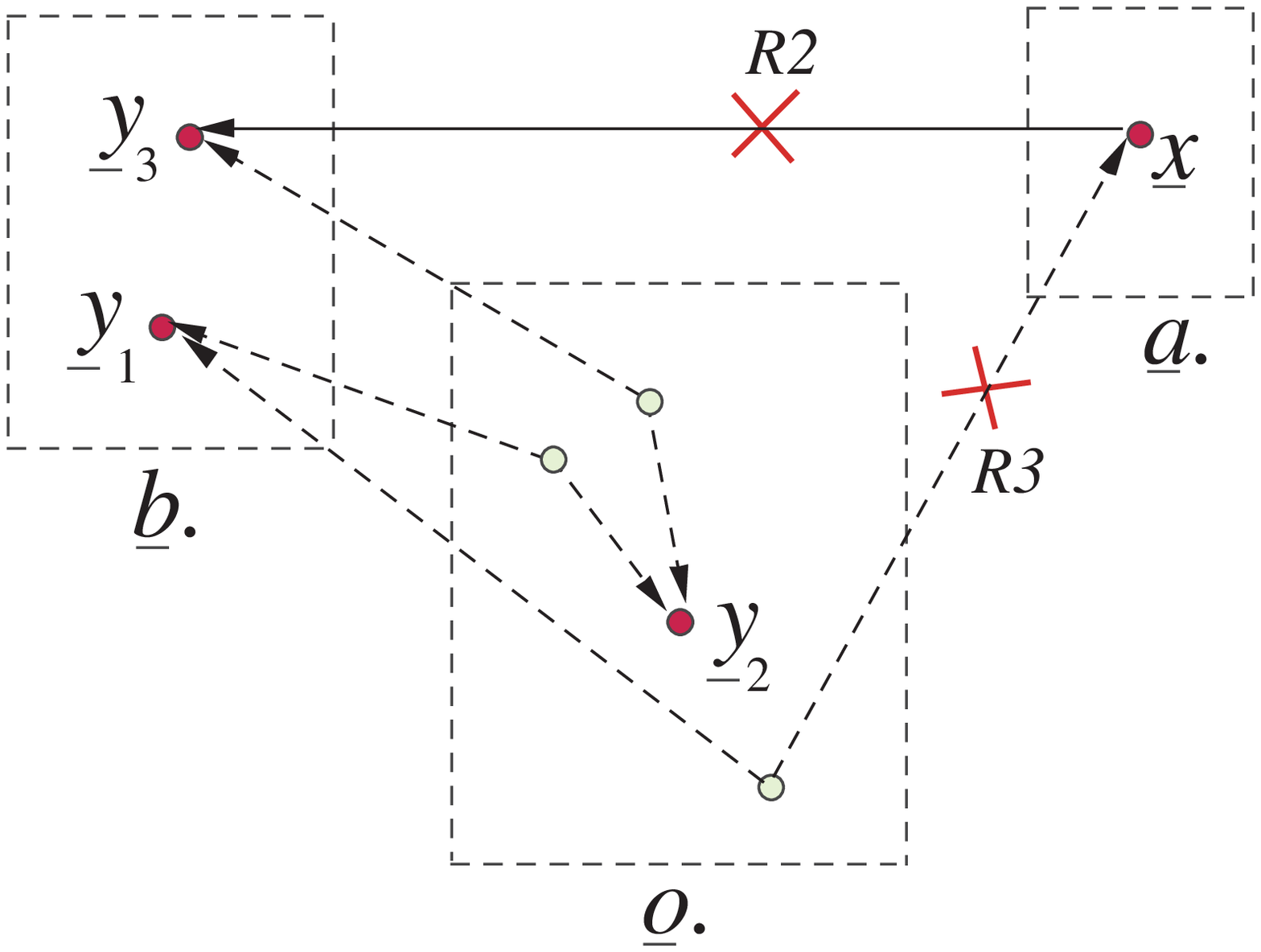, height=2in}
    \caption{A portrait of
    $G_{\myvee{\rva}.}$ for Rule 2 and
    $G_{\myhat{\rva}.}$ for Rule 3,
    alluded to in Claim \ref{cl-miss-teeth-r23-fail}.
    }
    \label{fig-miss-teeth-r23-fail}
    \end{center}
\end{figure}

See Ref.\cite{Tuc-intro}
where the 3 Rules
of Judea Pearl's do-calculus
are stated. Using the notation there,
let
$\rvb.=\rvy_{3,1},
\rva.=\rvx,
\rvh.=\emptyset,
\rvi.=\emptyset,
\rvo.=(\rvy_2,\rvu.) $.
One can see from Fig.\ref{fig-miss-teeth-r23-fail}
that
there exists an unblocked path
from $\rva.$ to $\rvb.$ at fixed $(\rvh.,\rvi.)$
in
$G_{\myhat{\rvh}.,\myvee{\rva}.}=
G_{\myvee{\rva}.}$ (resp.,
$G_{\myhat{\rvh}.,(\rva.^-)^\wedge}=
G_{\myhat{\rva}.}$) so Rule 2 (resp., Rule 3)
cannot be used.
\qed

\begin{figure}[h]
    \begin{center}
    \epsfig{file=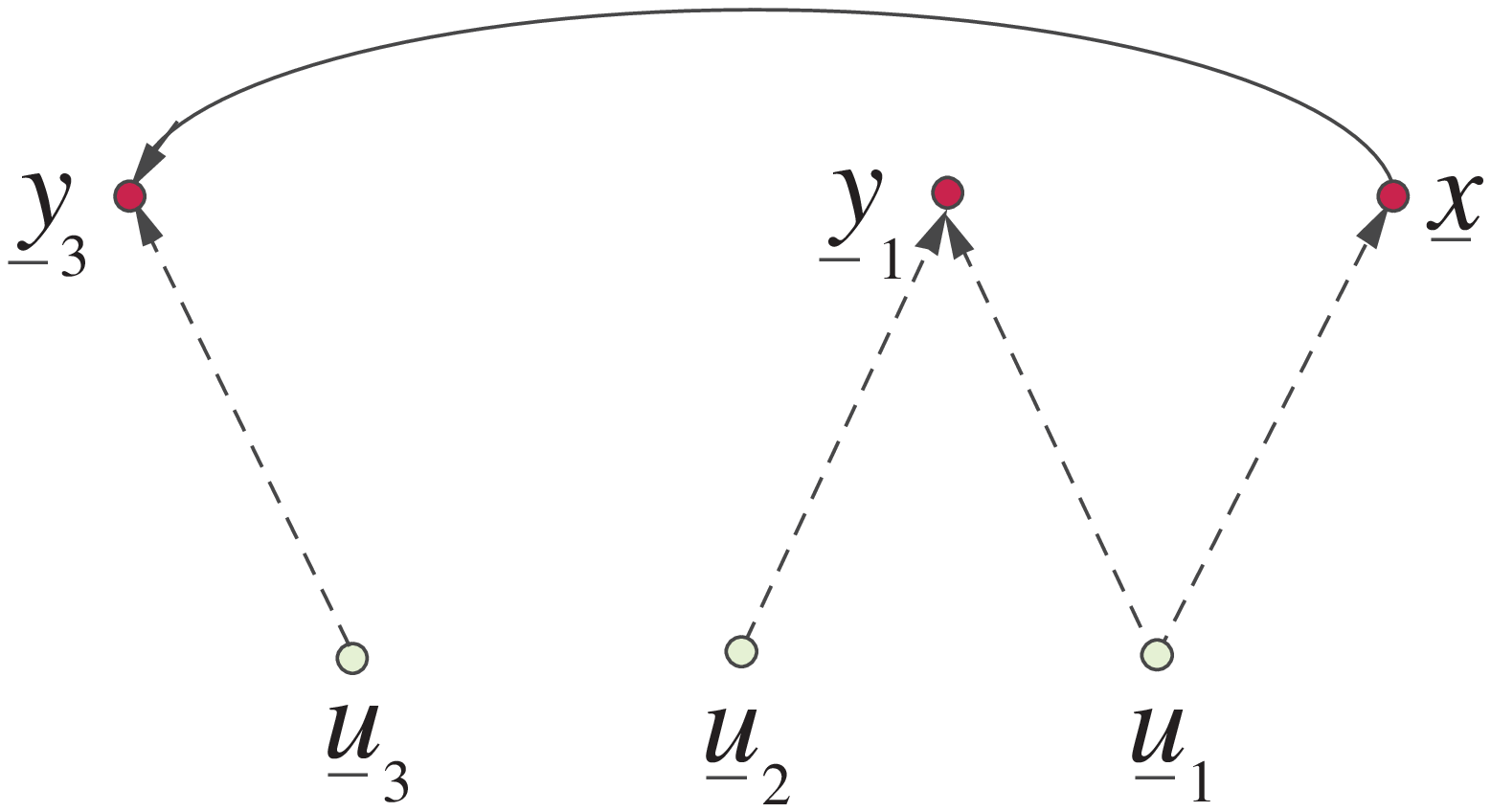, height=1.25in}
    \caption{
    Graph $G_{\rvv.^-}$ for Section \ref{sec-miss-teeth}.
    }
    \label{fig-miss-teeth-pruned}
    \end{center}
\end{figure}

At this point, instead of giving up,
we prune
the graph $G_{\rvv.}$
of Fig.\ref{fig-miss-teeth}
to $G_{\rvv.^-}$
where $\rvv.^- = \ol{\rv{an}}(\rvy_{3,1}\cup\rvx, G_{\rvv.})$
to obtain the graph of Fig.\ref{fig-miss-teeth-pruned}.

For this new graph,
the following table applies.

\beq
\begin{array}{c||c|c||c||}
&\multicolumn{2}{c||}{\ul{\calv}.=\rvvcc{0}}
&\rvvcc{1}
\\ \cline{2-4}
 &\rvx&\rvy_1&\rvy_3
\\ \hline
\rvt&\checkmark & &
\\ \hline
\rvs.& &\checkmark&\checkmark
\\ \hline
\rvd.& &\checkmark&\checkmark
\\\hline
\end{array}
\;
\label{table-miss-teeth-pruned}
\eeq

One possible topological ordering for the
visible nodes $\rvv.$
of this graph is

\beq
\rvy_3\larrow\rvy_1\larrow\rvx
\;
\eeq

According to Claim \ref{cl-pv-cc},
\beq
P(v.)= \qq{y_1,x}
\underbrace{\qq{y_3}}_{=P(y_3|y_1,x)=P(y_3|x)}
\;,
\eeq
where

\beqa
\qq{y_1,x} &=&\av{
P(y_1|u_2,u_1)P(x|u_1)
}_{u.}
\\
&=&
P(y_1|x)P(x)
\;
\eeqa

Eq.(\ref{eq-under-upsilons})
can be specialized using
the data from table Eq.(\ref{table-miss-teeth-pruned})
to get the following values for the upsilon terms:

\beq
\Upsilon_1 = P(y_{3,1}|\myhat{x})
\;,
\eeq

\beq
\Upsilon_2 = 1
\;,
\eeq

\beq
\Upsilon_3= P(y_1|\myhat{y_3},\myhat{x})
\;,
\eeq

\beq
\Upsilon_4=P(y_3|x)
\;.
\eeq

\begin{claim}\label{cl-miss-teeth-r3}
\beq
P(y_1|\myhat{y_3},\myhat{x})
=P(y_1|\myhat{y_3})
\;.
\label{eq-miss-teeth-r3}
\eeq
\end{claim}
\proof

See Ref.\cite{Tuc-intro}
where the 3 Rules
of Judea Pearl's do-calculus
are stated. Using the notation there,
let
$\rvb.=\rvy_1,
\rva.=\rvx,
\rvh.=\rvy_3,
\rvi.=\emptyset,
\rvo.=\rvu. $.
Note that
$\rva.^-=\rva.-
\rv{an}(\rvi.,G_{\myhat{\rvh}.})=\rva.$
so
$G_{\myhat{\rvh}.,(\rva.^-)^\wedge}=
G_{\myhat{\rvh}.,\myhat{\rva}.}$.
Fig.\ref{fig-d-sep-tp-fig3} portrays
$G_{\myhat{\rvh}.,\myhat{\rva}.}$.
Apply Rule 3
to that figure.
\qed

\begin{figure}[h]
    \begin{center}
    \epsfig{file=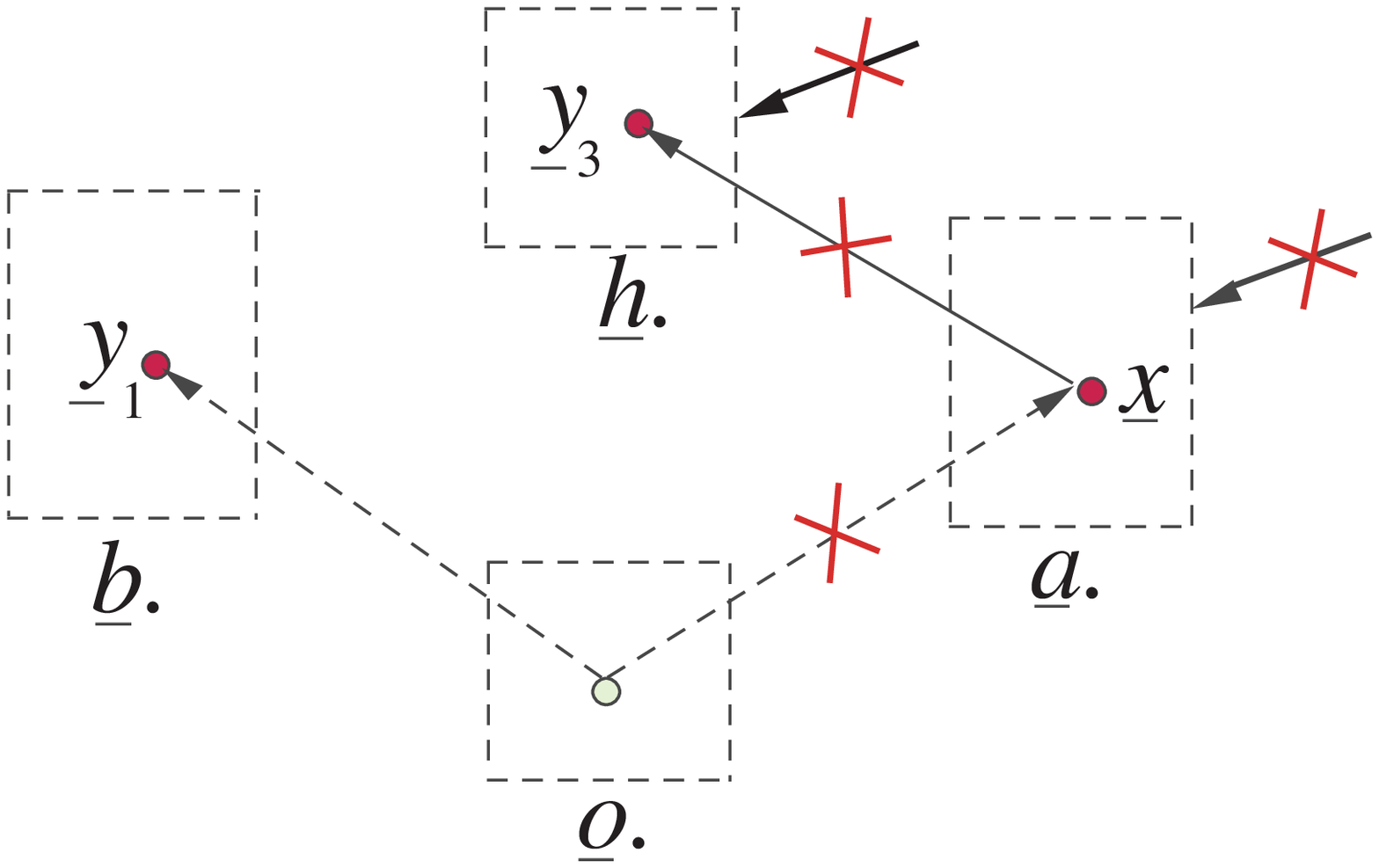, height=1.75in}
    \caption{A portrait of
    $G_{\myhat{\rvh}.,\myhat{\rva}.}$,
alluded to in Claim \ref{cl-miss-teeth-r3}.
    }
    \label{fig-miss-teeth-r3}
    \end{center}
\end{figure}

Note that
\beqa
P(y_1|\myhat{y_3})&=&
\sum_x \qq{y_1,x}
\\
&=&
P(y_1)
\;.
\label{eq-fin-miss-teeth-r3}
\eeqa

Combining the
upsilon values given, Eq.(\ref{eq-miss-teeth-r3})
and Eq.(\ref{eq-fin-miss-teeth-r3}), we conclude that
Eq.(\ref{eq-under-upsilons}), when
fully specialized to this example,
becomes

\beq
P(y_{3,1}|\myhat{x})=
P(y_1) P(y_3|x)
\;.
\eeq

\section{Appendix- Examples of \\
Non-identifiable probabilities}
\label{app-not-id}

In this appendix, we present
several examples of non-identifiable
uprooted probabilities $P(s.|\myhat{t})$.
For each example, we will
give two specific models which have the same
probability of visible nodes $P(v.)$
but which yield different $P(s.|\myhat{t})$,
thus proving  that
$P(s.|\myhat{t})$ is
not $P_{\rvv.}$ expressible,
and, thus,
not identifiable.

One of our examples,
the one in Section \ref{sec-tp-fig9},
is claimed erroneously by
Ref.\cite{R290L}
to be an example of an identifiable probability.
In Section \ref{sec-tp-fig9},
we prove that
the probability being sought in that case
is really not identifiable
but the algorithm of Ref.\cite{R290L}
somehow fails to detect this fact.

\subsection{One shark tooth graph}
\label{sec-1tooth}

\begin{figure}[h]
    \begin{center}
    \epsfig{file=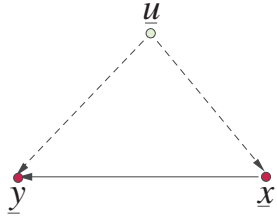, height=1.25in}
    \caption{Graph $G$ for Section \ref{sec-1tooth}.
    }
    \label{fig-1tooth}
    \end{center}
\end{figure}

In this example,
we show that $P(y|\myhat{x})$
is not identifiable
for the graph of Fig.\ref{fig-1tooth}.

For this example,
the following table applies.

\beq
\begin{array}{c||c|c||}
&\multicolumn{2}{c||}{\ul{\calv}.=\rvvcc{0}}
\\ \cline{2-3}
 &\rvx&\rvy
\\ \hline
\rvt&\checkmark &
\\ \hline
\rvs.& &\checkmark
\\ \hline
\rvd.& &\checkmark
\\\hline
\end{array}
\;
\label{table-1tooth}
\eeq

One possible topological ordering for the
visible nodes $\rvv.$
of this graph is

\beq
\rvy\larrow\rvx
\;
\eeq

According to Claim \ref{cl-pv-cc},
\beq
P(v.)=\qq{x,y}
\;,
\eeq
where

\beqa
\qq{x,y}&=& \av{P(y|x,u)P(x|u)}_u
\\
&=&
P(y|x)P(x)
\;.
\eeqa

Note that
\beq
P(y|\myhat{x})=
\av{P(y|x,u)}_u
\;,
\eeq
and

\beq
P(\cald.|\calv.^{c\wedge},\myhat{t})=
P(y|\myhat{x})
\;.
\eeq

\begin{claim}\label{cl-1tooth}
Rule 2 (resp., Rule 3) fails to prove that
$P(y|\myhat{x})$ equals
$P(y|x)$ (resp., $P(y)$).
\end{claim}
\proof

\begin{figure}[h]
    \begin{center}
    \epsfig{file=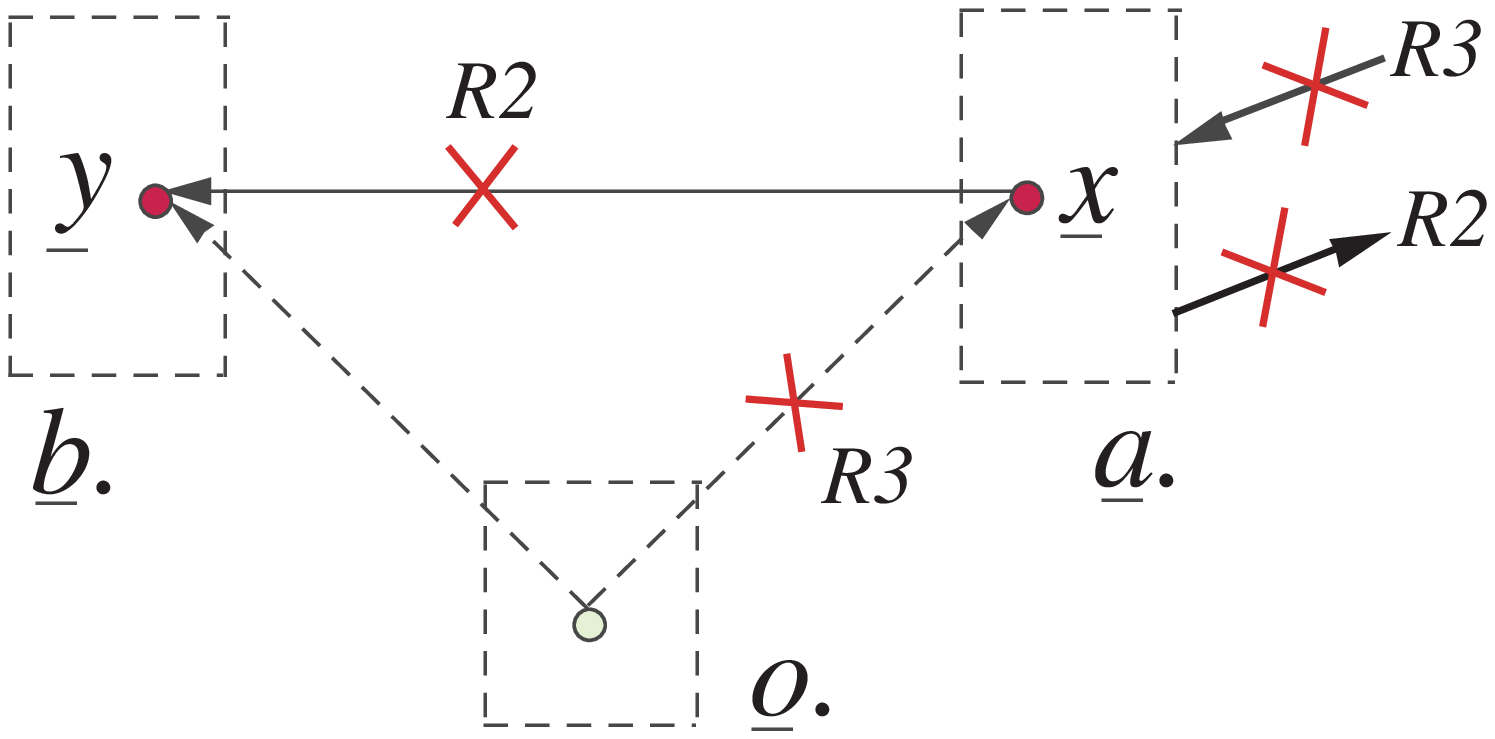, height=1.25in}
    \caption{A portrait of
    $G_{\myvee{\rva}.}$ for Rule 2 and
    $G_{\myhat{\rva}.}$ for Rule 3,
    alluded to in Claim \ref{cl-1tooth}.
    }
    \label{fig-d-sep-1tooth}
    \end{center}
\end{figure}

See Ref.\cite{Tuc-intro}
where the 3 Rules
of Judea Pearl's do-calculus
are stated. Using the notation there,
let
$\rvb.=\rvy,
\rva.=\rvx,
\rvh.=\emptyset,
\rvi.=\emptyset,
\rvo.=\rvu $.
One can see from Fig.\ref{fig-d-sep-1tooth}
that
there exists an unblocked path
from $\rva.$ to $\rvb.$ at fixed $(\rvh.,\rvi.)$
in
$G_{\myhat{\rvh}.,\myvee{\rva}.}=
G_{\myvee{\rva}.}$ (resp.,
$G_{\myhat{\rvh}.,(\rva.^-)^\wedge}=
G_{\myhat{\rva}.}$) so Rule 2 (resp., Rule 3)
cannot be used.
\qed

\begin{claim}\label{cl-id-1tooth}
$P(y|\myhat{x})$ for the
graph of Fig.\ref{fig-1tooth} is not identifiable
\end{claim}
\proof

Consider a model for the graph
 of Fig.\ref{fig-1tooth} with $y,x,u\in Bool$
 and

\beq
\left\{
\begin{array}{l}
P(y|x,u)=\delta_{y}^{x\wedge u}=\delta_{y}^{xu},
\\
P(x|u)=\delta_{x}^{u},
\\
P(u) =\frac{1}{2}
\end{array}
\right.
\;.
\label{eq-probs-1tooth}
\eeq
Note that for this model

\beq
P(v.)=P(x,y) = \frac{1}{2}\sum_u \delta_{y}^{xu}\delta_{x}^{u}
=
\frac{\delta_{y}^{x}}{2}
\;,
\eeq
and

\beq
P(y|\myhat{x})=
\frac{1}{2}\sum_u\delta_{y}^{xu}
= \frac{\delta_{y}^{0} + \delta_{y}^{x}}{2}
\;.
\eeq
One can
define a second model
$P'$ with

\beq
\left\{
\begin{array}{l}
P'(y|x,u)=P(\ol{y}|\ol{x},u)
\\
P'(x|u) = P(\ol{x}|u)
\\
P'(u) = P(u)
\end{array}
\right.
\;.
\eeq
Note that $P'(v.) = \frac{\delta_{y}^{x}}{2} =
P(v.)$ but
$P'(y|\myhat{x})=
\frac{\delta_{y}^{1} + \delta_{y}^{x}}{2}
\neq P(y|\myhat{x})$.
Hence, there exist
two models for the graph of
Fig.\ref{fig-1tooth} that
have the same $P(v.)$ but
different $P(y|\myhat{x})$.
Thus, $P(y|\myhat{x})$
is not $P_{\rvv.}$ expressible.
\qed

\begin{claim}
There exists a model for the
graph of Fig.\ref{fig-1tooth}
for which $H(\rvy:\myhat{\rvx})<0$.
\end{claim}
\proof

Consider a model for the graph
 of Fig.\ref{fig-1tooth} with
 the same node transition probabilities
 as those given by Eq.(\ref{eq-probs-1tooth}), except for the
 following change
\beq
P(y|x,u)=\delta_{y}^{x\oplus u}
\;.
\eeq
Note that for this model

\beq
P(v.) =P(x,y)= \frac{1}{2}\sum_u \delta_{y}^{x\oplus u}
\delta_{x}^{u}
=
\frac{\delta_{y}^{0}}{2}
\;,
\eeq
and

\beq
P(y|\myhat{x})=
\frac{1}{2}\sum_u\delta_{y}^{x\oplus u}
= \frac{1}{2}
\;.
\eeq
Therefore,

\beq
H(\rvy:\myhat{\rvx})=
\sum_{x,y}P(x,y)\ln\frac{P(y|\myhat{x})}{P(y)}
=\sum_{x,y}
\frac{\delta_{y}^{0}}{2}
\ln \frac{\frac{1}{2}}{\delta_{y}^{0}}
=
-\ln 2 <0
\;.
\eeq
\qed

\subsection{3 shark teeth graph
(see Appendix A of Ref.\cite{R290L})}
\label{sec-3teeth}

\begin{figure}[h]
    \begin{center}
    \epsfig{file=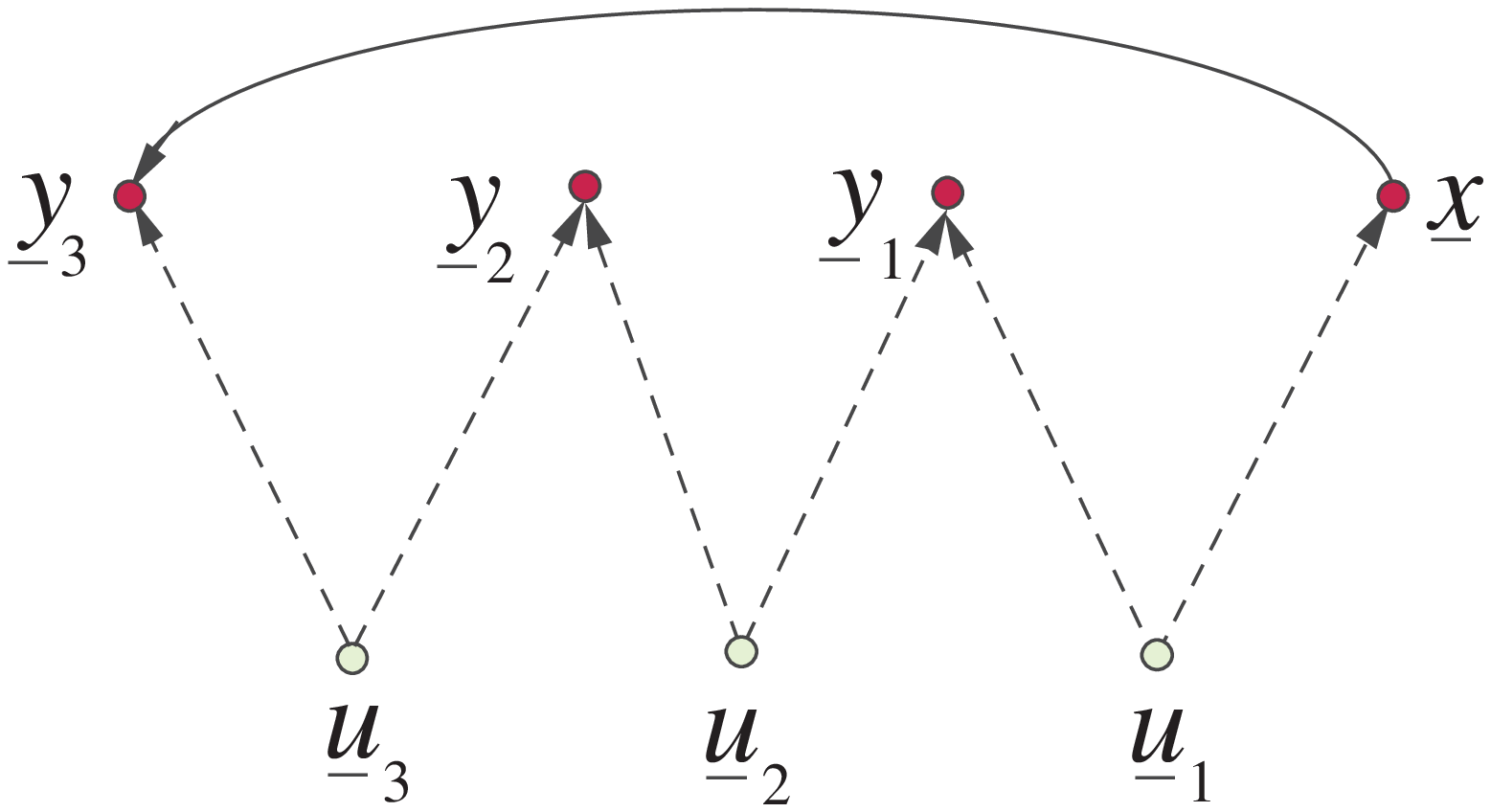, height=1.25in}
    \caption{
    Graph $G$ for Section \ref{sec-3teeth}.
    }
    \label{fig-3teeth}
    \end{center}
\end{figure}

In this example,
we show that $P(y.|\myhat{x})$
is not identifiable
for the graph of Fig.\ref{fig-3teeth}\footnote{
Fig.\ref{fig-3teeth} is identical to Fig.\ref{fig-miss-teeth},
but we repeat it here for convenience.}.
This section generalizes
the results of the previous section from
a graph with ``one tooth" to a graph
with ``3 teeth".
It will become clear
as we proceed that the results of this
section generalize easily to
a graph with an arbitrary number
$N\geq 1$ of teeth.

For this example,
the following table applies.

\beq
\begin{array}{c||c|c||c||c||}
&\multicolumn{4}{c||}{\ul{\calv}.=\rvvcc{0}}
\\ \cline{2-5}
 &\rvx&\rvy_3&\rvy_2&\rvy_1
\\ \hline
\rvt&\checkmark &&&
\\ \hline
\rvs.& &\checkmark&\checkmark&\checkmark
\\ \hline
\rvd.& &\checkmark&\checkmark&\checkmark
\\\hline
\end{array}
\;
\label{table-3teeth}
\eeq

One possible topological ordering for the
visible nodes $\rvv.$
of this graph is

\beq
\rvy_3\larrow\rvy_2\larrow\rvy_1\larrow\rvx
\;
\eeq

According to Claim \ref{cl-pv-cc},
\beq
P(v.)=\qq{y.,x}
\;,
\eeq
where

\beqa
\qq{y.,x}&=& \av{
P(y_3|x,u_3)
P(y_2|u_3,u_2)
P(y_1|u_2,u_1)
P(x|u_1)}_{u.}
\\
&=&
P(y.|x)P(x)
\;.
\eeqa

Note that
\beq
P(y.|\myhat{x})=
\av{
P(y_3|x,u_3)
P(y_2|u_3,u_2)
P(y_1|u_2,u_1)
}_{u.}
\;,
\eeq
and

\beq
P(\cald.|\calv.^{c\wedge},\myhat{t})=
P(y.|\myhat{x})
\;.
\eeq

\begin{claim}\label{cl-3teeth}
Rule 2 (resp., Rule 3) fails to prove that
$P(y.|\myhat{x})$ equals
$P(y.|x)$ (resp., $P(y.)$).
\end{claim}
\proof

\begin{figure}[h]
    \begin{center}
    \epsfig{file=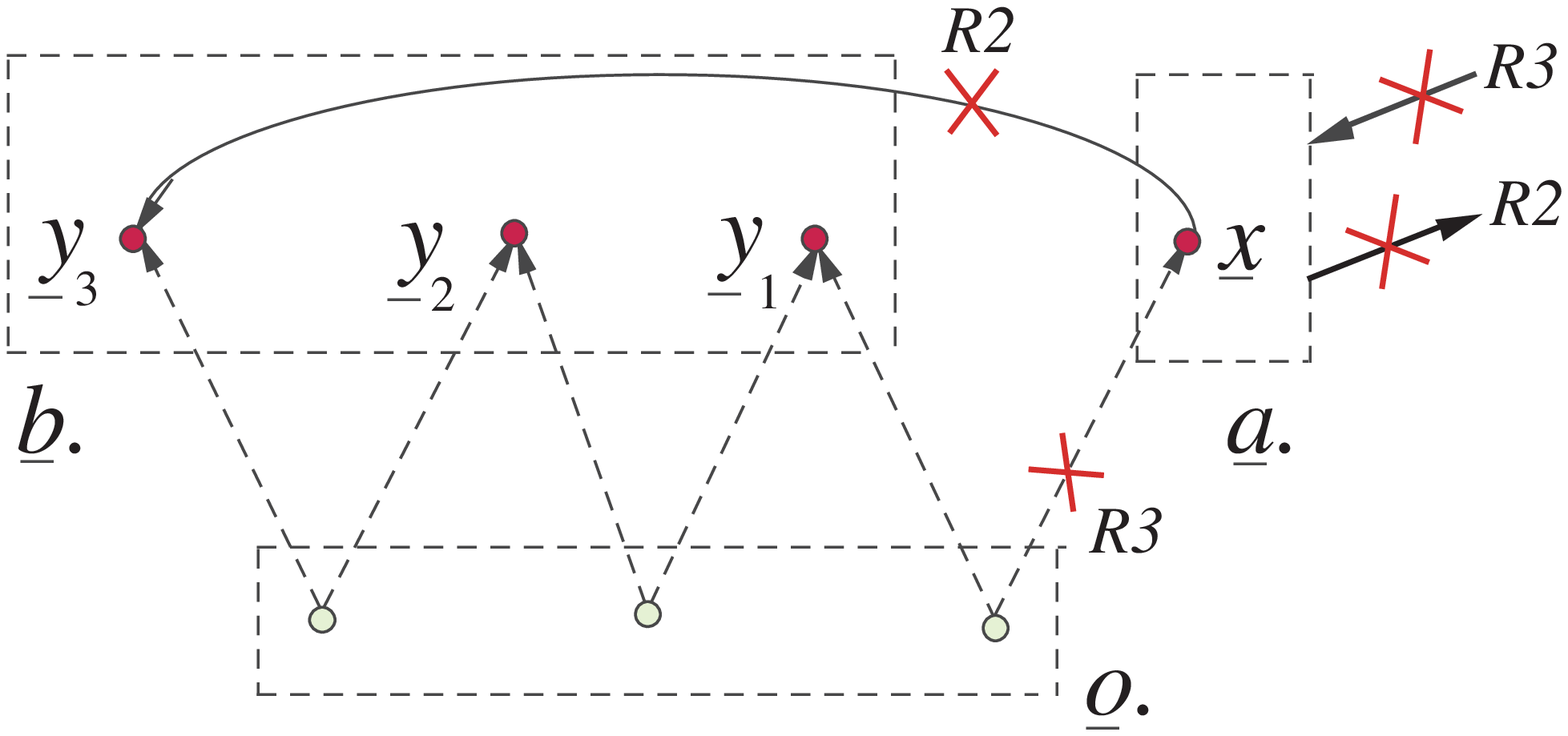, height=1.5in}
    \caption{A portrait of
    $G_{\myvee{\rva}.}$ for Rule 2 and
    $G_{\myhat{\rva}.}$ for Rule 3,
    alluded to in Claim \ref{cl-3teeth}.
    }
    \label{fig-d-sep-3teeth}
    \end{center}
\end{figure}

See Ref.\cite{Tuc-intro}
where the 3 Rules
of Judea Pearl's do-calculus
are stated. Using the notation there,
let
$\rvb.=\rvy.,
\rva.=\rvx,
\rvh.=\emptyset,
\rvi.=\emptyset,
\rvo.=\rvu. $.
One can see from Fig.\ref{fig-d-sep-3teeth}
that
there exists an unblocked path
from $\rva.$ to $\rvb.$ at fixed $(\rvh.,\rvi.)$
in
$G_{\myhat{\rvh}.,\myvee{\rva}.}=
G_{\myvee{\rva}.}$
(resp., $G_{\myhat{\rvh}.,(\rva.^-)^\wedge}=
G_{\myhat{\rva}.}$)
 so Rule 2 (resp., Rule 3)
cannot be used.
\qed

\begin{claim}\label{cl-not-id-3teeth}
$P(y.|\myhat{x})$ for the
graph of Fig.\ref{fig-3teeth} is not identifiable
\end{claim}
\proof

For the graph
 of Fig.\ref{fig-3teeth},
 we have

 \beq
 P(v.)=
 P(y.,x)=
 \av{
 P(y_3|x,u_3)
 P(y_2|u_3,u_2)
 P(y_1|u_2,u_1)
 P(x|u_1)
 }_{u.}
 \;,
 \eeq
 and

 \beq
 P(y.|\myhat{x})=
 \av{
 P(y_3|x,u_3)
 P(y_2|u_3,u_2)
 P(y_1|u_2,u_1)
 }_{u.}
 \;.
 \eeq

Consider $u_j,x,y_j\in Bool$.
Let

\beq
P(u_j)=\frac{1}{2}
\;
\eeq
for $j=1,2,3$
and

\beq
P(x|u_1) = \delta_x^{u_1}
\;.
\eeq

Also let

\beq
P(y_3|x,u_3)=
\left[
M_3(y_3)
\right]_{x,u_3}
\;,
\eeq
where\footnote{The definition of the matrices $\Omega$ and
$\cala$ and some of their properties
are given in Section \ref{sec-basic}.}

\beq
M_3(y_3)=
\Omega
\left[
\begin{array}{cc}
(-1)^{y_3}&0\\
(-1)^{y_3}g_0&1
\end{array}
\right]
\Omega^T
\;.
\label{eq-lower-triangular-m}
\eeq
We will assume that
$g_0$ is a real number that is
much smaller than 1 in absolute value.
Note that
$\sum_{y_3} M_3(y_3) = 2\cala =
\left[
\begin{array}{cc}
1&1\\
1&1
\end{array}
\right]
$
as expected
since $P(y_3|x,u_3)$
is a probability
distribution.

Also let

\beq
P(y_2|u_{3},u_2)=
\left[
M_2(y_2)
\right]_{u_{3},u_2}
\;,
\eeq
where

\beq
M_2(y_2)=
\Omega
\left[
\begin{array}{cc}
-(1)^{y_2}&0\\
0&1
\end{array}
\right]
\Omega^T
\;.
\label{eq-diag-m}
\eeq
Note that
$
\sum_{y_2} M_2(y_2) = 2\cala
$
as expected
since $P(y_2|u_{3},u_2)$
is a probability
distribution.

Also let

\beq
P(y_1|u_2,u_1)=
\left[
M_1(y_1)
\right]_{u_2,u_1}
\;,
\eeq
where

\beq
M_1(y_1)=
\Omega
\left[
\begin{array}{cc}
(-1)^{y_1}&(-1)^{y_1}(-g_0)\\
0&1
\end{array}
\right]
\Omega^T
\;.
\label{eq-upper-triangular-m}
\eeq
Note that
$
\sum_{y_1} M_1(y_1) = 2\cala
$
as expected
since $P(y_1|u_2,u_1)$
is a probability
distribution.

When dealing with $N>3$
teeth $\rvy_N, \rvy_{N-1}, \ldots, \rvy_1$, one can use

\begin{itemize}
\item
$M(y_N)=\Omega(\mbox{ lower triangular matrix })\Omega^T$, as we did for $M(y_3)$
in Eq.(\ref{eq-lower-triangular-m}).
\item
For $j\in\{N-1,N-2,\ldots,2\}$,
$M(y_j)=\Omega(\mbox{ diagonal matrix })\Omega^T$, as we did for $M(y_2)$
in Eq.(\ref{eq-diag-m}).
\item
$M(y_1)=\Omega(\mbox{ upper triangular matrix })\Omega^T$, as we did for $M(y_1)$
in Eq.(\ref{eq-upper-triangular-m}).
\end{itemize}

If we define
 \beq
 M(y.)=M_3(y_3) M_2(y_2) M_1(y_1)
 \;,
 \eeq
 then

 \beq
 P(y.,x)=\frac{1}{2^3}
 [M(y.)]_{x,x}
 \;,
 \eeq
and

\beq
P(y.|\myhat{x})=\frac{1}{2^3}
 \sum_{u_1} [M(y.)]_{x,u_1}
 =
 \frac{1}{2^3}
 \left([M(y.)]_{x,x}
 +
 [M(y.)]_{x,\ol{x}}
 \right)
 \;.
 \eeq
 (As usual, $\ol{x} = 1-x$).
Let

\beq
\sigma = (-1)^{y_1+y_2+y_3}
\;.
\eeq
Then

\beqa
M(y.)&=&
 \Omega
 \left[
 \begin{array}{cc}
\sigma& \sigma(-g_0) \\
\sigma g_0 & 1 -\sigma g_0^2
 \end{array}
 \right]
 \Omega^T
 \\
 &=&
(1-\sigma g_0^2)\cala
+
\sigma
\left[
\begin{array}{cc}
0 & -g_0\\
g_0 & 0
\end{array}
\right]
+
\frac{\sigma}{2}
\left[
\begin{array}{cc}
1 & -1\\
-1 & 1
\end{array}
\right]
\;,
\eeqa
so

 \beq
 P(y., x)=
 \frac{1}{2^4}
 [1 + \sigma(1-g_0^2)]
 \;,
 \eeq
 and

  \beq
 P(y.|\myhat{x})=
 \frac{1}{2^3}
 \left[1 - \sigma g_0((-1)^x + g_0)\right]
 \;.
 \eeq
Let $0<|g_0|<<1$. To first
order in $g_0$,
when we change $g_0$,
$P(y.,x)$ remains fixed but
$P(y.|\myhat{x})$ changes.\footnote{
Note that in order to prove that
$P(y.|\myhat{x})$
is not identifiable
for the $N$-shark teeth graph,
Appendix A of Ref.\cite{R290L}
attempts to find a
model for which $P(y.,x)$
is the same for all $(y.,x)$.
I wasn't able to prove non-identifiability
making that assumption.
The above proof does not make that very strong assumption.}
Thus, $P(y|\myhat{x})$
is not $P_{\rvv.}$ expressible.
\qed

\begin{claim}\label{cl-neq-info-3teeth}
There exists a model for the
graph of Fig.\ref{fig-3teeth}
for which $H(\rvy.:\myhat{\rvx})<0$.
\end{claim}
\proof

Consider a model for the graph
 of Fig.\ref{fig-3teeth} with
 $u_j,y_j,x\in Bool$ and

\beq
\left\{
\begin{array}{l}
P(u_j)=\frac{1}{2}
\mbox{ for }j=1,2,3
\\
P(x|u_1)=\delta_x^{u_1}
\\
P(y_3|x,u_3) = \delta_{y_3}^{x\oplus u_3}
\\
P(y_j|u_{j+1},u_j)=
\delta_{y_j}^{u_{j+1}\oplus u_j}
\mbox{ for }j=2,1
\end{array}
\right.
\;.
\eeq
Note that for this model

\beqa
P(v.)= P(y.,x) &=&
\frac{1}{2^3}\sum_{u.}
\delta_{y_3}^{x\oplus u_3}
\delta_{y_2}^{u_3\oplus u_2}
\delta_{y_1}^{u_2\oplus u_1}
\delta_{x}^{u_1}
\\
&=&
\frac{\delta_{y_3\oplus y_2\oplus y_1}^{0}}{2^3}
\;,
\eeqa
and

\beqa
P(y.|\myhat{x})
&=&
\frac{1}{2^3}\sum_{u.}
\delta_{y_3}^{x\oplus u_3}
\delta_{y_2}^{u_3\oplus u_2}
\delta_{y_1}^{u_2\oplus u_1}
\\
&=&
\frac{1}{2^3}
\;.
\eeqa
Therefore,

\beqa
H(\rvy.:\myhat{\rvx})
&=&
\sum_{y.,x}P(y.,x)\ln\frac{P(y.|\myhat{x})}{P(y.)}
\\
&=&
\sum_{y.,x}
\frac{\delta_{y_3\oplus y_2 \oplus y_1}^{0}}{2^3}
\ln \frac{
\frac{1}{2^3}
}{
\frac{1}{2^2}\delta_{y_3\oplus y_2 \oplus y_1}^{0}
}
\\
&=&
-\ln(2)
\sum_{y.}
\frac{\delta_{y_3\oplus y_2 \oplus y_1}^{0}}{2^2}
=
-\ln 2 <0
\;.
\eeqa
\qed

\begin{figure}[h]
    \begin{center}
    \epsfig{file=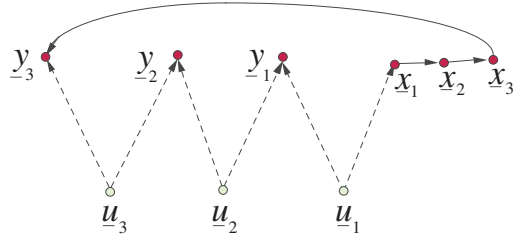, height=1.25in}
    \caption{
    Modified 3 shark teeth graph
    $G_{mod}$
    mentioned in Section \ref{sec-3teeth}.
    }
    \label{fig-3teeth-mod}
    \end{center}
\end{figure}

The results of this section
concerning the non-identifiability
of $P(y.|\myhat{x})$
for the N shark teeth graph of
Fig.\ref{fig-3teeth}
apply as well
to what I call ``modified
N shark teeth" graphs,
an example of which is given
in Fig.\ref{fig-3teeth-mod}.
For the graph $G_{mod}$
of Fig.\ref{fig-3teeth-mod},
$P(y.|\myhat{x_3})$
is not identifiable.
To show this one can
use the same models
that we used in the unmodified
case, but with $P(x_3|x_2)=\delta_{x_3}^{x_2}$,
and
$P(x_2|x_1)=\delta_{x_2}^{x_1}$.

\subsection{Example from Ref.\cite{R290L}-Fig.9}
\label{sec-tp-fig9}

\begin{figure}[h]
    \begin{center}
    \epsfig{file=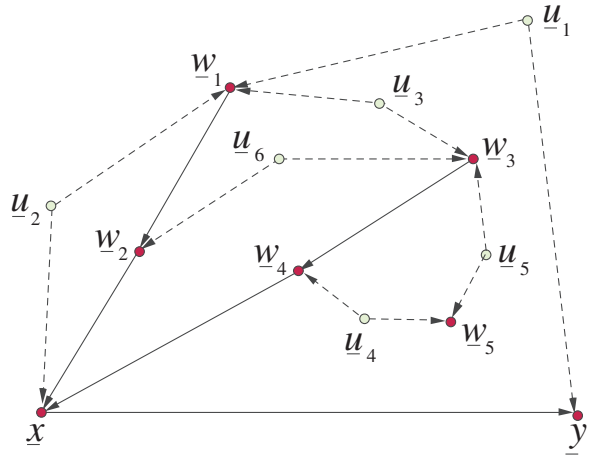, height=2.5in}
    \caption{
    Graph $G$ for Section \ref{sec-tp-fig9}.
    }
    \label{fig-tp-fig9}
    \end{center}
\end{figure}

In this example,
we show that $P(y|\myhat{x})$
is not identifiable
for the graph of Fig.\ref{fig-tp-fig9}.

For this example,
the following table applies.

\beq
\begin{array}{c||c|c|c||}
&\multicolumn{3}{c||}{\ul{\calv}.=\rvvcc{0}}
\\ \cline{2-4}
 &\rvx&\rvy&\rvw.
\\ \hline
\rvt&\checkmark &&
\\ \hline
\rvs.& &\checkmark&
\\ \hline
\rvd.& &\checkmark&
\\\hline
\end{array}
\;
\label{table-tp-fig9}
\eeq

One possible topological ordering for the
visible nodes $\rvv.$
of this graph is

\beq
\rvy\larrow\rvx
\larrow\rvw_2
\larrow\rvw_4
\larrow\rvw_1
\larrow\rvw_3
\larrow\rvw_5
\;
\eeq

According to Claim \ref{cl-pv-cc},
\beq
P(v.)=\qq{y,x,w.}
\;
\eeq
where

\beqa
\qq{y,x,w.}
&=&
\av{
\begin{array}{l}
P(y|x,u_1)
P(x|w_{2,4},u_2)
P(w_2|w_1,u_6)
P(w_4|w_3,u_4)\\
P(w_1|u_{1,2,3})
P(w_3|u_{3,5,6})
P(w_5|u_{4,5})
\end{array}
}_{u.}
\\
& =&
P(y,x|w.)P(w.)
\;.
\eeqa
Note that

\beq
P(\cald.|\calv.^{c\wedge},\myhat{t})=
P(y|\myhat{x})
\;.
\eeq

\begin{claim}\label{cl-tp-fig9}
Rule 2 (resp., Rule 3) fails to prove that
$P(y|\myhat{x})$ equals
$P(y|x)$ (resp., $P(y)$).
\end{claim}
\proof

\begin{figure}[h]
    \begin{center}
    \epsfig{file=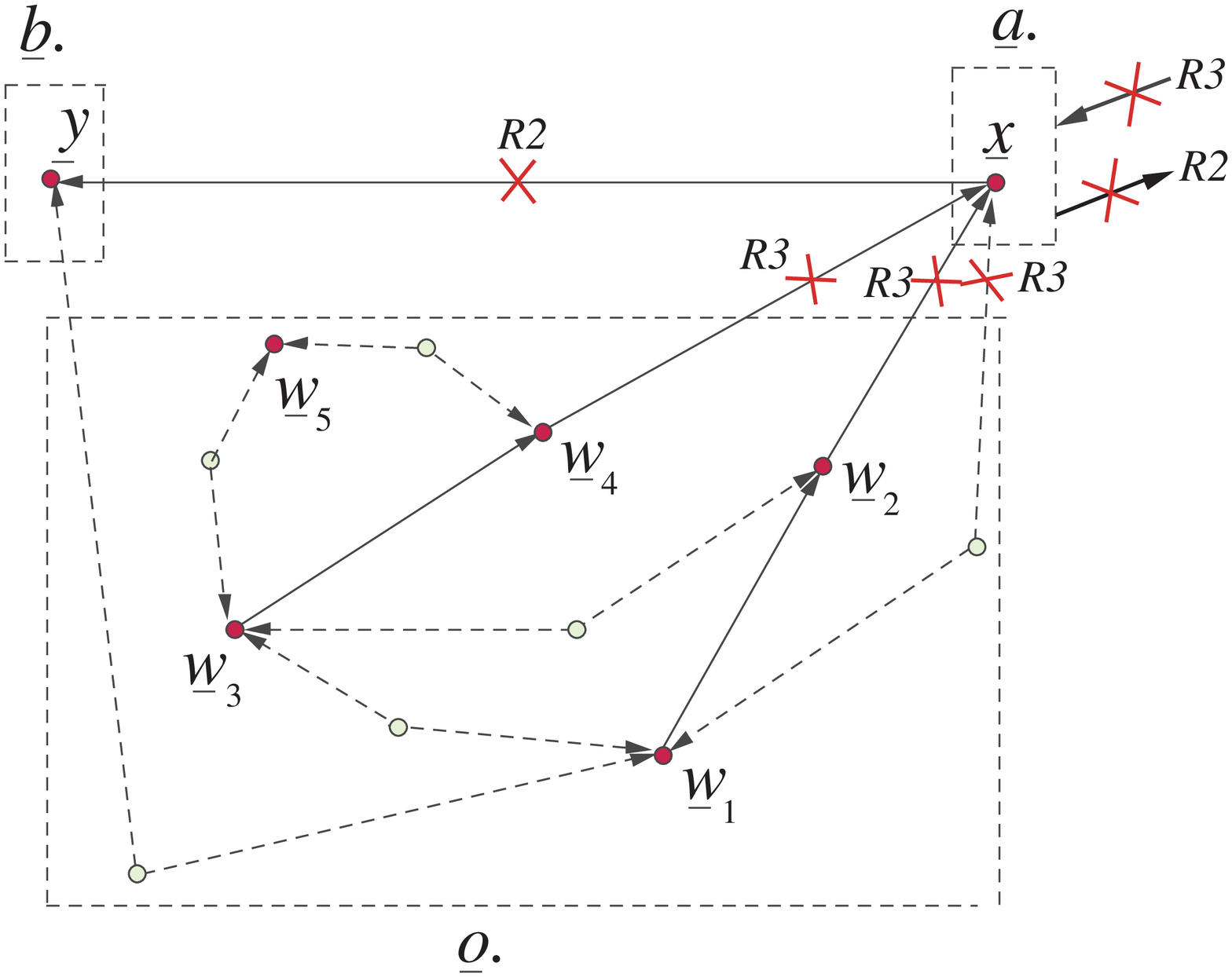, height=3.0in}
    \caption{A portrait of
    $G_{\myvee{\rva}.}$ for Rule 2 and
    $G_{\myhat{\rva}.}$ for Rule 3,
    alluded to in Claim \ref{cl-tp-fig9}.
    }
    \label{fig-d-sep-tp-fig9}
    \end{center}
\end{figure}

See Ref.\cite{Tuc-intro}
where the 3 Rules
of Judea Pearl's do-calculus
are stated. Using the notation there,
let
$\rvb.=\rvy,
\rva.=\rvx,
\rvh.=\emptyset,
\rvi.=\emptyset,
\rvo.=(\rvu.,\rvw.) $.
One can see from Fig.\ref{fig-d-sep-tp-fig9}
that
there exists an unblocked path
from $\rva.$ to $\rvb.$ at fixed $(\rvh.,\rvi.)$
in
$G_{\myhat{\rvh}.,\myvee{\rva}.}=
G_{\myvee{\rva}.}$ (resp.,
$G_{\myhat{\rvh}.,(\rva.^-)^\wedge}=
G_{\myhat{\rva}.}$) so Rule 2 (resp., Rule 3)
cannot be used.
\qed

\begin{claim}\label{cl-no-id-tp-fig9}
$P(y|\myhat{x})$ for the
graph of Fig.\ref{fig-tp-fig9} is not identifiable
\end{claim}
\proof

Consider a model for the graph
 of Fig.\ref{fig-tp-fig9}
such that

\beq
\left\{
\begin{array}{l}
P(w_j|pa(\rvw_j))=
P(w_j)\mbox{ for }j=3,4,5
\\
P(w_2|pa(\rvw_2))=
\delta_{w_2}^{w_1}
\\
P(w_1|u_{1,2,3})=P(w_1|u_1)
\\
P(x|pa(\rvx))=
\delta_x^{w_2}
\end{array}
\right.
\;.
\label{eq-probs-tp-fig9}
\eeq
For such a model,

\beq
P(x,y,w.)=
\av{
P(y|u_1,x)\delta_{x}^{w_2}
P(w_1|u_1)\delta_{w_2}^{w_1}
P(w_{3,4,5})
}_{u_1}
\;
\eeq
so

\beq
P(x,y)=
\av{
P(y|u_1,x)P(\rvw_1=x|u_1)
}_{u_1}
\;.
\eeq
This is the same $P(x,y)$
that we obtained in the one shark
tooth example
that we considered in Section
\ref{sec-1tooth}. In that
section we learned that
$P(y|\myhat{x})$
for that graph
is not identifiable.
\qed

\begin{claim}
There exists a model for the
graph of Fig.\ref{fig-tp-fig9}
for which $H(\rvy:\myhat{\rvx})<0$.
\end{claim}
\proof

In the previous claim,
we showed that for the
graph of Fig.\ref{fig-tp-fig9},
one can define a special type
of model for which
$P(x,y)$
corresponds to the
one shark tooth example of
Section
\ref{sec-1tooth}. In that section we gave a model for which
$H(\rvy:\myhat{\rvx})=-\ln(2)$.
\qed

\end{document}